\documentclass[fleqn,usenatbib]{mnras}

\usepackage{newtxtext,newtxmath}

\usepackage[T1]{fontenc}

\DeclareRobustCommand{\VAN}[3]{#2}
\let\VANthebibliography\thebibliography
\def\thebibliography{\DeclareRobustCommand{\VAN}[3]{##3}\VANthebibliography}

\usepackage{threeparttable}
\usepackage{graphicx}	
\usepackage{amsmath}	
\usepackage{subcaption}
\usepackage{scalerel}

\def\rhalf{\mbox{$r_{1/2}$}}

\def\Msun{\, M_{\odot}}
\def\Msunpc2{\,\rm M_{\odot}\,pc^{-2}}
\def\Zsun{\, Z_{\odot}}

\def\fb{f_{\rm b}}

\def\Rtwoh{R_{\rm 200c}}
\def\Mtwoh{M_{\rm 200c}}

\def\Msol{\,M_\odot}

\def\yZ{y_{\rm Z}}

\def\Gyr{\,\rm Gyr}

\def\etaw {\eta_{\rm w}}
\def\zetaw{\zeta_{\rm w}}
\def\sfr{\dot{\mathcal{M}}_{\star}}
\def\SFR{\dot{M}_{\star}}
\def\Rloss{\mathcal{R}}
\def\oneRloss{(1-\Rloss)}
\def\sfrloss{\oneRloss\,\sfr}
\def\dotMgin{\dot{M}_{\rm g,in}}
\def\dotMg{\dot{M}_{\rm g}}
\def\Mh{M_{\rm h}}
\def\Mhp{M_{\rm h,peak}}

\def\Ms{M_{\star}}
\def\Mg{M_{\rm g}}
\def\Rd{R_{\rm d}}
\def\HH{\rm H_2}
\def\Sigg{\Sigma_{\rm g}}
\def\Sigsfr{\dot{\Sigma}_{\star}}
\def\SigHH{\Sigma_{\rm H_2}}
\def\SigHI{\Sigma_{\rm HI}}
\def\MHI{M_{\rm HI}}
\def\MHH{M_{\rm H_2}}

\def\dotMh{\dot{M}_{\rm h}}
\def\dotMout{\dot{M}_{\rm g,out}}
\def\dotMs{\dot{M}_{\star}}
\def\Zigm{Z_{\rm IGM}}
\def\Zg{Z_{\rm g}}
\def\Zs{Z_\star}
\def\Zw{Z_{\rm w}}
\def\MZg{M_{Z,\rm g}}
\def\MZs{M_{Z,\star}}
\def\dotMZg{\dot{M}_{Z,\rm g}}
\def\dotMZs{\dot{M}_{Z,\star}}
\def\UMW{U_{\rm MW}}
\def\DMW{D_{\rm MW}}
\def\SigHIth{\Sigma_{\rm HI,th}}
\def\fHH{f_{\rm H_2}}
\def\tauHH{\tau_{\rm dep,H_2}}

\def\epsin{\epsilon_{\rm in}}
\def\epsincoldhot{\epsilon_{\rm in,ch}}
\def\epsinuv{\epsilon_{\rm in,UV}}
\def\epsinprev{\epsilon_{\rm pr}}
\def\chid{\chi_{\rm d}}
\def\zpeak{z_{\rm peak}}
\def\Msten{M_{\star,10}}
\def\etapow{\eta_{\rm p}}
\def\etanorm{\eta_{\rm norm}}
\def\etac{\eta_{\rm c}}
\def\zrei{z_{\rm rei}}
\def\Mhalf{\mathcal{M}_{50}}

\title[GRUMPY: a simple dwarf galaxy modelling framework]{GRUMPY: a simple framework for realistic forward-modelling of dwarf galaxies}

\author[A. Kravtsov \& V. Manwadkar]{
Andrey Kravtsov$^{1,2,3}\thanks{E-mail: kravtsov@uchicago.edu}$ and Viraj Manwadkar$^1$
\\
$^{1}$Department of Astronomy  \& Astrophysics, The University of Chicago, Chicago, IL 60637 USA\\
$^{2}$Kavli Institute for Cosmological Physics, The University of Chicago, Chicago, IL 60637 USA\\
$^{3}$Enrico Fermi Institute, The University of Chicago, Chicago, IL 60637 USA
}

\date{Accepted XXX. Received YYY; in original form ZZZ}

\pubyear{2020}

\begin{document}
\label{firstpage}
\pagerange{\pageref{firstpage}--\pageref{lastpage}}
\maketitle

\begin{abstract}
We present a simple regulator-type framework designed specifically for modelling formation of dwarf galaxies.
Despite its simplicity, when coupled with realistic mass accretion histories of haloes from simulations and reasonable choices for model parameter values, the framework can reproduce a remarkably broad range of observed properties of dwarf galaxies over seven orders of magnitude in stellar mass.  In particular, we show that the model can {\it simultaneously} match observational constraints on the stellar mass--halo mass relation, as well as observed relations between stellar mass and gas phase and stellar metallicities, gas mass, size, and star formation rate, as well as general form and diversity of  star formation histories (SFHs) of observed dwarf galaxies. The model can thus be used to predict photometric properties of dwarf galaxies hosted by dark matter haloes in $N$-body simulations, such as colors, surface brightnesses, and mass-to-light ratios and to forward model observations of dwarf galaxies. We present examples of such modelling and show that colors and surface brightness distributions of model galaxies are in good agreement with observed distributions for dwarfs in recent observational surveys. We also show that in contrast with the common assumption, the absolute magnitude-halo mass relation is generally predicted to have a non-power law form in the dwarf regime, and that the fraction of haloes that host detectable ultrafaint galaxies is sensitive to reionization redshift ($\zrei$) and is predicted to be consistent with observations for $\zrei\lesssim 9$. 
\end{abstract}




\section{Introduction}
\label{sec:intro}

Studies of dwarf galaxies are at the current frontier of studies of galaxy formation and are an important testing ground for galaxy formation models and properties of dark (DM) particles \citep[see, e.g.,][for a review]{bullock_boylankolchin17}. Their faintness and low surface brightness make them difficult to detect, but modern detection and followup strategies allowed observers to discover extremely low luminosity \citep[e.g.,][]{Koposov.etal.2015,Drlica_Wagner.etal.2015} and low surface brightness \citep[e.g.,][]{Torrealba.etal.2019} systems. 

Indeed, the last two decades resulted in many discoveries of new dwarf galaxies and surprising findings about many of their properties, such as extremely low luminosities and metallicities of the ultrafaint dwarfs (UFDs) that make some of them barely distinguishable from small-mass star clusters \citep[e.g., see][for a review]{simon19}. Most of these discoveries were done within the Local Group \citep[hereafter LG, e.g.,][for a recent census]{drlica_wagner_etal20}, but recent \citep[][]{Geha.etal.2017,Carlsten.etal.2020,Mao.etal.2021} and upcoming observational surveys and facilities, such as the Vera C. Rubin Observatory\footnote{\href{https://www.lsst.org/}{https://www.lsst.org/}} \citep[][]{Ivezic.etal.2019}, will greatly expand the volume in which ultra-faint galaxies are detectable.  

Shallow potential wells of dwarf galaxies imply that effects of feedback processes known to play a key role in shaping properties of more massive galaxies should be particularly pronounced in the dwarf regime \citep[e.g.,][]{Dekel.Silk.1986,Efstathiou.1992,Efstathiou.2000,MacLow.Ferrara.1999,Pontzen.Governato.2012}. Observed properties of dwarf galaxies can thus be used to test and constrain the physics of galaxy star formation and feedback \citep[e.g.,][]{Tassis.etal.2012,Munshi.etal.2019,Agertz.etal.2020}. 
Given that many aspects of dwarf galaxy populations are still being discovered, models also have a unique opportunity to make {\it pre}dictions in this regime, such as successful prediction of the existence of the ultrafaint dwarf (UFD) population of galaxies \citep{Ricotti.Gnedin.2005,Gnedin.Kravtsov.2006,Bovill.Ricotti.2009}.

There was a tremendous recent progress in the ability of simulations to reproduce observed properties of galaxies \citep[see][for a review]{Naab.Ostriker.2017}. Even more impressively, several studies showed that cosmological simulations with star formation and feedback calibrated to reproduce properties of $\gtrsim L_\star$ galaxies also produce reasonably realistic 
population of dwarf galaxies \citep[e.g.,][]{Wang.etal.2015,Wetzel.etal.2016,Wheeler.etal.2019,Simpson.etal.2018,Font.etal.2020,Font.etal.2021,Munshi.etal.2021,Applebaum.etal.2021}.
At the same time, it was shown that galaxy formation simulations face a number of uncertainties and numerical challenges in the dwarf galaxy regime \citep{Munshi.etal.2019,Mina.etal.2020}.
This is not surprising both because we are not yet certain how to model many of the relevant processes and because the dynamic range required to model processes relevant for ISM structure, star formation and feedback of dwarf galaxies forming in a cosmological volume is formidable. 
In addition, simulations are computationally expensive and require a significant effort to run and analyze. This makes modelling large samples of galaxies or tailoring model predictions for interpretation of a particular observational survey difficult. 

Phenomenological ``semi-empirical'' models or halo occupation distribution based models of galaxy-halo connection \citep[see][for a review]{Wechsler.Tinker.2018} are another versatile and inexpensive tool that is often used to interpret observations of dwarf galaxies using large $N$-body simulations. Such approaches proved to be particularly useful in modelling abundance of dwarf galaxies as a function of luminosity in the Local Group \citep[see, e.g.,][for recent examples]{Jethwa.etal.2018,Graus.etal.2019,Nadler.etal.2020}. However, models of this kind rely on  availability of luminosity  function or specific star formation rate scaling with stellar mass to inform mapping observable properties to properties of haloes  \citep[e.g.,][for a recent example]{Wang.etal.2021}.  The usability and predictive power of these models is thus limited by lack of calibration data in the dwarf galaxy regime at both $z=0$ and earlier epochs. 

Semi-analytic models are a traditional tool for modelling dwarf galaxies \citep[][]{Kauffmann.etal.1993,Somerville.2002,Benson.etal.2002a,Benson.etal.2002b} and avoid both issues of complexity and computational expense of galaxy formation simulations and the lack of calibration data of semi-empirical models by modelling evolution of key galaxy properties using a system of differential equations \citep[see][for reviews]{Benson.2010,Somerville.Dave.2015}. Such models were used productively to model dwarf satellites observed in the Local Group using modern generation of high-resolution zoom-in $N$-body simulations \citep[e.g.,][]{starkenburg_etal13,bose_etal18}. 

During the last decade, a number of studies introduced and used much simplified ``bathtub'' or ``regulator'' versions of semi-analytic model that strip down modelling down to the minimal number of equations following mass balance of gas in the interstellar medium (ISM) of galaxies \citep[e.g.,][]{Finlator.Dave.2008,Bouche.etal.2010,Krumholz.Dekel.2012,Lilly.etal.2013,Feldmann.2013,Peng.Maiolino.2014,Birrer.etal.2014,Furlanetto.etal.2017}. Despite their simplicity, models of this type proved to be successful in matching many key observational scaling relations and their evolution \citep[see,e.g.,][for a review]{Tacconi.etal.2020}. Several studies have shown that such framework can also be useful for interpretation of halo mass--stellar mass, stellar mass--metallicity correlations, and star formation histories  in the dwarf galaxy regime \citep[][]{Tassis.etal.2012,Ledinauskas.Zubovas.2018,Ledinauskas.Zubovas.2020}, especially if mass accretion histories extracted from cosmological simulations are used to model evolution of halo mass and provide environmental information about distance to the host and tidal interactions \citep[e.g.,][]{Kravtsov.etal.2004,Orban.etal.2008}. 

In this study we describe a regulator-type model, which uses input halo mass accretion rate and models evolution of four components: ISM gas mass, stellar mass, and metallicity of the ISM and stars. We complement the standard features of such model with a model of star formation based on molecular hydrogen, as well as effects of UV heating of IGM and feedback-driven outflows, that allows one to make a larger number of quantitative predictions. We show that the model integrated using mass accretion histories extracted from the ELVIS simulations \citep[][]{GarrisonKimmel.etal.2014} reproduces a broad range of observational properties and constraints in the dwarf galaxy regime down to the faitest dwarf galaxies detected observationally, including $\Mh-\Ms$, stellar mass--metallicity relations for both ISM and stars, ISM gas fractions and star formation rates and histories, etc. 

The paper is organized as follows. We describe the model in Section~\ref{sec:model} and motivate different modelling and parameterization choices. We present comparisons with various observations in Section~\ref{sec:comparisons}, discuss our results and compare to other relevant studies in Section \ref{sec:discussion}, and summarize our results and conclusions in Section~\ref{sec:summary}. In the Appendix~\ref{app:mc_oka} we present justification for the expression for the characteristic mass for gas accretion suppression due to UV heating, while in  Appendix~\ref{app:mhh2} we show additional results illustrating sensitivity of gas masses and star formation rates to details of molecular hydrogen modelling.

\section{Galaxy evolution model}
\label{sec:model}

The basic approach of our model is similar to that of other previous regulator models \citep[e.g.,][]{Krumholz.Dekel.2012,Lilly.etal.2013,Feldmann.2013}, but with some important differences which will be outlined below. As in all ``regulator''-type models, galaxy evolution is modelled using a small set of coupled differential equations accounting for mass conservation of different components and their conversions into each other. 

For instance, the evolution of the ISM gas mass, $M_{\rm g}$, is governed by its rate of change due to the inflow of new gas, $\dotMgin$, rate of gas conversion into stars, $\dotMs$, and the rate of outflows, $\dotMout$, driven by feedback:
\begin{equation}
\dotMg = \dotMgin - \dotMs - \dotMout.
\label{eq:mgevo}
\end{equation}
We will discuss modelling of these three terms in Sections~\ref{sec:mhalo}-\ref{sec:zevo}.
In particular, the model for $\dotMs$ provides an additional model equation governing evolution of galaxy stellar mass. We will complement the equations for $\dotMg$ and $\dotMs$ with two additional equations modelling evolution of mass of heavy elements in the ISM gas and stars, $\dotMZg$ and $\dotMZs$, respectively (Section~\ref{sec:zevo}). 


\subsection{Modelling evolution of halo mass}
\label{sec:mhalo}

The backbone of the model is the rate of change of the total gravitating mass of the halo (baryons and DM): $\dotMh$. In this study we use the rates $\dotMh(t)$ extracted for a sample of haloes from cosmological simulations of halo formation.

Halo mass evolution exhibits systematic trends with mass and redshift, which can be described by relatively simple analytic approximations \citep[e.g.,][]{Krumholz.Dekel.2012}.
However, such average evolution does not take into account diversity of mass accretion histories of real objects and in particular significant and rapid changes in the mass accretion rate associated with mergers and/or periods of high mass accretion rate. The diversity of halo mass accretion histories (MAHs) is a basic source of scatter in galaxy properties, which can be particularly large in the ultra-faint regime due to suppression of gas accretion and star formation after the epoch of reionization \citep[e.g.,][]{Rey.etal.2019,Katz.etal.2020}. 

To account for the diversity of MAHs of dwarf galaxies, in this study we use the MAHs extracted from the ELVIS suite of high-resolution simulations \citep{GarrisonKimmel.etal.2014}. These simulations were following a ``zoom-in'' high-resolution regions of $\sim 1 $ Mpc around twenty four isolated MW-sized haloes and twelve pairs of MW-sized haloes separated by $\sim 800$ kpc using GADGET-2 and GADGET-3 codes \citep{Springel.2005} in a box of 70.4 Mpc on a side. Simulations were initialized using  MUSIC code \citep{hahn_abel11} with adopted flat $\Lambda$CDM model with the following parameters: $\Omega_{\rm m} = 0.266$, $\Omega_\Lambda =
0.734$, $n_s = 0.963$, $\sigma_8 = 0.801$, and $h = H0/100=0.71$. With these parameters the particle mass within that high-resolution volume is $m_{\rm p}=1.9\times  10^5\, M_\odot$, while the
Plummer-equivalent force softening was held fixed in physical coordinates at $z<9$ at 141 pc. Three of the isolated halos were also simulated with eight times
more particles in the high-resolution regions ($m_{\rm p} = 2.35\times 10^4\,M_\odot$, and Plummer softening of $70.4$ pc) and we use two of these as HR simulations in this paper.

Although in the future studies employing the model presented here, it will be important to model different aspects of host halo environment and/or effects of central galaxy disk, in this study we focus on the evolution of dwarf haloes in the vicinity of isolated Milky Way-sized haloes \citep[i.e., the suite of isolated simulations of][]{GarrisonKimmel.etal.2014}.

In practice, we use evolution tracks of haloes and subhaloes that exist at $z=0$ in the vicinity of a Milky Way-sized halo extracted from simulations using the Consistent Trees code \citep{Behroozi.etal.2013}. The tracks are in the form of several halo properties, such as its virial mass, scale radius, maximum circular velocity, etc., measured at a series of redshifts from the first epoch at which progenitors to $z=0$. 

The halo mass in the evolutionary tracks is defined using the ``virial'' density contrast relative to the mean density, $\Delta_{\rm vir}$, derived using tophat spherical collapse model and equal to $\simeq 18\pi^2$ for epochs when $\Omega_{\rm m}\simeq 1$  and is progressively larger for smaller values \citep[e.g.,][]{Lahav.etal.1991}. This mass in this definition can change significantly due to the evolution of the mean background density even when halo itself does not accrete any new mass -- the mass {\it pseudo-evolution} \citep[see][for a detailed discussion]{Diemer.etal.2013}. Given that the rate of halo mass change is used to define the overall accretion rate of baryons, as described below, to minimize the effects of such mass pseudo-evolution, we convert 
the halo mass to the mass within radius enclosing the density contrast of $\Delta_{\rm c}=200$ relative to {\it the critical density} of the universe at the corresponding epochs. This mass, $\Mh=M_{200,\rm c}$ in this definition will be used throughout in our model. $M_{200\rm c}$ suffers much less from the pseudo-evolution effect, because critical density in the $\Lambda$CDM evolves relatively little at low redshifts. The conversion from $M_{\rm vir}$ to $M_{200\rm c}$ at each epoch is done assuming NFW profile for halo mass distribution with the scale radius, $r_{\rm s}$, and $M_{\rm vir}$ provided by the halo track at that epoch. 

To provide a continuous representation of the evolution of halo mass for numerical model integration, we first  approximate $\log M_{\rm 200c}$ as a function of $\log t$ using the values of mass at the $t_i$ available in the track and interpolating using cubic spline. We then evaluate numerical derivative  $d\ln M_{\rm 200c}/d\ln t$ and enforce the monotonicity condition $d\ln M_{\rm 200c}/d\ln t\geq 0$ at the times $t_i$ to ensure that monotonic evolution of mass. Although mass of real haloes can decrease during periods of mergers of tidal stripping, we only use the mass evolution rate to model the baryon accretion rate, which should become zero when halo mass does not grow, but should not be negative during such stages of evolution. Finally, we approximate the $d\ln M_{\rm 200c}/d\ln t$ subject to the monotonicity condition as a function of cosmic time $t$ using the cubic spline. This spline approximation is used during the model integration to compute $\dotMh=(\Mh(t)/t)\,d\ln \Mh/d\ln t$, where we will assume $\Mh(t)\equiv M_{200c}(t)$ throughout the rest of the paper. 

\subsection{Modelling evolution of gas inflow}
\label{sec:mgin}

Although the processes that govern cooling and inflow of gas onto the ISM of galaxies can be complex, we only include approximate treatments of the most important processes: the cold mode accretion of gas, suppression of gas accretion due to UV heating, and slow down or complete suppression of inflow of gas due to formation of hot gaseous halo once the total mass of the object exceeds the transition mass between cold and hot accretion modes. 

Specifically, gas inflow rate onto galaxy is assumed to be proportional to the total mass accretion rate, $\dotMh$:
\begin{equation}
\dot{M}_{\rm g,in} = \epsin\,\frac{\Omega_{\rm b}}{\Omega_{\rm m}}\, \dotMh
\label{eqn:Mgin}
\end{equation}
where $\Omega_{\rm b}$ and $\Omega_{\rm m}$ are the mean baryon and total matter densities in units of the critical density, total mass accretion rate, $\dotMh$, is computed as described above, and the factor $\epsin$ parameterizes the fraction of the univeral baryon mass fraction, $\Omega_{\rm b}/\Omega_{\rm m}$, that is accreted by the object in gaseous form.  

The factor $\epsin$ encapsulates effects of different processes suppressing gas inflow mentioned above. We model it as a product of several factors to account for these processes:
\begin{equation}
\epsin=\epsincoldhot\,\epsinuv\,\epsinprev,
\label{eq:epsin}    
\end{equation}
where $\epsincoldhot$ is a fraction of gas accreted onto halo in the ``cold-mode'' and ``hot-mode'' regimes, $\epsinuv$ is to account suppression of gas accretion due to UV heating after reionization, and $\epsinprev$ accounts for any gas accretion suppression due to ``preventative'' feedback \citep[e.g.,][]{Mo.Mao.2002,Mo.Mao.2004,Oh.Benson.2003}. 

The effect of preventative feedback is generally expected to be manifested in a suppression of 
gas accretion, but the mass dependence of such suppression is poorly understood. In this study we simply model $\epsinprev$ as a constant with the fiducial value of one, although we also explored models with $\epsinprev=0.6$ and other values.  
Below we discuss factors $\epsincoldhot$ and $\epsinuv$ in more detail. 

\subsubsection{Suppression of gas inflow in the hot accretion mode}

The cold accretion is expected to be the main mode of gas accretion in dwarf galaxy regime because dwarf-scale objects are not expected to maintain hot gaseous haloes and cooling of gas is efficient for objects with virial temperatures $\lesssim 10^5$ K. In this mode, gas accreted by the halo accretes onto galaxy's ISM on a time scale close to the free-fall time at the virial radius \citep[see, e.g.,][]{Rosdahl.Blaizot.2012}. In this regime we thus simply assume that $\epsincoldhot=1$.

In the ``hot-mode'' accretion we assume that accreted gas is shock heated and settles into equilibrium in hot gaseous halo. Transition to the hot mode-regime is thought to be accompanied with the onset of rapid growth of the central black hole which then continuously heats the halo gas via outflows \citep[e.g.,][]{Bower.etal.2017}, thereby shutting down accretion onto the galaxy ISM. Thus, in the fully hot-mode accretion $\epsincoldhot=0$.  

We assume a simple model for the transition from the cold to hot accretion regime, where the $\epsin$ factor is modulated by a soft step function of the form a fixed mass threshold, $M_{\rm hot}$:
\begin{align}
\epsincoldhot &= 1-s(\mu_{\rm hot},\phi); \label{eq:epsincoldhot} \\
\mu_{\rm hot}&=\Mh/M_{\rm hot};\\
s(x,y) &=\left[1+\left(2^{y/3}-1\right)x^y\right]^{-3/y}\label{eq:sstep}
\end{align}
with $\phi=4$ and $M_{\rm hot}=4\times 10^{11}\Msun$. This threshold is close to the value estimated in analytical models and cosmological simulations of galaxy formation, which only weakly depends on redshift \citep[e.g.,][]{Birnboim.Dekel.2003,Keres.etal.2005,Keres.etal.2009,Dekel.Birnboim.2006,Dekel.etal.2009}. This approximation is sufficient for the purposes of modelling dwarf galaxies with $\Mh<M_{\rm hot}$.

\subsubsection{Suppression of gas inflow due to UV heating}
\label{ssec:uv_heating}

The effect of UV heating after reionization is modelled using approximation to the results of cosmological simulations of \citet[][cf. also \citealt{Hoeft.etal.2006}, \citealt{Naoz.etal.2013}, \citealt{Noh.McQuinn.2014}, and \citealt{Dawoodbhoy.etal.2018}]{Okamoto.etal.2008}. This process was not included in the KD12 model, but is important for proper modelling of dwarf galaxies \citep[e.g.][]{Efstathiou.1992,Bullock.etal.2000}, especially in the ultra-faint regime \citep[e.g.,][]{Ricotti.Gnedin.2005,RodriguezWimberly.etal.2019,Hutter.etal.2020}. 

UV heating results in the suppression of gas inflow rate, thereby quenching star formation in dwarf galaxies and also affecting the faint end of the galaxy luminosity function \citep[e.g.,][]{Efstathiou.1992,Bullock.etal.2000}. The $\epsinuv$ parameter in eq.~\ref{eqn:Mgin} contains information on the suppression of gas inflow rate due to UV heating. To calculate it, we first consider the baryonic fraction $\fb$ evolution in haloes of different mass.

According to \citet{Gnedin.2000}, the baryonic fraction $\fb$ in haloes can be expressed as a function of their mass $\Mh$ and redshift $z$ as 
\begin{equation}
\fb (\Mh, z) =\frac{\Omega_{\rm b}}{\Omega_{\rm m}}\,s\left(\mu_{\rm c},\omega\right);\ \ \ \mu_{\rm c}=\Mh/M_{\rm c}(z)
\label{eqn:gnedin_fb}
\end{equation}
where $M_{\rm c}(z)$ is the characteristic mass scale at which baryon fraction is suppressed by a factor of two relative to the universal value. Haloes with masses larger than $M_{\rm c}(z)$ have baryonic fractions closer to the universal baryonic fraction $\Omega_{\rm b}/\Omega_{\rm m}$. While, baryonic fractions for haloes with masses smaller than $M_{\rm c}(z)$ rapidly fall to zero as a function of decreasing halo mass. \citet{Okamoto.etal.2008} showed that the functional form of this dependence in their simulations is well approximated by the function $s$ in eq.~\ref{eq:sstep} with $\omega=2$. 

The data in Figure 3 of \citet{Okamoto.etal.2008} for the characteristic mass scale $M_c(z)$ can be approximated as an exponential modified by a sigmoid function to account for the sharp increase in characteristic mass scale at reionization:
\begin{equation} 
M_{\rm c}(z) =  1.69 \times 10^{10}\,\Msol\,\frac{\exp(-0.63 z)}{1 + \exp([z/8.7]^{15})}  
\label{eqn:Mcz}
\end{equation}
The functional form of eq.~\ref{eq:sstep} with this characteristic mass also approximates results of reionization simulations of \citet[][see their Fig. 7]{Gnedin.Kaurov.2014}.  

Note that in the simulations of \citet{Okamoto.etal.2008}, reionization is assumed to occur at $z = 9$. To make our model applicable for any redshift of reionization $\zrei$, we apply the constraint that the characteristic mass scale at $z=\zrei$  should be the same as in the simulations of \citet{Okamoto.etal.2008} at $z=9$, while we keep $M_{\rm c}(z=0)$ to be the same for all $\zrei$ values. In addition to this, we introduce a parameter controlling how fast $M_{\rm c}$ increases at $z\approx\zrei$. Such parameter can be used to model reionization that occurs faster than reionization by the mean cosmic UV background, as could occur by an onset of a strong local UV source. We thus use a general expression for characteristic mass with reionization redshift $\zrei$ and sharpness parameter $\gamma$
\begin{align}
    M_c(z | \zrei, \gamma ) &= 1.69\times10^{10}\,\Msun\, \frac{\exp(-0.63 z)}{1 + \exp([z/\beta]^{\gamma})}
    \label{eq:Mc_zre}
\end{align}
where
\begin{equation}
    \beta = \zrei \left[  \ln\left( 1.82\times 10^3\,\exp(-0.63\zrei) - 1 \right) \right]^{-1/\gamma}
    \label{eq:betarei}
\end{equation}
Evolution of the characteristic mass for different reionization redshifts given by these questions and its comparison with cosmological simulations of \citet{Okamoto.etal.2008}, in which $\zrei=9$ was assumed, is shown in Figure~\ref{fig:mc_oka} in Appendix~\ref{app:mc_oka}.

To solve for $\epsinuv$ in eq.~\ref{eqn:Mgin} that results in agreement with the baryonic fractions $\fb(\Mh, z)$ described by eq.~\ref{eqn:gnedin_fb}, we differentiate $\Mg(z) = \fb (\Mh, z) \Mh$ with respect to time. Note that the simulations in \citet{Okamoto.etal.2008} only considered gas (no stars) as the baryonic component. So, the equation $M_g(z) = \fb(\Mh, z) \Mh$ is valid and differentiating it indeed does give us the effect of UV heating on gas inflow rate. 

The combination of the above effects results in the combined equation for gas inflow rate: 
\begin{multline}
\epsinuv = \max\bigg( 0,s\left(\mu_{\rm c},\omega\right)\Big[(1+X) - \\
2\epsilon(z,\gamma)\, \frac{\Mh}{\dotMh}\, X (1+z) H(z)  \Big] \bigg),
\label{eq:epsinuv}
\end{multline}
where again $\mu_{\rm c}(z)=\Mh/M_{\rm c}(z)$, $\beta$ is given by eq.~\ref{eq:betarei} and 
\begin{align}
\epsilon(z,\gamma) &= \frac{0.63}{1 +e^{(z/\beta)^{\gamma}}} + \frac{\gamma z^{\gamma-1}}{\beta^{\gamma}} \frac{e^{(z/\beta)^{\gamma}} }{(1 +e^{(z/\beta)^{\gamma}})^2} \\
X &= \frac{3 c_{\omega} M_{\omega} }{ 1 + c_{\omega} M_{\omega}} \\
c_{\omega} &= 2^{\omega/3} - 1 \quad\text{,}\quad M_{\omega} = \left( \frac{M_c(z)}{M_h}  \right)^{\omega} \quad\text{,}\quad \omega = 2
\end{align}
and $H(z)$ is the Hubble function given by 
\begin{equation}
H(z) = H_0\,\sqrt{\Omega_{\rm m} (1+z)^3 + \Omega_\Lambda}.
\label{eq:Ez}
\end{equation}
Note that in eq.~\ref{eq:epsinuv}, we have $\epsinuv \geq 0$ only as we do not allow the gas to photo-evaporate from haloes once accreted. Physical motivation for this assumption is that gas cooling in small-mass haloes is expected to be fast and we assume that gas joins the galaxy ISM shortly after it is accreted. We assume that once the gas is in the ISM, it cannot be evaporated by UV background. Note that although evaporation was detected in simulations of \citet{Okamoto.etal.2008}, these simulations did not include cooling and thus did not model gas cooling and its accretion onto the ISM. 

\begin{figure}
    \centering
    \includegraphics[width=0.5\textwidth]{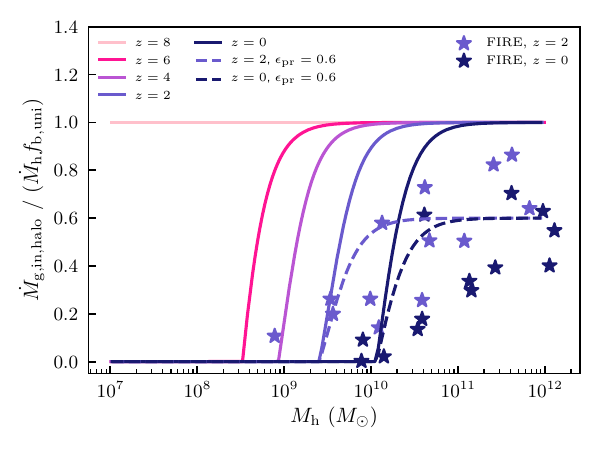}
    \caption{Ratio of the gas accretion rate, $\dotMgin$, and total mass accretion rate normalized by the universal baryon fraction through the virial radius for haloes of different mass at a number of redshifts in our model for $\epsinprev=1$ and $\epsinprev=0.6$ and compares them to measurements of this ratio the FIRE-2 simulations \citep[][]{Pandya.etal.2020}}
    \label{fig:mgin_ratio}
\end{figure}

Figure~\ref{fig:mgin_ratio} shows the ratio of the gas accretion rate, $\dotMgin$, and total mass accretion rate normalized by the universal baryon fraction through the virial radius for haloes of different mass at a number of redshifts in our model for $\epsinprev=1$ and $\epsinprev=0.6$ and compares them to measurements of this ratio the FIRE-2 simulations \citep[][]{Pandya.etal.2020}. 

\begin{table*}
	\centering
	\caption{Key model components and related parameters.}
	\label{tab:paramtable}
	\begin{threeparttable}
	\begin{tabular}{p{0.17\linewidth}p{0.19\linewidth}p{0.5\linewidth}}
		\hline\hline
		Model/parameters & Fiducial value\tnote{(a)} & Description\\
		\hline
		\textit{Gas Inflow Model} &&   \\
		$\epsincoldhot$ & Equation~\ref{eq:epsincoldhot}\tnote{ (b)} & fraction of gas accretion rate that accretes directly onto the ISM in the cold-mode regime   \\
		$M_{\rm hot}$ & $4\times10^{11}\Msol$ & characteristic halo mass scale of the transition from the cold- to hot-mode accretion regime   \\
		$\epsinuv$ & Equation~\ref{eq:epsinuv}\tnote{ (c)} & Factor accounting for gas accretion suppression due to UV heating after reionization  \\
		$\epsinprev$ & $1.0$ & Factor accounting for gas accretion suppression due to preventative feedback \\
		$\epsin=\epsincoldhot\,\epsinuv\,\epsinprev$ & Equation~\ref{eq:epsin} & Fraction of $\Omega_{\rm b}/\Omega_{\rm m}\,\dotMh$ accreted by the galaxy ISM \\[1mm]
		\textit{Reionization Model} &&  \\
		$\zrei$ & $6$ & redshift of reionization \\
		$\gamma$ & $15$ & controls the sharpness of the increase in characteristic mass scale at time of reionization \\[1mm]
		\textit{Gas Disk Model}&& \\
		$\chid$ & $0.06$ & Proportionality constant between gas disk scale length $\Rd$ and $\Rtwoh$.  \\
		$\log_{10}(r_{\rm pert})$ & $\mathcal{N}(0, \sigma_{\log R}^2)$; $\sigma_{\log R}=0.25$  & Perturbation factor of the disk size for individual haloes from the mean $R_{\rm d}\propto \Rtwoh$ relation; constant for each halo during evolution  \\
		$\SigHIth$ & $0.2\,\Msun\,\rm pc^{-2}$ & Threshold surface density value of neutral hydrogen above which it becomes self-shielding to UV radiation \\[1.5mm]
		\textit{Star Formation Model} && \\
		$\tauHH$ & $2.5\Gyr$ & Molecular hydrogen gas depletion time\\
		$\Rloss$ & $0.44$ & Mass fraction of newly formed stars instantaneously returned to the ISM \\[1.5mm]
		\textit{Galactic Outflows Model} && \\
		$\etaw$ & Equation~\ref{eq:etamodel} & Mass loading factor  \\
		$\etanorm$ & $3.6$ & Normalization of the mass loading factor \\
		$\etapow$ & $-0.35$ & Power-law exponent in the power law dependence of mass loading factor on $\Ms$ \\
		$\etac$ & $4.5$ & Constant term in the mass loading factor expression  \\
		\textit{Chemical evolution model} &&  \\[1.5mm]
		$\Zigm$ & $10^{-3}\Zsun$ & Characteristic metallicity of heavy elements accreted by the galaxy \\
		$\zetaw=Z_{\rm wind}/\Zg$ & $=1.0$ & Wind metallicity enhancement factor \\
		$\yZ$ & $=0.06$ & Nucleosynthetic yield of heavy elements per unit star formation rate \\
		\hline
	\end{tabular}
	\begin{tablenotes}\footnotesize
    \item[(a)] numerical values indicate that parameter is constant with the value specified in the table. 
	\item[(b)] a function of $\Mh$
	\item[(c)] a function of $\Mh$, $z$, $\zrei$
    \end{tablenotes}
	\end{threeparttable}
\end{table*}

\subsection{Model for the surface density profile of gas}
\label{sec:sigrmodel}

To estimate star formation rate of galaxies and their atomic and molecular content, we model distribution of the ISM gas assuming the exponential profile
\begin{equation}
\Sigg = \Sigma_0\,\exp\left(-\frac{R}{\Rd}\right).
\end{equation}
The approximately exponential distribution of gas disk is broadly expected in theoretical models \citep[e.g.,][]{Bullock.etal.2001} and is consistent with observed distribution of gas in late type galaxies \citep[][]{Bigiel.Blitz.2012,Kravtsov.2013}.

The mass within a given radius for the exponential profile is given by:
\begin{eqnarray}
M(<R)&=&2\pi\Sigma_0\int\limits_0^R R^\prime\exp(-R^\prime/\Rd) dR^\prime\nonumber\\
&=&2\pi\Sigma_0 \Rd^2\left[1-\frac{R}{\Rd}e^{-R/\Rd}-e^{-R/\Rd}\right]
\label{eq:mrexp}
\end{eqnarray}
and thus the total mass ($R=\infty$) is $\Mg=2\pi\Sigma_0 \Rd^2$ so that the central surface density is
\begin{equation}
\Sigma_0=\frac{\Mg}{2\pi \Rd^2}.
\end{equation}

The mass $\Mg$ in the above equations is the total ISM gas mass tracked by the model eq.~\ref{eq:mgevo}. To specify $\Sigma(R)$, we assume that at 
each redshift $z$ during evolution $\Rd$ is proportional to $\Rtwoh(z)$:
\begin{equation}
\Rd = \chid \Rtwoh(z) =  162.63\,{\,\rm kpc}\,\chid\,M_{\rm h,12}^{1/3}\,H^{-2/3}(z),  
\label{eq:rdmodel}
\end{equation}
where $\chid$ is the proportionality constant -- a parameter of the model, $M_{\rm h,12}$ is halo $\Mtwoh$ mass at $\zpeak$ in units of $10^{12}\,\Msun$, $H(z)=H_0E(z)$ with $E(z)$ is given by eq.~\ref{eq:Ez} above. 

The assumption of proportionality $\Rd\propto\Rtwoh$ has some theoretical motivation, if galaxy sizes reflect the angular momentum acquired by objects during their collapse \citep[e.g.,][]{Fall.Efstathiou.1980,Ryden.Gunn.1987,Mo.etal.1998}. Support for such theoretical scenario exists in results of galaxy formation simulations with efficient feedback  \citep{Scannapieco.etal.2008,Zavala.etal.2008,Sales.etal.2010,Agertz.Kravtsov.2016,Sokolowska.etal.2017} and in observational scaling of specific angular momentum of stars with the stellar mass of galaxies.  Even more direct motivation for such linear relation is provided by the fact that the observed sizes of stellar distributions in galaxies are consistent with such scaling at both $z=0$ \citep[][]{Kravtsov.2013} and higher redshifts \citep{Shibuya.etal.2015,Huang.etal.2017,Somerville.etal.2018}. 

Although these indications were largely derived for sizes of stellar distribution, a similar linear relation exists for the scale lengths of gaseous disks for galaxies in the THINGS sample \citep[][]{Kravtsov.2013} with the scale length of gas is about three times larger than that of the stellar disk. 

The equation~\ref{eq:rdmodel} is thus adopted as our model for the gas disk scale length. The mean value of $\chid$ is a single parameter adopted for {\it all} objects and we choose its value to be $\chid=0.06$ \citep[e.g.,][]{Mo.etal.1998}. In addition, given that disk sizes are expected to have substantial scatter in the models and that in observations there is also a significant scatter of sizes at a given stellar mass, we introduce a random scatter of 0.25 dex for the population of haloes that we model. The scatter 
is introduced by drawing a normal random number $r_{\rm pert}$ with zero mean and the rms of $\sigma=0.25$ and multiplying $\chid$ by $10^{r_{\rm pert}}$. We use a single value of $\chid 10^{r_{\rm pert}}$ for each object throughout its evolution, but draw different $r_{\rm pert}$ values for different objects at the beginning of their integration. 

We use the exponential model for gas surface density profile to estimate the neutral gas mass using a simple assumption that gas becomes self-shielding against cosmic ionizing UV radiation (and thus neutral) at surface densities above a given threshold value: $\Sigg\geq\SigHIth\approx 0.1-0.2\,\Msun\,\rm pc^{-2}$ \citep[e.g.,][see also \citealt{Bland.Hawthorne.2017} and references therein]{Sunyaev.1969,Felten.Bergeron.1969,Bochkarev.Sunyaev.1977,Maloney.1993}.

For an exponential disk of total mass $\Mg$ and disk scale length $\Rd$, mass within a given $R$ is given by eq.~\ref{eq:mrexp}. Denoting $x=R/\Rd$, the radius at which surface density exceeds the threshold is $x_{\rm HI}=-\ln\left(\SigHIth/\Sigma_0\right)$ and thus $M_{\rm HI}$ is given by the above equation evaluated at $x_{\rm HI}$.

\subsection{Modelling molecular hydrogen}
\label{sec:h2model}

In addition to tracking the total ISM mass $\Mg$, we include a model to track mass of molecular and atomic phases of the ISM, $\MHH$ and $\MHI$. These quantities can be used for comparisons with observations, which probe primarily these neutral phases. Most importantly, modelling $\MHH$ allows us to model SFR more robustly, as described below.

To estimate $\MHH$  we use the model presented by \citet{Gnedin.Draine.2014}. This model was calibrated using results of numerical calculations that include radiative transfer and a chemistry network within the ISM of a simulated galaxy. The results of the model are presented for a range of averaging scales and thus the model accounts both for the $\HH$ chemistry in the ISM and the structure of the ISM itself on sub-kpc scales. Importantly, the model also was optimized for dwarf galaxy regime. 

Specifically, we use the equations for $R=\SigHH/\SigHI$ 
in the erratum to the original \citet{Gnedin.Draine.2016} paper, which presents the corrected approximation of equations in the original paper: 
\begin{eqnarray}
R&=&\frac{\SigHH}{\SigHI}\approx\left(\frac{\Sigg}{\Sigma_{R=1}}\right)^\alpha;\\
\Sigma_{R=1}&=&\,\frac{50\,\Msun\,\rm pc^{-2}}{g}\,\frac{u^{1/2}}{1+1.69u^{1/2}};\\
g &=& \left(\DMW^2+D_*^2\right)^{1/2};\label{eq:g}\\
u &=& 0.001 + 0.1\UMW;\label{eq:u}\\
\alpha &=& 0.5 + \left[1+\left(\frac{\UMW\DMW^2}{600}\right)^{1/2}\right]^{-1},
\end{eqnarray}
where $\UMW$ and $\DMW$ are the far UV flux in the Lyman-Werner bands in units of the Draine value (see below) and dust abundance in units of the Milky Way value, respectively. They also provide an improved approximation:
\begin{eqnarray}
R&=&\frac{\SigHH}{\SigHI} \approx q\,\frac{1+\xi q}{1+\xi};\\  q&=&\left(\frac{\Sigg}{\Sigma_{R=1}}\right)^\alpha;\\
\Sigma_{R=1}&=&\,\frac{40\,\Msun\,\rm pc^{-2}}{g}\,\frac{u^{0.7}}{1+u^{0.7}};\\
\alpha&=& 1+ 0.7\frac{u^{0.35}}{1+u^{0.7}},
\end{eqnarray}
where $g$ and $u$ are given by eqs.~\ref{eq:g} and \ref{eq:u} above and parameter $\xi$ is 0 when model results are applied on kpc scales and $0.25$ when it is applied at $\lesssim 500$ pc scales. 

The fiducial value of $D_*$ advocated by \citet{Gnedin.Draine.2014} is $D_*=0.17$ and we adopt it here. We model dust mass fraction as $\DMW=f_{\rm d} \MZg/\Mg/\Zsun,$ where $\Mg$ is the current total mass of gas in the ISM, $\MZg$ is the current mass of heavy elements in the ISM modelled as described in Section~\ref{sec:zevo}, $\Zsun=0.015$ is the metal mass fraction in the solar neighborhood consistent with measurements from solar spectroscopy \citep[e.g.,][]{Asplund.etal.2009,Lodders.2019}, although values of $\approx 0.019-0.02$ are advocated based on the solar wind composition and helioseismology \citep[e.g.,][]{VonSteiger.Zurbuchen.2016}. We choose $f_{\rm d}=0.5$ to account for the fact that about half of the ISM heavy elements is locked in dust. The specific value of $f_{\rm d}$ does not affect results of the model significantly. 

We estimate the flux of far UV (FUV) $91.2-111$ nm far interstellar radiation field in the Milky Way units, $\UMW$, using average surface density of star formation rate, $\langle\dot{\Sigma}_{\star}\rangle$, estimated as described below (Section~\ref{sec:sfmodel}) normalized to the surface density of star formation at solar radius in the Milky Way, $\dot{\Sigma}_{\star,\odot}\approx 2.5\,\Msun\,\rm pc^{-2}\,yr^{-1}$ \citep[e.g., see Fig. 7 in][]{Kennicutt.Evans.2012}: $\UMW = \langle\dot{\Sigma}_{\star}\rangle/\dot{\Sigma}_{\star,\odot}$.

Given $R=\SigHH/\SigHI$, the molecular fraction is given by 
\begin{equation}
\fHH(\Sigg) = \frac{R}{1+R},    
\end{equation}
and the total $\HH$ mass is computed by integrating over the exponential $\Sigg(R)$ profile assumed in our model: 
\begin{equation}
\MHH = 2\pi\,\int\limits_0^{R_{\rm HI}}\fHH(\Sigg)\Sigg(R)RdR, 
\label{eq:MH2}
\end{equation}
where $R_{\rm HI}$ is the radius corresponding to the assumed self-shielding threshold $\SigHIth$ described above in Section \ref{sec:sigrmodel}.

\subsection{Star formation model and evolution of stellar mass}
\label{sec:sfmodel}
Star formation rate in our model is calculated using $\MHH$ computed with the $\HH$ model (Section~\ref{sec:h2model}, eq.~\ref{eq:MH2}) and assuming a constant depletion time of molecular gas, $\tauHH$:
\begin{equation}
\sfr = \frac{\MHH}{\tauHH}.
\label{eq:sfr}
\end{equation}

However, we also assume the instantaneous recycling or that a fraction $\Rloss$ of the gas converted stars is returned immediately to the ISM, so that the actual rate of gas decrease due to star formation in eq.~\ref{eq:mgevo} is
\begin{equation}
\dotMs=\sfrloss.
\label{eq:msevo}
\end{equation}
This equation also serves as equation for the evolution of mass in stars. We do not include any contribution from stellar mass brought in by mergers. Although doing so is relatively straightforward using either merger trees extracted from simulation outputs or constructed using an extended Press-Schechter method \cite[see, e.g.][]{Krumholz.Dekel.2012}, given that we focus on dwarf galaxies this is an unnecessary complication because mergers are expected to contribute a negligible fraction of stellar mass in dwarf galaxies \citep[see][]{fitts_etal18}, and contribute only $\lesssim 1\%$ of stellar mass in galaxies such as Milky Way \citep[e.g.,][]{Purcell.etal.2007}. Note, however, that mergers may be important for modelling of detailed properties of dwarf galaxies, such as stellar population gradients \citep[e.g.,][]{tarumi_etal21,chiti_etal21a} or when modelling detailed properties of the stellar population of the Milky Way \citep[e.g.,][]{chiti_etal21}.

Although in general $\tauHH$ can depend on redshift and galaxy properties, in this pilot study we adopt a constant fiducial value of $\tauHH=2.5$ Gyrs consistent with measurements in nearby galaxies 
\citep[e.g.,][]{Bigiel.etal.2008,Bigiel.etal.2011,Bolatto.etal.2011,Rahman.etal.2012,Leroy.etal.2013}. There are indications that molecular depletion time scale can be somewhat smaller in the dwarf galaxy regime \citep[][]{Saintonge.etal.2011,Saintonge.etal.2016,Saintonge.etal.2017,Bolatto.etal.2017,Dou.etal.2020}, so exploration of smaller values of $\tauHH$ is warranted. Finally, we adopt $\Rloss=0.44$, which is close to the values expected for the \citet{Chabrier.2003}  initial mass function of stars \citep[e.g.,][]{Leitner.Kravtsov.2011, Vincenzo.etal.2016}.

\begin{figure*}
    \centering
    \includegraphics[width=0.9\textwidth]{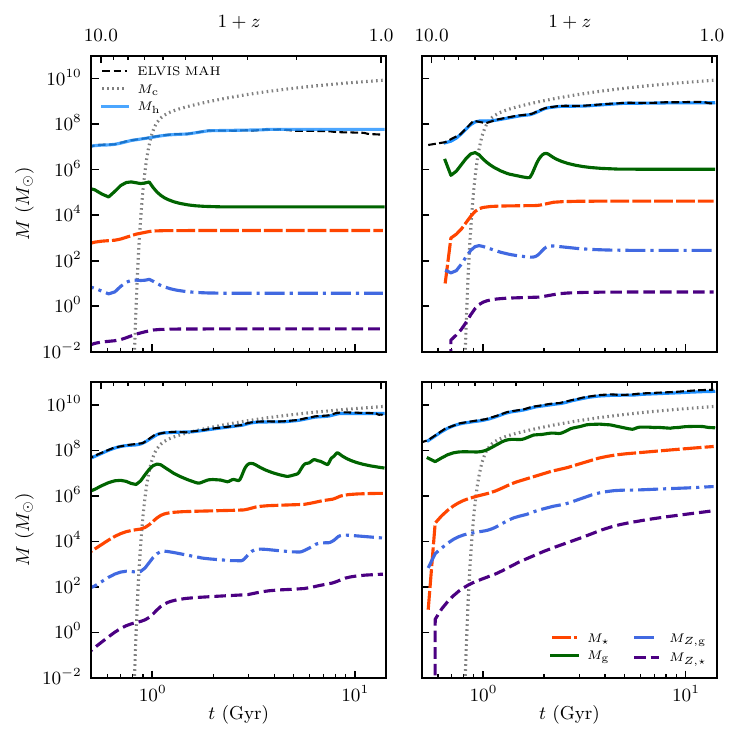}
    \caption{The time evolution of halo mass ($M\textsubscript{h}$), stellar mass ($\Ms$, orange long-dashed line), gas mass ($\Mg$, solid green line) and heavy element mass in gas ($\MZg$, blue dot-dashed line) and stars ($\MZs$, indigo dashed line) for different mass ELVIS haloes according to the model with fiducial parameters and $\zrei = 6$. The dashed-black shows the actual MAH extracted from the ELVIS suite and agrees well with the integrated halo mass (solid light-blue line) that is done according to procedure described in Section~\ref{sec:mhalo}. } 
    \label{fig:tracks}
\end{figure*}

\subsection{Modelling the half-mass radius of stellar distribution}
\label{sec:rhalfmodel}

Given that stars in our model are assumed to form from molecular gas and radial surface density profile of H$_2$ is a part of the model, we can use these parts of the model to estimate the half-mass radius of stars of our model galaxies. We do this in the post-processing stage using the tabulated evolution of $\Mh$, $\Ms$, $\Mg$, and $\MZg$ to reconstruct the surface density of $\Sigma_{\HH}(R)=f_{\rm H_2}\Sigg(R)$ and estimate the half-mass radius of H$_2$. 
As discussed above, observations indicate that $\tauHH=\SigHH/\Sigsfr$ is approximately independent of surface density, which implies that the radial distribution of molecular gas and 
young stars is similar as indeed is observed in nearby galaxies  \citep[][]{Bolatto.etal.2017}.

Assuming that $\tauHH$ in the model is constant throughout the evolution and radial redistribution of stars after their formation is negligible, the half-mass radius of stellar distribution
can be estimated as follows. We compute $M_{\HH}(<R,t_i)$ profiles for a dense grid of cosmic times, $\{t_i\}$, with constant $\Delta t$ from the beginning of the evolutionary track to $z=0$ by integrating $\Sigma_{\HH}(R)$ profile up to a given $R$. The total stellar mass profile formed throughout evolution is then approximately  $\Ms(<R)\approx \Delta t/\tauHH\,\Sigma_i M_{\HH}(<R,t_i)$ and we estimate the half-mass radius of the stellar distribution using this profile.

\subsection{Modelling of galactic outflows}
\label{sec:windmodel}

The third term in eq.~\ref{eq:mgevo} accounts for the feedback-driven outflows, which are expected to be especially strong in dwarf galaxies due to their shallow potential wells \citep{Dekel.Silk.1986}. In our model we adopted a commonly used parametrization of the outflow rate: 
\begin{equation}
\dotMout = \etaw \sfr,
\label{eq:moutwind}
\end{equation}
where $\etaw$ is the mass loading factor and $\sfr=\MHH/\tauHH$ is star formation rate not accounting for stellar mass loss. 

We adopt a model for $\etaw$ motivated by results of cosmological galaxy formation simulations of \citet{Muratov.etal.2015}, who found that at all redshifts $\etaw$ of dwarf galaxies can be approximated by a power law $\etaw=3.5\, M^{-0.35}_{\star,10}$, where $\Msten$ is stellar mass of the galaxy at the corresponding epoch in units of $10^{10}\,\Msun$. At the same time, \citet{Muratov.etal.2015} could not measure a detectable outflow in hosts with stellar masses close to that of the Milky Way indicating that mass loading in such galaxies should be small \citep[see also][]{AnglesAlcazar.etal.2017,Pandya.etal.2021}. To account for this we adopt parametrization in which $\etaw$ decreases to zero above a certain mass:
\begin{equation}
    \etaw = \max\left(0, 2000, \etanorm \Msten^{-\etapow}-\etac\right)
\label{eq:etamodel}
\end{equation}
We use the fiducial values close to those derived by \citet{Muratov.etal.2015}:  $\eta_{\rm norm}=3.6$, $\etapow=0.35$, $\etac=4.5$. The latter corresponds to $\etaw$ decreasing to zero for $M_\star\gtrsim 5\times 10^9\,\Msun$. The maximum of $\etaw=2000$ is somewhat arbitrary and is included to prevent extremely small time steps during integration. However, there is some physical basis to limit efficiency of stellar winds in the smallest galaxies, as their small stellar mass to halo ratio and small expected number of supernovae prevents them from driving winds effectively \citep[see, e.g.,][]{bullock_boylankolchin17}. 

Although the specific parametrization is motivated by a fit to the simulation results, an approximately power law dependence of $\etaw$ on stellar mass is expected in the dwarf galaxy regime. For example, basic assumptions about the wind driving imply power law scaling of $\etaw$ with halo mass $\etaw\propto \Mh^\sigma$, where $\sigma=-2/3$ and $\sigma=-1/3$ for the energy- and momentum-driven winds, respectively \citep[see, e.g.,][]{Furlanetto.etal.2017}. Given that the stellar mass function in the dwarf galaxy regime is close to the power law, the $\Ms-\Mh$ relation in the dwarf galaxy regime is also expected to be close to a power law $\Ms\propto\Mh^\delta$, where $\delta\approx 2\pm 0.5$ \citep[see, e.g.,][]{Kravtsov.2010}. Therefore, mass loading factor can be expected to scale as $\etaw\propto \Ms^{\sigma/\delta}$ with $\sigma/\delta\approx -0.13\div -0.45$ with the value obtained by \citet{Muratov.etal.2015} closest to the energy-driven winds and $\delta= 2$: $\sigma/\delta\approx -2/3/2\approx -0.33$. 

At the same time, we note that results of the galaxy formation simulations with respect to wind driving processes and mass loading factor scalings are still quite uncertain  \citep[see, e.g., Fig. 14 in][and associated discussion]{Mitchell.etal.2020}. Recent analysis of the FIRE-2 simulations by \citet{Pandya.etal.2021} shows that mass loading factors are generally consistent with the \citet{Muratov.etal.2015} scaling adopted above with scaling $\eta\propto\Ms^{\etaw}$ independent of redshift and $\etaw\approx 0.35-0.45$, but normalization can be lower by a factor of two due to ambiguities of which gas is included in the outlfow. One should thus potentially allow a range of possible scalings. Nevertheless, all simulations qualitatively agree with the expected simple wind models in that they all predict increase of mass loading factor with decreasing halo or stellar mass for $\Mh\lesssim 10^{12}\,\Msun$. Thus, there is a strong theoretical motivation for the mass dependent $\etaw$ models. 

\subsection{Evolution of heavy elements abundance}
\label{sec:zevo}

The model includes differential equations for the rate of change of mass in heavy elements in the ISM and stars. 

The rate of change of mass of heavy elements in the interstellar gas is due to accretion of gas from the IGM, production of new heavy elements by stars, and by removal of heavy elements from the ISM when it is converted into stars
or carried away in outflows: 
\begin{eqnarray}
&\dotMZg = \dot{M}_{\rm Z,in} + \yZ\sfr - \Zg\sfrloss - \etaw \Zw\sfr\nonumber\\
&=\Zigm\dotMgin +\left[\yZ - \left(1-\mathcal{R}+\zetaw\etaw\right)\,\Zg\right]\sfr,\label{eq:mzgevo}
\label{eq:mzgmodel}
\end{eqnarray}
where $\Zg=\MZg/\Mg$ is the mass fraction of heavy elements in the ISM (or ISM metallicity), $\Zigm$ is the characteristic metallicity of heavy elements accreted by the galaxy, and $\Zw$ is characteristic metallicity of the gas carried out in winds. 
We also define the wind metallicity enhancement factor $\zetaw=\Zw/\Zg$.  The factor allows to account for differences in the metallicity of wind and the ISM. For example, metallicity of the wind can be larger than that of the ISM if a fraction of newly produced heavy elements produced by young massive stars is efficiently lost as the superbubbles break out of the ISM before they have a chance to mix with the rest of the gas, thereby enhancing wind metallicity relative to metallicity of the ISM. 

Note that in the above equation the definition of $\zetaw$ is different from the parameterization used in DK12, who modelled this via the parameter $\zeta$ defined as the fraction of newly produced metals immediately lost in winds. We use the above parameterization because $\zeta$ in the DK12 definition is not well known, while $\zetaw$ is  closely related to the outflow metallicity that can be probed in observations \citep[e.g.][]{Chisholm.etal.2018} and measured in simulations \citep[e.g.,][]{Muratov.etal.2017}. Note also that \citet{Peeples.Shankar.2011} used $\zetaw$ notation to define a related, but different quantity, which in our notation is given by $\zetaw\etaw$.

Note also that in the eq.~\ref{eq:mzgmodel} we implicitly assume that heavy elements mix with {\it all} of the ISM gas, not just its neutral part. Such mixing is expected due to ubiquitous turbulence in the ISM, as well as re-accretion and mixing of fountain outflows in the outer regions of galaxies \citep[see, e.g., discussion in][]{Tassis.etal.2008}.

The parameter $\yZ$ is the yield of heavy elements per unit star formation rate  
produced by young massive stars and dispersed by supernovae and AGB stars; it has the value of $y_Z\approx 0.06$ for the \citet{Chabrier.2003} IMF \citep[e.g.,][]{Vincenzo.etal.2016} where details and dependence of this parameter on assumptions is explored. Note that in practice this parameter 
simply affects the overall normalization of the metallicity-mass relation and is degenerate with the normalization of the wind mass loading factor $\etaw$ scaling with stellar mass.

Our adopted fiducial values for these parameters are: $\Zigm=10^{-3}\Zsun$, $\yZ=0.06$, $\zetaw=1$ (i.e., $\Zw=\Zg$). Although, there are  indications that metallicity of outflows can generally be larger than that of the ambient ISM (i.e., $\zetaw>1$) from both models \citep[e.g.,][]{MacLow.Ferrara.1999,Muratov.etal.2017,Forbes.etal.2019} and from observations \citep[e.g.,][]{Chisholm.etal.2018}, however, the characteristic value and scaling of $\zetaw$ with system mass are currently uncertain. 
At the same time, we find that 
a reasonably good match to observed metallicity scalings can be obtained with $\zetaw=1$, as long as a mass-dependent scaling of $\etaw$ is adopted. We discuss this issue further in Section~\ref{sec:discussion}. 
The adopted $\zetaw=1$ is close to results of the FIRE galaxy formation simulations, which find $\zetaw\approx 1.2$ independent of galaxy mass \citep[][]{Muratov.etal.2017}, albeit with a sizeable scatter. Note that  \citet{Pandya.etal.2021} also showed that $\zetaw\approx 1$ in the FIRE-2 simulations in dwarf-scale haloes.

The non-zero value of $\Zigm$ is attributed to pre-enrichment of IGM by the Population III and early Population II stars \citep[e.g.,][]{Greif.etal.2010, Wise.etal.2012} and is strongly indicated by the relatively large metallicities of the smallest dwarf galaxies \citep[e.g., see][for discussion]{Tassis.etal.2012}. 

To track the mass of heavy elements locked in long-lived stars, $\MZs$ we use an additional equation which is simply:
\begin{equation}
\dotMZs = \Zg\sfrloss.
\label{eq:mzsevo}
\end{equation}

\subsection{Summary of the model}

The main components of the model and related parameters with their fiducial values are summarized in Table~\ref{tab:paramtable}. We implemented this model in a Python package \texttt{GRUMPY} (Galaxy formation with RegUlator Model in PYthon).\footnote{The code will be made publicly available after this manuscript is accepted.}

Integration of the model is carried out from an initial time $t_i$ corresponding to the first redshift available in the halo track to $t_U$ corresponding to $z=0$ using the system of coupled ordinary differential equations given by eqs.~\ref{eq:mgevo}, \ref{eq:msevo}, \ref{eq:mzgevo}, \ref{eq:mzsevo} along with the terms and factors described above in this section.

Figure~\ref{fig:tracks} shows representative examples of evolution of the modelled properties of galaxies in haloes of different masses in our fiducial model. The figure shows that haloes that fall far below $M_{\rm c}(z)$ at high $z$ form most of their stars at high redshifts. These haloes retain gas but its surface density is too low to form molecular hydrogren and stars. Note that the object shown in the top right panel forms stars in several episodes at $z\gtrsim 3$ even though its halo mass is smaller than $M_{\rm c}$ at these redshifts. This is because the degree of suppression of gas accretion around $M_{\rm c}$ is
a smooth function of $\Mh$ and even haloes with $\Mh\lesssim M_{\rm c}$ can continue to accrete some gas. Likewise, the object shown in the bottom left panel continues to form stars throughout its evolution even though its halo mass is somewhat below $M_{\rm c}$ most of the time. 

Another notable feature of the evolution is decrease of mass of metals in gas following episodes of significant star formation (manifested as epochs of sharp increase of stellar mass) in the three objects with the lowest final $\Mh$. This loss is due to heavy elements carried away by outflows, which have large mass loading factors for such small-mass systems in the fiducial model. The most massive object shown in the bottom right panel, however, does not show such decreases due to the small mass-loading factor for objects of such stellar mass. 
We can also note that 
decrease of $\MZg$ is smaller than decrease of $\Mg$ due to continuing injection of heavy elements by young stars into the ISM. 
The gas-phase metallicity thus still grows, but by an amount smaller than in the absence of outflows. This illustrates that modulating effect the outflows have on the gas-phase metallicity.

\begin{figure*}
    \centering
    \includegraphics[width=\textwidth]{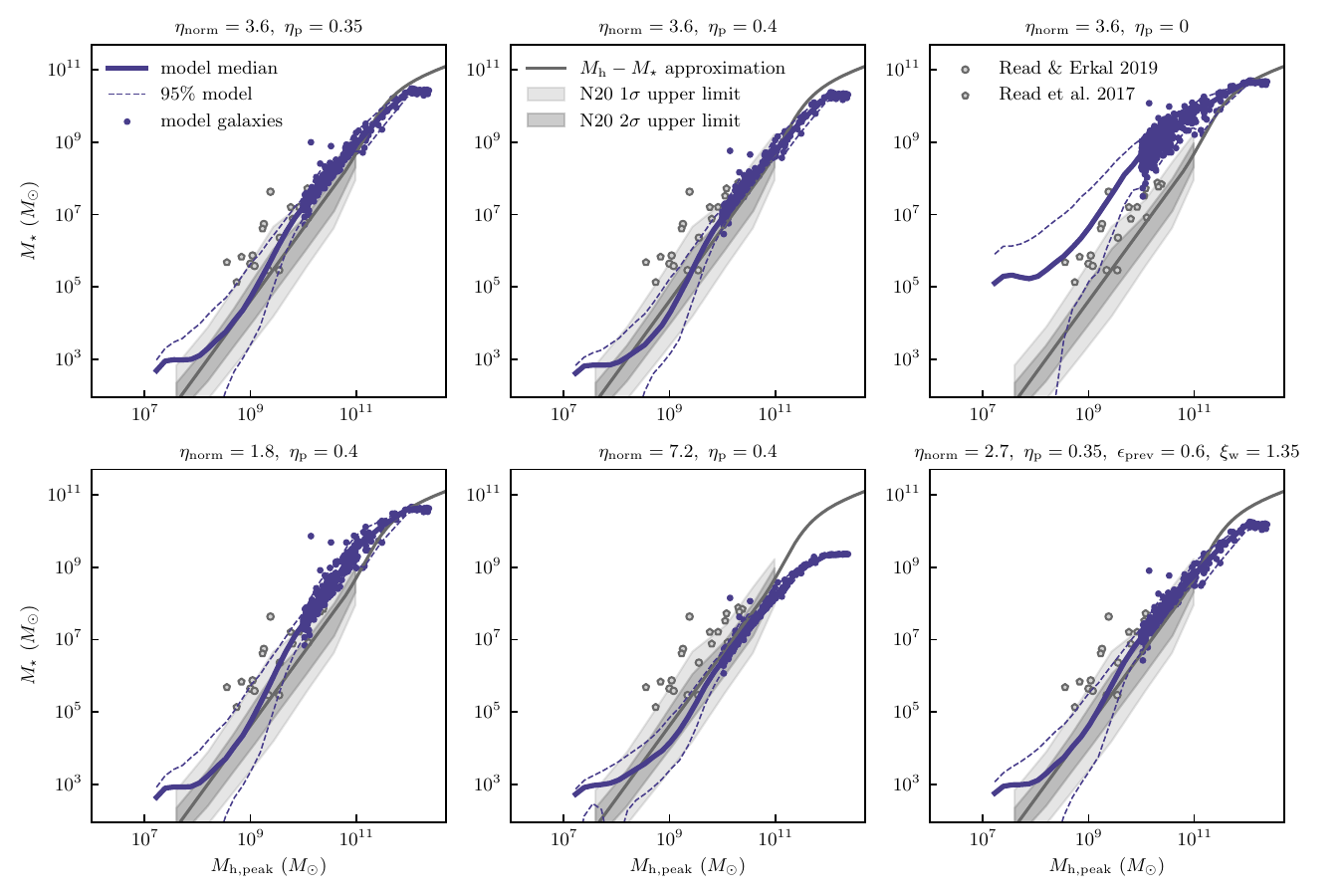}
    \caption{Relation between galaxy stellar mass, $\Ms$, and peak halo mass throughout its evolution, $\Mhp$, derived statistically \citep[gray bands,][]{Nadler.etal.2020} or individually \citep[gray circles and pentagons][]{Read.etal.2017,Read.Erkal.2019} for the Milky Way satellites and from the abundance matching (solid line; see Section~\ref{sec:msmh} for details) compared to the results of the model run with different parameters (all other parameters are kept at their fiducial values. Note that the gray bands indicate $1\sigma$ and $2\sigma$ {\it upper limits} around the mean relation derived by \citet{Nadler.etal.2020}. The solid line shows the median for model galaxies at masses where there are more than 10 galaxies per bin, while dashed lines show the region enclosing $95\%$ of the model galaxies. At larger masses individual model galaxies are shown as dark blue points. }
    \label{fig:msmhwind}
\end{figure*}

\begin{figure*}
    \centering
    \includegraphics[width=\textwidth]{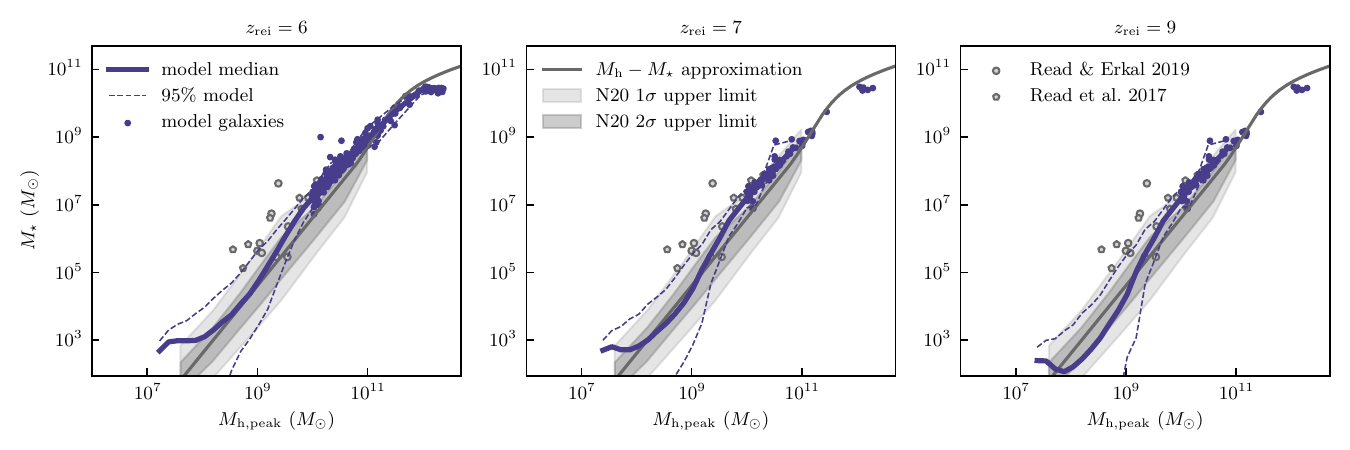}
    \caption{Relation between galaxy stellar mass, $\Ms$, and peak halo mass throughout its evolution, $\Mhp$. Notation and symbols are as in Fig.~\ref{fig:msmhwind}, but this figure shows results of the model for the fiducial parameter choice and reionization redshifts $\zrei=6, 7, 9$ (from left to right, as labeled at the top of each panel). }
    \label{fig:msmhzrei}
\end{figure*}

\begin{figure*}
    \centering
    \includegraphics[width=\textwidth]{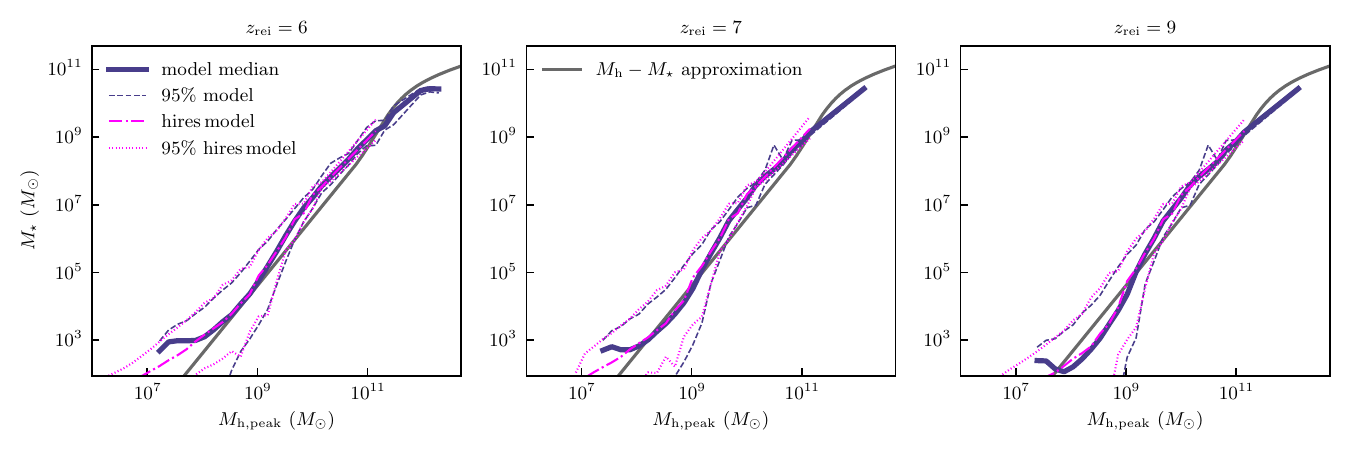}
    \caption{Comparison of the $\Ms-\Mhp$ relations in the fiducial (median and 95\% range shown by the thick slid blue and dashed lines, respectively) and high-resolution (median and 95\% range shown by magenta dot-dashed and dotted lines) ELVIS simulations for three different reionization redshifts using the same reionization model as in Figure~\ref{fig:msmhzrei}. The figure shows that the model calculations using halo tracks from the higher resolution simulations are consistent with the model calculations with fiducial simulations for $\Mhp\gtrsim 5\times 10^7\,\Msun$ and $\Ms\gtrsim 500-1000\,\Msun$.}
    \label{fig:msmhres}
\end{figure*}

\begin{figure}
    \centering
    \includegraphics[width=0.5\textwidth]{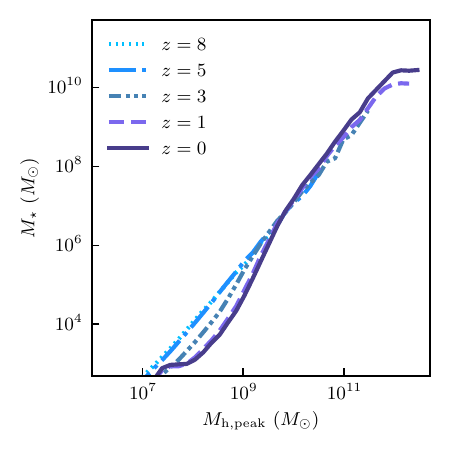}
    \caption{Evolution of the $\Mhp-\Ms$ relation in the model with $\etanorm=3.6$, $\etapow=0.35$, and $\zrei=6$. The evolution at small masses occurs due to evolving characteristic halo mass, $M_{\rm c}(z)$, below which gas accretion is suppressed due to the UV background heating (see Section~\ref{sec:mgin} and eq.~\ref{eqn:Mcz}). At larger masses galaxies evolve along the narrow $\Mhp-\Ms$ relation shaped by the wind model. }
    \label{fig:msmhevo}
\end{figure}

\section{Comparisons of model results with observations}
\label{sec:comparisons}

\begin{figure*}
    \centering
    \includegraphics[width=\textwidth]{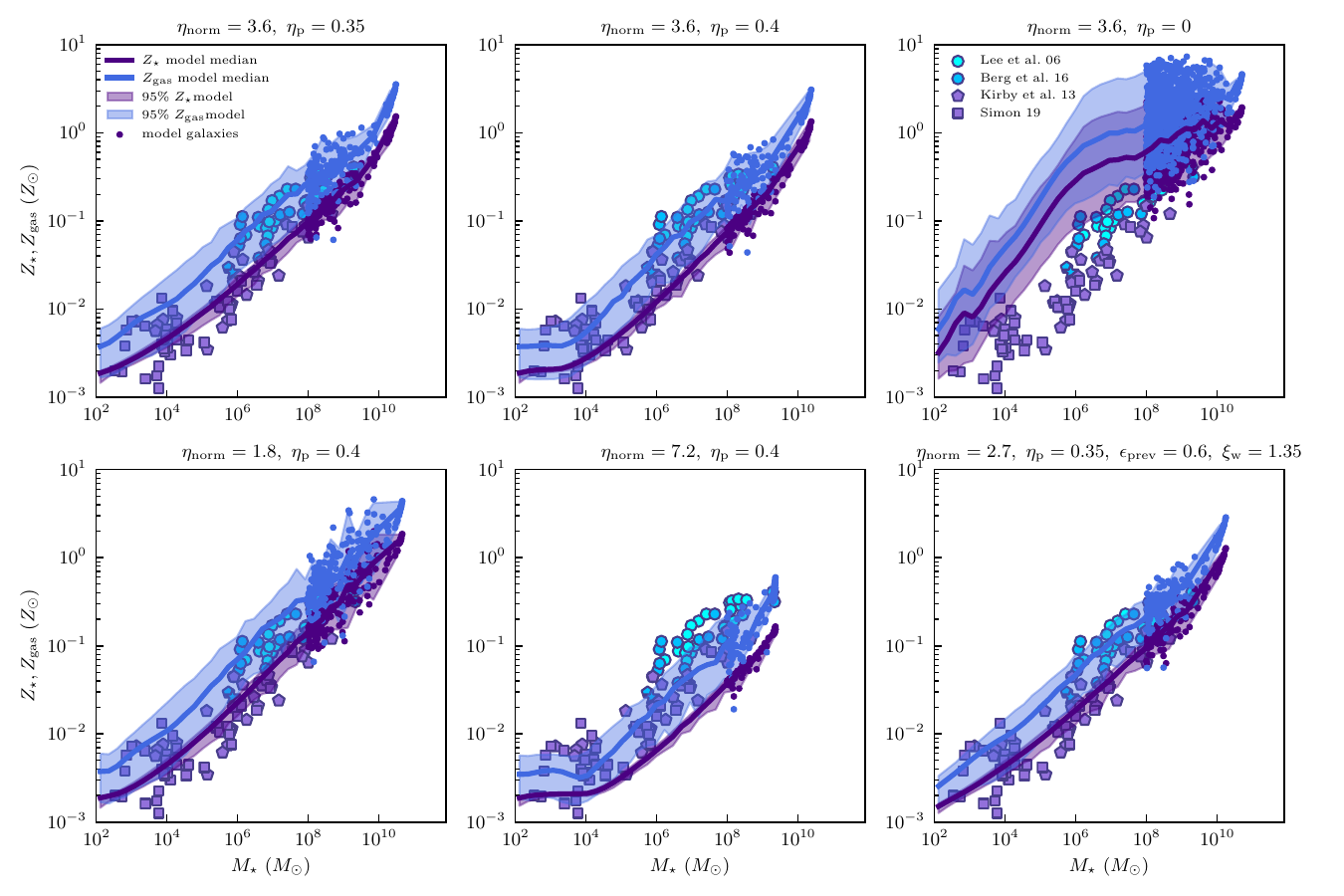}
    \caption{Relation between galaxy stellar mass, $\Ms$, and gas-phase ($Z_{\rm gas}$) and stellar ($Z_\star$) metallicity predicted by the model with different assumptions about outflows and observations (all models were run with the parameters indicated in the panel titles and other parameters fixed to their fiducial values). The median gas phase metallicity of model galaxies is shown by the light blue solid line for bins with more than 5 galaxies; for masses where there are few model galaxies we show individual model galaxies as dots. The light blue dashed lines show the region enclosing 95\% of model galaxies. Median stellar metallicity is shown by the dark blue line while 95\% region around it is shown by the dotted lines. Observed gas phase metallicities are shown for the samples of  \citet{Lee.etal.2006} and \citet{Berg.etal.2016} that are based on the oxygen abundance and for assumed solar abundance of $\rm [O/H]_\odot=8.7$ \citep{Lodders.2019}. We also show examples of extremely low metallicity galaxies, Leo P \citep[][]{McQuinn.etal.2015} and Leoncino \citep[][]{McQuinn.etal.2020} by light blue star and pentagon symbols, respectively. Observed stellar metallicities are from compilations by \citet[][pentagons]{Kirby.etal.2013} and \citet[][squares]{simon19}.}
    \label{fig:mszwind}
\end{figure*}

\subsection{Stellar Mass - Halo Mass relation}
\label{sec:msmh}

Relation between stellar and halo mass is one of the most important aspects of the galaxy-halo connection and we thus start with examining this relation. Specifically, we consider relation between stellar mass of model galaxies and the maximum halo mass, $\Mhp$, achieved by each halo during its evolution. The halo mass here is defined within the radius enclosing the density contrast of 200 with respect to the critical density of the universe at the redshift of analysis. 

Figure~\ref{fig:msmhwind} shows $\Ms-\Mhp$ relation for five models with different assumptions about normalization of the wind mass loading factor, $\etanorm$, and the slope of its scaling with stellar mass, $\etapow$ (see eq.~\ref{eq:etamodel}) and one model that includes preventative feedback and small wind metallicity enhancement (bottom left panel). The model results are shown as the median $\Ms$ in bins of $\Mhp$ constructed using individual model galaxies and the 95\% range of $\Ms$ around the median. In the bins where the number of galaxies is small we also show individual model galaxies as dark dots. 

Model results are compared to the stellar mass-halo mass relation statistically constrained using the full current census of detected Milky Way satellites  by \citet{Nadler.etal.2020}, shown by gray shaded regions, and for a number of nearby individual dwarf galaxies \citep[][]{Read.etal.2017,Read.Erkal.2019} shown by gray circles and pentagons. Note that these observational constraints were obtained for satellite galaxies in the Milky Way or M31, while model galaxies include both satellite galaxies at $z=0$ and galaxies outside the virial radius. Note also that halo mass in \citet{Nadler.etal.2020} was defined using ``virial'' density contrast rather than 200 times critical density. Thus, this comparison is approximate, although the effect of this difference should be small.

The gray solid line is the $\Ms-\Mhp$ relation derived using abundance matching of the halo mass function to the stellar mass function of \citet{Bernardi.2013} at $\Ms>10^9\,\Msun$ \citep{Kravtsov.2018}, but with parameters that modify the relation at $\Ms<10^9\,\Msun$ to be consistent with the relation constrained by \citet{Nadler.etal.2020}. This line is given by equation 3 of \citet{Behroozi.etal.2013c} with parameters $\log_{10}M_1=11.39$, $\log_{10}\epsilon=-1.69$, $\alpha=-2$, $\delta=4.34$, $\gamma=0.53$ and represents a relation that is consistent with observations in the full range of stellar masses: $\Ms\in 10^2-10^{12}\,\Msun$. 

Note that there is some tension between estimates for individual galaxies and statistical constraint of \citet{Nadler.etal.2020}.
However, the latter study assumed a power law relation between $\Ms$ and $\Mhp$ with normalization fixed at $\Mhp=10^{11}\,\Msun$. Thus, these constraints can be consistent with the non-power law form of the relation in the 
model with $\etanorm=3.6$, $\etapow=0.35$ (upper left panel), because this relation matches the \citet{Nadler.etal.2020} relation both at $\Mhp\approx 10^{10}-10^{11}\,\Msun$ and at $\Mhp\sim 10^8-10^9\,\Msun$ and is consistent with most estimates for individual galaxies. In addition, to 
compare individual measurements with statistical inference of \citet{Nadler.etal.2020} one needs to take into account selection effects for observed galaxies properly \citep[see, e.g.,][]{Jethwa.etal.2018}. Indeed, as we show in the follow up paper (Manwadkar \& Kravtsov in prep.) this model provides a good match to the observed luminosity function of the Milky Way satellites, which was used to derive constraints on the $\Ms-\Mhp$ relation by \citet{Nadler.etal.2020}.

Figure~\ref{fig:msmhwind} shows that the normalization and slope of the model relation depends on the wind model parameters. 
The amplitude of $\Ms-\Mhp$ relation is directly related to the normalization of the mass-loading factor, $\etanorm$, while its slope at $\Ms\gtrsim 10^6\,\Msun$ depends on the slope of the dependence of mass loading factor on stellar mass $\etapow$. This is not surprising because basic theoretical considerations (e.g., energy-driven vs momentum-driven wind) indicate that both the wind mass loading factor and the $\Ms-\Mh$ relation depend on the physical mechanism of wind driving. 
First, we note that the model with constant mass loading factor (lower left panel) results in normalization and slope of the $\Ms-\Mhp$ relation inconsistent with observational constraints. This model also results in the largest scatter in the relation. 

Interestingly, the model with $\etanorm=3.6$
and $\etapow=0.35-0.4$---parameters very close to those derived from the FIRE simulations \citep[][]{Muratov.etal.2015}---provide the best match to the relations derived from observations. This model predicts the scatter that is quite small ($\lesssim 0.1$ dex) for $\Mhp\gtrsim 10^{10}\,\Msun$ but increases with decreasing halo mass. In the ultra-faint dwarf galaxy regime ($\Ms<10^5\,\Msun$) the scatter reaches $\approx 0.3-0.4$ dex and is roughly consistent with scatter constraints of \citet{Nadler.etal.2020}. Note that the latter are upper limits, not measurements of the scatter. 

The small scatter of the $\Ms-\Mhp$ relation implies that model galaxies evolve mostly along the relation and that the evolution of the normalization of the relation with redshift is small. This is  consistent with constraints from observed evolution of high-$z$ luminosity function of galaxies \citep[][]{Mirocha.2020}.

The increased scatter at small masses is due {\it almost entirely} to effects of gas accretion suppression due to UV background heating in haloes with masses smaller than characteristic mass $M_{\rm c}(z)$ (see Section~\ref{ssec:uv_heating}). As $M_{\rm c}(z)$ increases rapidly by orders of magnitude at $z>\zrei$, the scatter in $\Ms$ for a given halo mass arises as some objects assembled their mass earlier and thus could form more stars before reionization than other objects. For objects with halo masses close to characteristic mass, different evolutionary mass assembly histories will lead to different fraction of time with $\Mh>M_{\rm c}$ and thus different amounts of gas accretion and star formation (e.g., see some representative evolutionary tracks of our model galaxies shown in Fig.~\ref{fig:tracks}). Interestingly, 
the increase of scatter in $\Ms$ for small $\Mhp$ is similar to results of high-resolution galaxy formation simulations reported by \citet{Munshi.etal.2021}.

The bottom right panel of Figure~\ref{fig:msmhwind} shows that similar $\Ms-\Mh$ relation can be obtained with suppression of mass accretion due to preventative feedback ($\epsinprev=0.6$), but smaller wind normalization (see Section~\ref{sec:discussion} for further discussion).

Figure~\ref{fig:msmhzrei} shows $\Ms-\Mhp$ relations in the models with fiducial parameter values but different $\zrei$. As the redshift of reionization decreases from 9 to 6, an increasing fraction of haloes is able to accrete gas at $z\gtrsim\zrei$ and form stars resulting in a less pronounced steepening of the relation at small masses. 
Note that due to a limited resolution of the simulation tracks the evolutionary histories of the smallest haloes in the catalog can be subject to biases. We check this by comparing model results in the ELVIS simulations with standard resolution with the high-resolution resimulations of the two host haloes in Figure~\ref{fig:msmhres}. The figure shows that model relations are robust for $\Mhp\gtrsim 5\times 10^7\,\Msun$ or $\Ms\gtrsim 500-1000\,\Msun$.

The overall effect of UV heating on $\Ms-\Mhp$ relation is shown in Figure~\ref{fig:msmhevo} that shows evolution of this relation in the fiducial model ($\zrei=6$). In this case the $\Ms-\Mhp$ relations at $z>5$ are close to a power law, while at lower redshifts relation at $\Mh\lesssim 10^9\,\Msun$ steepens as the characteristic suppression mass grows and accretion is suppressed in progressively more massive haloes.

\begin{figure}
    \centering
    \includegraphics[width=0.5\textwidth]{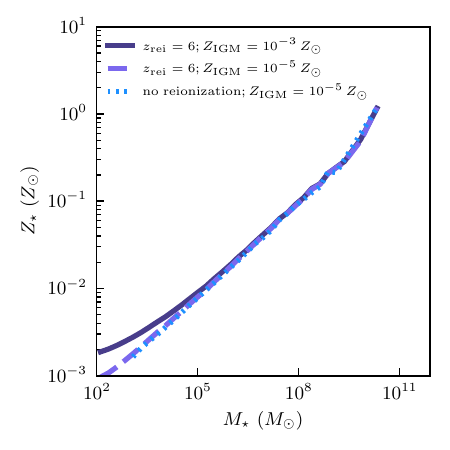}
    \caption{The $\Ms-\Zs$ relation in the fiducial model with $\Zigm=10^{-3}\,\Zsun$ and $\zrei=6$ (solid line) and in models where $\Zigm=10^{-5}\,\Zsun$ (dashed line) and $\Zigm=10^{-5}\,\Zsun$ and no suppression of gas accretion due to UV background heating. The figure shows that the $\Ms-\Zs$ relation is not affected by the latter process and is only mildly sensitive to the floor in $\Zigm$ at the smallest masses.}
    \label{fig:msz_rei_zigm}
\end{figure}


\subsection{Stellar Mass - Metallicity relation}
\label{sec:msz}

Another scaling relations that traditionally is used as a diagnostic of chemical evolution and inflows and outflows in galaxies \citep[e.g.,][]{Garnett.2002,Peeples.Shankar.2011,DeLucia.etal.2020} and as a key test of feedback in galaxy formation models \citep[e.g.,][]{Finlator.Dave.2008} is correlation of galaxy stellar mass and metallicity.  Much observational effort in the last two decades has been aimed towards probing this correlation in the dwarf galaxy regime, with heavy element abundances measured both for stars using absorption lines in stellar emission \citep[e.g., see][for a recent review]{simon19} and for gas using emission lines from HII regions in galaxies with gas and ongoing star formation \citep[][]{Lee.etal.2006,Berg.etal.2012,Berg.etal.2016}.

Figure~\ref{fig:mszwind} compares results of the model for gas phase, $\Zg$, and stellar, $\Zs$, metallicities for galaxies of different stellar mass with observations. The figure shows the same models as in Figure~\ref{fig:msmhwind} and uses the solid lines to show the median relation of model galaxies and shaded regions to show 95\% spread around the median; in bins with small number of galaxies we also show individual model galaxies as dots of the corresponding color. As expected, the model $\Ms-Z$ relation depends quite strongly on the assumptions about the outflow mass loading rate: larger values of mass loading factor normalization, $\etanorm$, or steeper dependence of $\etaw$ on stellar mass (i.e. larger $\etapow$) result in lower metallicities and steeper $\Ms-Z$ relation. 

Interestingly, the model reproduces a small offset in the gas-phase and stellar metallicity that can be seen in observed galaxies.
Here again the models with $\etanorm=3.6$ and $\etapow=0.35-0.4$ provide the best match to observations. 
Although the model with $\etapow=0.4$ underestimates average observed stellar metallicities somewhat, this discrepancy can potentially be offset by normalization of the mass loading factor $\etanorm$ and/or wind metallicity factor $\zetaw$. 

The scatter in the gas-phase metallicities is also comparable to observations and the model reproduces a population of extremely metal poor galaxies studied extensively in observations \citep[e.g.,][and references therein]{McQuinn.etal.2015,McQuinn.etal.2020}, although observations also show existence of outliers from this relation. 

The scatter in $\Ms-\Zs$ relation, on the other hand, is generally small. The only exception is the model with constant mass-loading factor ($\etapow=0$), which does not reproduce the observed correlation, but results in significant scatter in both $\Zg$ and $\Zs$. 

The only sources of scatter in the model are the diversity of halo MAHs and scatter in the gas disk size-halo radius relation. However, the scatter in these quantities affects stellar and gas metallicities very differently: $\Zs$ is average metallicity in the cumulative stellar mass and generally increases motononically, while gas metallicity is affected by star formation and associated metal enrichment, outflows, and inflows that control {\it instantaneous} gas mass \citep[e.g.,][]{Finlator.Dave.2008,Lilly.etal.2013,Torrey.etal.2019,VanLoon.etal.2021}. In the regulator-type models of the kind used here the equilibrium gas-phase metallicity is given by \citep[e.g., Appendix C in][]{Peeples.Shankar.2011}:
\begin{equation}
    \Zg = y_Z\,\left(\zeta_a + \zetaw\etaw +\alpha \Mg/\Ms\right)^{-1}
    \label{eq:ps11eq}
\end{equation}
where $y_Z$ is the nucleosynthetic yield, $\zeta_a=\dotMgin/\SFR$, $\zetaw=\Zw/\Zg$ is metallicity enhancement factor of outflows relative to the ISM metallicity, $\etaw=\dotMout/\sfr$ is the wind mass loading factor, and $\alpha$ is a factor of order unity related to the slopes of the $\Ms-\Mg$ and $\Ms-\Zg$ relations.
The equation shows that metallicity depends on both the wind model and the ratio of the ISM gas mass to stellar mass, $\Mg/\Ms$. As we will show below, $\Mg/\Ms$ in model and observed galaxies has substantial scatter, which translates into scatter in the $\Ms-\Zg$ relation. Note also the model $\Zg$ and $\Zs$ values do not include observational uncertainties which do contribute to the scatter of relations of observed galaxies. 

However, the small {\it intrinsic} scatter of the model $\Ms-\Zs$ relation clearly indicates that the model relation does not evolve significantly with redshift in the dwarf galaxy regime.\footnote{The model does predict evolution in haloes of $\Mh\gtrsim 10^{11}\,\Msun$.}

We note that the models with $\etanorm=3.6$
and $\etapow=0.35-0.4$ match observed $\Ms-\Zs$ correlation down to the smallest stellar masses in the UFD regime. This success is only in part due to the assumed metallicity floor of $\Zigm=10^{-3}\,\Zsun$ in these models motivated by the expected pre-enrichment from the Population III stars (see Section~\ref{sec:zevo}). 
Note, for example, that the $\Ms-\Zs$ relation flattens at the value $\Zigm\approx 0.002\,\Zsun$ -- twice larger than the assumed metallicity floor. However, Figure~\ref{fig:msz_rei_zigm} shows that if the metallicity floor is set at $\Zigm=10^{-5}\,\Zsun$ the relation changes only mildly at the smallest masses and the model still produces metallicities of the smallest galaxies quite close to the metallicities of observed UFDs. The figure also shows that the relation is not sensitive to the gas suppression due to UV heating after reionization as the relation for the model where this suppression is not included is the same as when it is included. This implies that the $\Ms-\Zs$ relation is shaped by the outflow model rather than by the assumed $\Zigm$ floor or gas inflows. The flattening of $\Ms-\Zs$ relation at small $\Ms$ is induced by the ceiling on the outflow mass loading factor of 2000 we adopt in our model (see Section~\ref{sec:windmodel}). 

A number of recent studies based on galaxy formation simulations argued that the metallicities of the UFDs with the smallest stellar masses are too low if  pre-enrichment by Pop III supernovae is not included via the metallicity threshold  \citep[e.g.,][]{Tassis.etal.2012,Wheeler.etal.2019,Agertz.etal.2020}. However, results of our model indicate that such floor may be less relevant and the discrepancy between simulation results and observations may be due to modelling of feedback and corresponding scaling of outflow mass loading factor with stellar mass. 

Finally, we note that the bottom right panel of Figure~\ref{fig:mszwind} shows results of the model with suppression of mass accretion due to preventative feedback ($\epsinprev=0.6$), but smaller wind normalization ($\etanorm=2.7$) and small wind metallicity enhancement ($\zetaw=1.35$). The comparison shows that this model reproduces observed $\Ms-\Zg$ and $\Ms-\Zs$ relations as well as models without preventative feedback. There is thus degeneracy of the model predictions with respect to the uniform accretion suppression  (see Section~\ref{sec:discussion} for further discussion).

\begin{figure}
    \centering
    \includegraphics[width=0.5\textwidth]{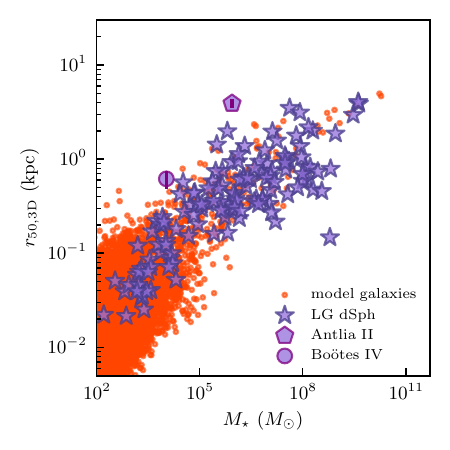}
    \caption{Comparison of model (dots) and observed Local Group dwarf galaxies shown by star symbols, and pentagon and circle for the lowest surface stellar density galaxies Antlia II \citep[][]{Torrealba.etal.2019} and Bo\"otes IV \citep[][]{Homma.etal.2019} in the plane of stellar mass and 3D stellar half mass radius. The half mass radius in the model is estimated as described in Section~\ref{sec:rhalf}. }
    \label{fig:msrhalf}
\end{figure}

\subsection{Half-Mass Radius - Stellar Mass relation}
\label{sec:rhalf}

One of the key parameters of galaxies is half-mass radius of the stellar distribution, $\rhalf$. The size of stellar extent of galaxies is related to surface brightness of galaxies and thus to their detectability. Studies of dwarf galaxies detected in the Local Group over the past decade revealed a wide range of galaxy sizes, from $\approx 15-25$ pc \citep[][]{Willman.etal.2011,Koposov.etal.2015,Drlica-Wagner.etal.2015,Laevens.etal.2015} comparable to sizes of some globular clusters to $\approx 2-3$ kpc \citep[e.g.,][]{Torrealba.etal.2019} comparable to sizes of the Milky Way-sized galaxies. 

Figure~\ref{fig:msrhalf} shows half-mass radii of the model galaxies computed as described in Section~\ref{sec:rhalf} as a function of their stellar mass and compares them to observed dwarf galaxies in the Local Group. For the latter we convert the half-mass radius measured from the projected surface density profile of stars to its 3D equivalent by multiplying the former by $1.34$ appropriate for spheroidal systems. The figure shows good qualitative agreement between sizes of model and observed galaxies. The scatter in the sizes is due largely to the 0.25 dex scatter assumed in the gas disk size model (Section~\ref{sec:sigrmodel}), although at small masses additional scatter in $\Ms$ is produced by reionization model. 

The overall average trend of the $\rhalf-\Ms$ relation, on the other hand, is set by the following two mechanisms.
For galaxies of large masses that are relatively unaffected by UV heating, the overall extent of the $\HH$ disk and thus stellar distribution reflect the model assumption that gaseous disk scale length is proportional to $R_{\rm 200c}$. In fact, 
the model galaxies at large masses are consistent with the scaling $\rhalf\approx 0.02R_{\rm 200c}$ inferred from observed sizes and converting stellar mass to halo mass using abundancing matching $\Ms-\Mh$ relation \citep[][]{Kravtsov.2013}. The tight $\Ms-\Mh$ relation translates this linear relation to a non-linear $\rhalf-\Ms$ relation. 

For smaller galaxies ($\Mh\lesssim 10^{10}\,\Msun$) the scatter in the $\rhalf-\Ms$ relation increases and relation steepens slightly. This is mainly due to the increasing scatter of $\Ms$ at small mass haloes discussed in Section~\ref{sec:msmh}. 
We find that gas suppression due to UV heating has negligible effect on the mean $\rhalf-\Ms$ trend. This is because at halo masses where reionization steepens the slope $\alpha$ of the mean $\Ms-\Mh$ relation, it also steepens the slope of $\rhalf-\Rtwoh$ relation for the corresponding galaxies. 
Given that $R_{\rm 200c}\propto \Mtwoh^{1/3}\propto \Ms^{1/{(3\alpha)}}$ and thus $\rhalf\propto \Rtwoh^\beta\propto\Ms^{\beta/{(3\alpha)}}$, the increase of both $\beta$ and $\alpha$ due to reionization nearly cancels
in the $\rhalf-\Ms$ relation. In fact, the $\rhalf-\Ms$ relation does not change significantly even if we do not include gas accretion suppression at all.

\begin{figure}
    \centering
    \includegraphics[width=0.5\textwidth]{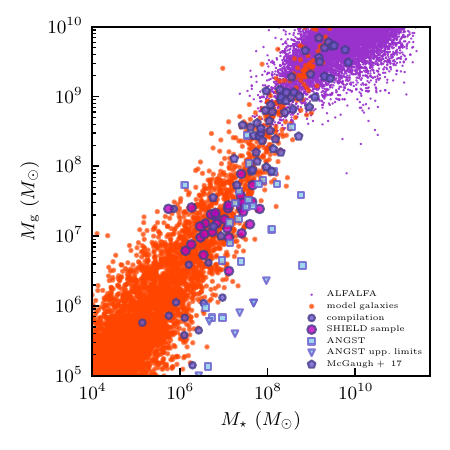}
    \caption{The $\Mg-\Ms$ relation in the fiducial model with $\etanorm=3.6$, $\etapow=0.35$ and $\zrei=6$ (shown by red dots) compared to observed galaxies from the ALFALFA survey \citep[][magenta dots]{Durbala.etal.2020}, ANGST survey \citep[stellar masses from Table 2 of][while HI masses are from Table for of \citealt{Ott.etal.2012}, squares and triangles for upper limits]{Weisz.etal.2011} that shows both measurements and upper limits for galaxies for which HI is not detected, sample of low surface brightness dwarf galaxies of \citep[][small pentagons]{McGaugh.etal.2017}, and a compilation of several dwarf galaxies with stellar masses $\Ms\sim 10^5-10^7\,\Msun$ from different sources shown as blue circles: LGS-3 \citep{Hidalgo.etal.2011}, UGC 4879 \citep{Kirby.etal.2012}, several small-mass galaxies (ESO410-005, ESO294-010, Leo T, Antlia, KKH 86, KKH 98, DDO210) using measurements compiled in Table 2 of \citet{Dooley.etal.2017}, Leo P \citep{McQuinn.etal.2015}, Leoncino \citep{McQuinn.etal.2020}, and Antlia B \citep{Hargis.etal.2020}.}
    \label{fig:msmgfid}
\end{figure}

\begin{figure*}
    \centering
    \includegraphics[width=0.49\textwidth]{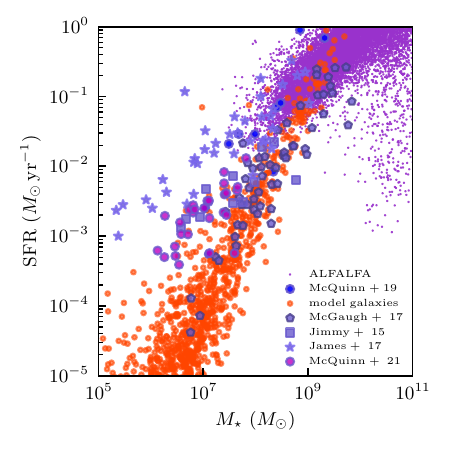}
    \includegraphics[width=0.49\textwidth]{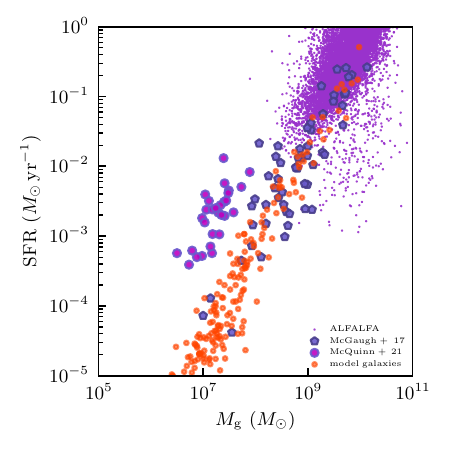}
    \caption{The $\Ms-\sfr$ (left panel) and $\Mg-\sfr$ (right panel) relations in the fiducial model with $\etanorm=3.6$, $\etapow=0.35$ and $\zrei=6$ (shown by red dots) compared to observed galaxies from the ALFALFA survey \citep[][magenta dots]{Durbala.etal.2020} and samples of \citet[][squares]{Jimmy.etal.2015}, \citet[][triangles]{Teich.etal.2016}, \citet[][pentagons]{McGaugh.etal.2017}, \citet[][stars]{James.etal.2017}, \citet[][blue circles]{McQuinn.etal.2019}, and galaxies in the SHIELD I and II samples \citep[][McQuinn et al. 2021, submitted]{McQuinn.etal.2015}. Note that we use $\sfr$ defined using eq.~\ref{eq:sfr} without accounting for stellar mass loss. }
    \label{fig:msfrfid}
\end{figure*}

\subsection{Gas Mass - Stellar Mass relation}
\label{sec:msmg}

Figure~\ref{fig:msmgfid} compares relation of stellar mass and the ISM gas mass of model galaxies in the fiducial model to existing measurements of these quantities in nearby dwarf galaxies.\footnote{Note that for the range of masses shown $\MHI$ estimated in our model is almost identical to the total gas mass in the disk. We therefore do not differentiate between $\Mg$ and $\MHI$ in this discussion.} 
There is a good overall agreement between average trends in the model and observations. The scatter of gas mass at a given $\Ms$ is comparable in the upward direction and is somewhat smaller than observed in the downward direction. This may be due to the fact that we did not include observational uncertainties into the model results or a missing source of scatter in the model.
For example, in the current model we do not include a model for gas stripping due to ram pressure or tides, although such a model can be included given that tracks of each halo relative to host halo are available in the ELVIS suite. 
Some of the observed outliers in the downward direction may be affected by gas stripping as some of these galaxies (e.g., galaxies in the ANGST sample) are located near massive galaxies.

The scatter also increases somewhat for larger reionization redshifts. However, overall we find that the $\Ms-\Mg$ relation is
remarkably insensitive to the parameters of the outflow model or UV gas suppression within the range of model variations discussed in the previous sections. Although the relative number of model galaxies in different parts of the correlation changes in different models, the overall trend and scatter remain almost unchanged. 

At the same time, we find that this relation is quite sensitive to the details of molecular gas modelling, especially at $\Ms\lesssim 10^8\,\Msun$. This is not surprising because molecular gas controls star formation rate and thus its integral -- the stellar mass -- and different models predict different abundance of molecular gas for a given gas mass and surface density profile. We illustrate this in the Figure~\ref{fig:msmgh2} in the Appendix~\ref{app:mhh2}, which shows similar comparison for the modified model for $f_{\rm H_2}$ from \citet{Gnedin.Draine.2014} and the $f_{\rm H_2}$ model used in DK12.

The $\Ms-\Mg$ relation probes gas fractions or gas-to-stellar mass ratios in model galaxies, but we choose to consider the $\Ms-\Mg$ relation directly because gas fraction is correlated with $\Ms$ and strong, distracting trends can be introduced in the plot of correlated quantities in the presence of large scatter.

\subsection{Star Formation Rates at $z=0$}
\label{sec:sfr}

An additional test of star formation prescription in model is comparison to the observed correlations between star formation rate (SFR) and $\Ms$ and $\Mg$ shown in Figure~\ref{fig:msfrfid} for the fiducial model. In the $\Ms-\sfr$ panel (left) we use a number of different dwarf galaxy samples, in which star formation rates were measured using different methods. For the ALFALFA sample we use $\Ms$ and SFRs estimates using GSWLC-2 method, which should  give the most reliable estimates \citep[][]{Durbala.etal.2020}, and we checked that results are qualitatively similar if we use SFR estimates based on the corrected near UV fluxes.

The figure shows that star formation rates in model galaxies of a given $\Ms$ and $\Mg$ are close to the distribution of observed star formation rates. The scatter of observed SFRs is  larger, which may partly be due to uncertainties in observational SFR estimates, different observational indicators probing star formation on different time scales \citep[e.g.,][]{Lee.etal.2009,McQuinn.etal.2015b}, and partly due to insufficient sources of scatter in the model. One such source of scatter is burstiness of star formation. For example, dwarf galaxies in the \citealt{McGaugh.etal.2017} sample were selected from a catalog of low surface brightness galaxies, which biases against systems with bright HII regions and recent bursts of star formation, and galaxies in that sample have lowest star formation rates at a given stellar mass. On the other hand, blue diffuse galaxies in the sample of \citet[][shown by stars in the left panel of Fig.~\ref{fig:msfrfid}]{James.etal.2017}  are selected using criteria that are sensitive to presence of bright HII regions and their star formation rates are measured using H$\alpha$ flux sensitive to star formation in the past $\lesssim 5$ Myrs \citep{Kennicutt.Evans.2012,Haydon.etal.2020,FloresVelazquez.etal.2021}. These galaxies thus are likely a subset of dwarfs that undergo a current burst of star formation. Indeed, they have systematically larger SFRs than other dwarf galaxies of similar stellar mass \citep[see also][]{Barat.etal.2020}. 

Galaxies in the SHIELD sample are selected based on their HI mass and tend to be gas rich for their stellar mass, which likely also biases the sample towards galaxies in a star formation burst state. If this interpretation is correct, this implies that dwarf galaxies undergo bursts in which their SFR increases by an order of magnitude on time scales $\sim 5-10$ Myrs. Such behavior is indeed observed in the FIRE simulations of dwarf galaxies \citep[][]{Sparre.etal.2017,Pandya.etal.2020}. This implies that to model objects such as blue diffuse galaxies or the full span of properties such as colors, a model for star formation burstiness should be included \citep[see, e.g.,][]{Faucher_Giguere.2018,Tacchella.etal.2020}. 
At the same time, the difference in the SFRs at a given $\Mg$ for the HI-selected SHIELD I and II samples and the low surface brightness galaxies in the \citet{McGaugh.etal.2017} sample that is shown by the right panel of Figure~\ref{fig:msfrfid} is striking and warrants a more detailed theoretical study. 

Similarly to the $\Ms-\Mg$ correlation, the $\Ms-\sfr$ and $\Mg-\sfr$ correlations in our model are not sensitive to the variations of the wind model parameters, which only affect the number of galaxies with a given $\Ms$, $\Mg$ and $\sfr$, but not the overall 
form of the correlation. Even the model with constant mass-loading factor has form close to observed correlations, although in this model scatter of SFRs is significantly smaller. This insensitivity simply reflects the fact that star formation is assumed to be proportional to molecular gas, which depends on total gas mass and characteristic size of radial gas distribution (i.e., gas surface density profile). Given that winds do not affect typical gas masses for galaxies of a given $\Ms$ and size of gas distribution is assumed to be independent of outflows in our model, $\MHH$ and SFR for galaxies with given $\Ms$ and $\Mg$ are also are not affected by outflows. 

Just like the $\Ms-\Mg$ correlation, on the other hand, correlations involving SFR are sensitive to the $\HH$ model used, as we illustrate in Figure~\ref{fig:mgsfrh2} in Appendix~\ref{app:mhh2}.  

\begin{figure}
\begin{center}
    \includegraphics[width=0.5\textwidth]{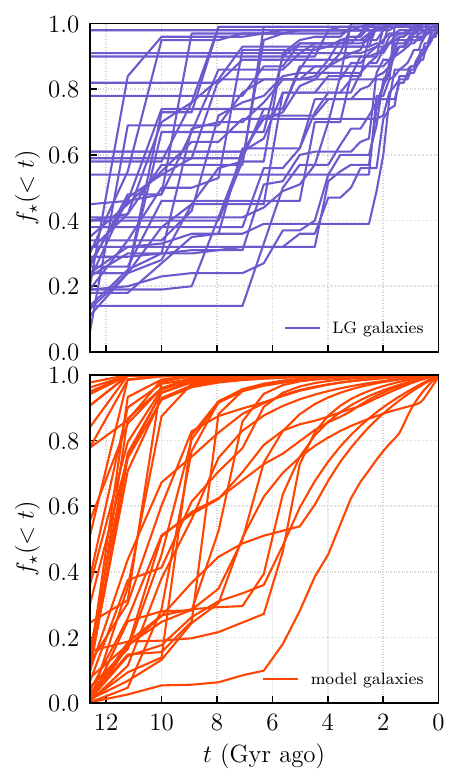}
\end{center}
    \caption{Comparison of cumulative star formation histories, $f_\star(<t)$ -- the fraction of the final stellar mass formed by a given lookback time $t$ -- of dwarf galaxies in the Local Group \citep[][top pael]{weisz_etal14} and model galaxies in the fiducial model with $\etanorm=3.6$, $\etapow=0.35$ and $\zrei=6$ (lower panel). To make sure that model galaxies have the same distribution of stellar masses as galaxies in the Local Group sample, in the lower panel we show a random subset of model galaxies equal in number to the number of observed galaxies in the upper panel with stellar masses of selected model galaxies within 0.1 dex of each observed galaxy. }
    \label{fig:fms_comp}
\end{figure}

\begin{figure*}
\begin{center}
    \includegraphics[width=\textwidth]{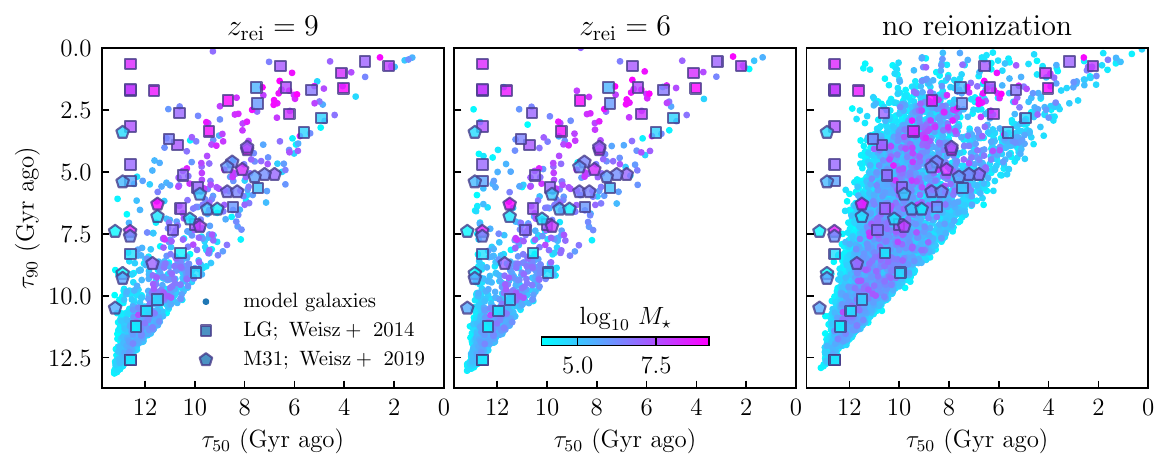}
\end{center}
    \caption{The lookback time at which galaxies formed $50\%$ of their final stellar mass, $\tau_{50}$, vs the time at which they formed $90\%$ of their final mass, $\tau_{90}$. The model galaxies are shown as small circles, while galaxies in the LG sample of \citet{Weisz.etal.2014}  are shown by squares and satellites of M31 \citep[][]{Weisz.etal.2019} are shown by pentagons. Both model and observed galaxies are colored by the logarithm of their stellar mass with mapping shown by the color bar in the middle panel. The three panels show models with the fiducial parameters, but different assumptions about reionization: $\zrei=9$ (left panel), $\zrei=6$ (middle panel), and a model with no UV heating included (right panel). The model galaxies are selected to have stellar masses $>7\times 10^3\,\Msun$ to match the minimum stellar mass of the LG sample. }
    \label{fig:t50t90_rei}
\end{figure*}

\subsection{Star Formation Histories}
\label{sec:sfh}

Although reasonable agreement with current star formation rates of observed dwarf galaxies is encouraging, photometric and spectroscopic properties of galaxies are determined by their entire star formation history (SFH). In this section we compare SFHs derived for galaxies in the Local Group and in our model. 

Figure~\ref{fig:fms_comp} shows cumulative SFHs for observed and model galaxies. To make comparison meaningful we match the number of observed galaxies in the sample of \citet{Weisz.etal.2014} shown in the upper panel (forty) in the sample of model galaxies. 
We also select each model galaxy to be within 0.1 dex of the logarithm of stellar mass of an observed galaxy, so that the two samples have similar distribution of stellar mass. The figure shows that model SFHs exhibit diversity qualitatively similar to the diversity of SFHs of LG dwarfs. In particular, there are both galaxies that form most of their stars at early times, $t>10$ Gyr, and ``late bloomers'' that build up a large fraction of their stellar mass at late epochs. Note, however, that observational SFHs have considerable uncertainties \citep[see][]{Weisz.etal.2014} that are not included in the model SFHs. The form and mass dependence of the model SFHs is also in good qualitative agreement with star formation histories of simulated dwarf galaxies in high-resolution FIRE simulations \citep[][]{Fitts.etal.2017,Garrison_Kimmel.etal.2019,Wheeler.etal.2019}.

To make a more quantiative comparison in Figure~\ref{fig:t50t90_rei} we show distributions of the lookback times at which galaxies formed $50\%$ of their final stellar mass, $\tau_{50}$, vs the time at which they formed $90\%$ of their final mass, $\tau_{90}$ for both the model galaxies and LG galaxies. The time $\tau_{50}$ can be thought of as a characteristic formation time of galaxy's stellar mass, while $\tau_{90}$ is an indicator of when most of the stars were assembled and can be a probe of the epoch of quenching of star formation.  

The three panels in Figure~\ref{fig:t50t90_rei} show models with the fiducial parameters, but different assumptions about reionization: $\zrei=9$, $\zrei=6$, and a model with no UV heating included. Not to overcrowd the figure, we only included model galaxies around five Milky Way-sized haloes in the ELVIS suite. The figure shows that models with UV heating after reionization reproduce the range of $\tau_{50}$ and $\tau_{90}$ distribution of observed galaxies remarkably well. They also qualitatively reproduce the trend  exhibited by observed galaxies for more massive galaxies to have smaller $\tau_{90}$. 

The model without reionization has many more galaxies in the plotted stellar mass range ($M_\star>7\times 10^3\,\Msun$) because gas accretion in many small-mass haloes that would be suppressed by UV heating proceeds unimpeded and results in extended star formation with many more haloes building up sizeable stellar mass. The distribution of $\tau_{50}$ and $\tau_{90}$ is also notably different in this model in that it misses the population of galaxies with $\tau_{50}>12$ Gyr and $\tau_{90}<12$ Gyr where a dozen observed galaxies are present. This is because the absence of UV heating shifts $\tau_{50}$ and $\tau_{90}$ in this model to smaller lookback time values. Thus, in the framework of our model the  area $\tau_{50}>12$ Gyr and $\tau_{90}<12$ Gyr of the $\tau_{50}-\tau_{90}$ distribution thus corresponds to galaxies that are uniquely affected by reionization. 

In the model with $\zrei=9$ this area is populated by a number of galaxies but this model lacks galaxies with $\tau_{50}>12$ Gyr and $\tau_{90}<5$ Gyr. Interestingly, in the $\zrei=6$ model this area does have several model galaxies and overall observed distribution is well matched.  

Another difference of models with and without reionization is that in the latter there are many galaxies that have $\tau_{90}<2$ Gyr, while in both observed samples and model galaxies in models with $\zrei=9$ and $\zrei=6$ this region is sparsely populated. 
This is likely due to the fact that characteristic mass below which gas accretion rate is suppressed increases with decreasing redshift and many of small-mass haloes have gas accretion suppressed at late epochs that quenches their star formation and results in $\tau_{90}\gtrsim 2$ Gyr. 

Effects of reionization are often assumed to correspond to a characteristic cosmic epoch around $\zrei$.  However, even after that epoch both the IGM temperature and UV heating rate evolve, which manifests in the redshift dependence of the characteristic mass $M_{\rm c}(z)$. This increase is substantial and affects evolution of systems of a wide range of masses even at late epochs. Comparison shown in Figure~\ref{fig:t50t90_rei} thus indicate that paucity of observed galaxies with $\tau_{90}<2$ Gyrs is a signature of the gas accretion suppression due to UV heating that affects dwarf galaxies at late epochs. 

At the same time, it is notable that $\tau_{90}$ of galaxies in the model without UV heating (right panel) span a wide range of values, $\sim 12.5-0$ Gyr. The only quenching process for these galaxies in the model is related to star formation process. In the model molecular gas is required for star formation, and molecular gas fraction depends on the gas surface density. When galaxy does not accrete new gas at a sufficient rate to replenish the gas consumed by star formation and lost to outlfows, gas surface density decreases until molecular fraction becomes negligible and star formation ceases. Thus, galaxy can be quenched simply due to low gas surface density during low mass accretion rate periods of evolution. Such periods may occur because halo mass does not grow significantly. The
``internal quenching'' of this kind may thus be a quenching channel for dwarf galaxies that is not related to environmental effects, such as ram pressure stripping. 

Overall, the comparisons shown in these figures indicate that our simple model with $\zrei\sim 6-9$ produces SFHs in qualtitative agreement with estimates derived for observed galaxies. It is not clear whether the relative {\it distribution} of galaxies with a given shape of SFH and thus in different parts of the $\tau_{50}-\tau_{90}$ diagram is sufficiently similar. For such comparison, observational selection effects and environmental selection (e.g., distance to the host) should be modelled carefully and we plan to carry out such comparison in future work.

\subsection{Photometric properties}
\label{sec:photometry}

\begin{figure}
\begin{center}
    \includegraphics[width=0.45\textwidth]{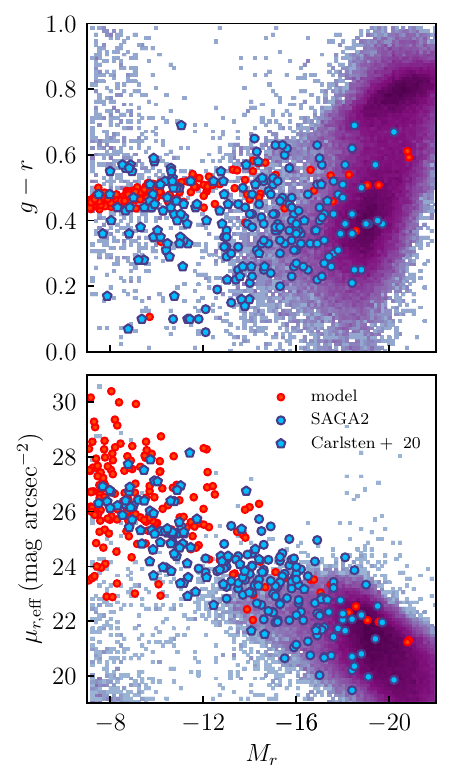}
\end{center}
    \caption{$g-r$ color (top panel) and average $r$-band surface brightness within half-light radius $\mu_{r,\rm eff}$ (bottom panel) vs $r$-band absolute magnitude, $M_r$. The model galaxies in the fiducial model are shown by red circles, while 2d histogram shows distribution for the SDSS galaxies at $z<0.07$ and blue circles and pentagons show dwarf galaxies around nearby $\sim L_\star$ galaxies in the SAGA sample \citep[][]{Geha.etal.2017,Mao.etal.2021} and sample of \citet{Carlsten.etal.2020}, respectively.  }
    \label{fig:grmur_comp}
\end{figure}

\begin{figure}
\begin{center}
    \includegraphics[width=0.45\textwidth]{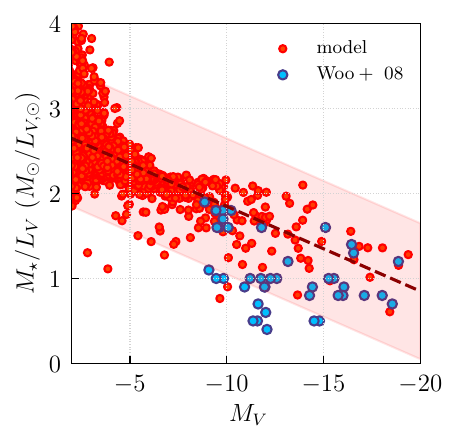}
\end{center}
    \caption{Stellar mass-to-light ratio in the $V$ band vs the $V$-band absolute magnitude. Model galaxies in the fiducial model (red points) are compared to estimates for the Local Group dwarf galaxies \citep{woo_etal08}.}
    \label{fig:mlv_comp}
\end{figure}

Comparisons presented in the previous sections show that the fiducial model matches most of the observed properties of dwarf galaxies quite well. Predictions of such model can thus be used to forward model observed properties of dwarf galaxies for comparisons with observations using quantities directly estimated from observations. Here we illustrate the usefulness of such forward-modelling capability using several interesting examples.

First, in Figure~\ref{fig:grmur_comp} we show comparison of the $g-r$ color and average surface brightness within half-light radius, $\mu_{r,\rm eff}$, vs $r$-band absolute magnitude distributions for model and observed galaxies. Namely, we compare to the distribution of these quantities for galaxies in the SDSS DR13 survey selected to have redshifts $z<0.07$ and reliable photometry and dwarf galaxies around nearby $\sim L_\star$ galaxies in the SAGA sample \citep[][]{Geha.etal.2017,Mao.etal.2021} and sample of \citet{Carlsten.etal.2020}. 

Magnitudes for model galaxies are computed using Flexible Stellar Population Synthesis model \citep[FSPS;][]{Conroy.etal.2009,Conroy.Gunn.2010}\footnote{\href{https://github.com/cconroy20/fsps}{\tt https://github.com/cconroy20/fsps}} using the SFHs and stellar metallicity evolution calculated by the model. 
We assume \citet{Chabrier.2003} IMF and default FSPS parameters. The figure shows that the model matches distribution of observed galaxies very well in the $\mu_{r,\rm eff}-M_r$ plane. The scatter in $g-r$ of observed galaxies is considerably larger than in the model, particularly for galaxies with $M_r\in [-19,-15]$. Model galaxies, match the reddest galaxies in the observed range, but there are many more galaxies with bluer colors among observed galaxies at these magnitudes than in the model. This is likely related to existence of galaxies with large specific star formation rates at these luminosities discussed above in Section~\ref{sec:sfr}, which we interpret as the evidence for burstiness of star formation in real dwarfs not included in our model. 

Figure~\ref{fig:mlv_comp} shows stellar mass-to-light ratio in the $V$ band, $M_\star/L_V$ for model galaxies and estimates for the Local Group galaxies from \citet{woo_etal08} their $B-V$ color. There are many model galaxies that can be considered counterparts of observed galaxies in this plane. However, the distribution of $M_\star/L_V$ of model galaxies is broader and exhibits a clear trend with $M_V$. The average trend can be described by 
\begin{equation}
M_\star/L_V=2.85 + 0.1\,M_V, 
\end{equation}
which is shown by the dashed line in the figure. Most model galaxies fall within $\pm 0.8$ in $M_\star/L_V$ around this trend (this area is shown by the shaded region in Fig.~\ref{fig:mlv_comp}), although the scatter increases significantly for model galaxies with $M_V\gtrsim -4$. This trend indicates that assumption of constant $M_\star/L_V$ often used in theoretical modelling of dwarf satellites is likely not accurate and much larger $M_\star/L_V$ values and scatter may exist for the ultra-faint dwarf galaxies. 

\begin{figure}
\begin{center}
    \includegraphics[width=0.5\textwidth]{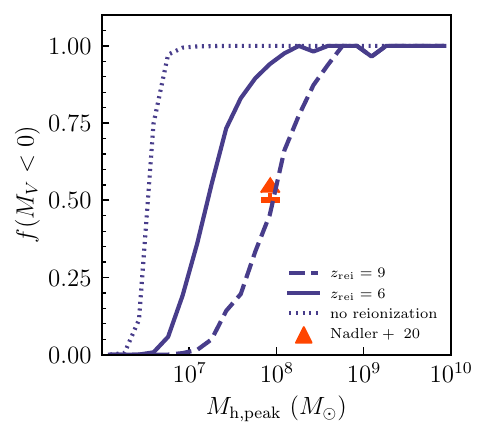}
\end{center}
    \caption{Fraction of haloes that host galaxies with $V$-band absolute magnitudes $M_V<0$ as a function of peak mass halo achieved during its evolution. Three lines show results of the fiducial model for three choices of reionization modelling: $\zrei=9$ (dashed line),  $\zrei=8$ (dot-dashed line), $\zrei=6$ (solid line), and no reionization (dotted line). 95\% confidence lower limit from analysis of abundance of ultra-faint Milky Way satellites by \citet[][]{Nadler.etal.2020} is shown by red arrow.  }
    \label{fig:fhl_zrei}
\end{figure}

\begin{figure}
\begin{center}
    \includegraphics[width=0.5\textwidth]{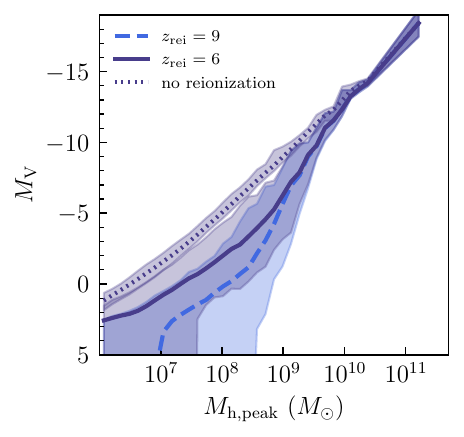}
\end{center}
    \caption{$V$-band absolute magnitude of model galaxies as a function of peak halo mass in the fiducial model for three choices of reionization modelling: $\zrei=9$ (dashed line), $\zrei=6$ (solid line), and no reionization (dotted line). The lines show medians for model galaxy distributions in bins of $\Mhp$, while shaded areas for $\zrei=6$ (dark blue) and $\zrei=9$ (light blue) show regions containing $95\%$ of galaxies. The $95\%$ region in the model with no reionization is much smaller ($\approx 0.5-1$ magnitudes) than in other models. The figure shows that reionization in our model has significant effect on the luminosities of ultrafaint galaxies with $M_V\gtrsim -7$ forming in haloes of a given mass, and has a more modest but still sizeable effect on galaxies of $-13\lesssim M_V\lesssim -7$.}
    \label{fig:mhmv_zrei}
\end{figure}
\subsection{Effect of reionization on the halo occupation fraction in the ultra-faint galaxy regime}
\label{sec:fhl}

Another illustration of the use of photometric predictions is the capability to predict the halo occupation fraction for galaxies with specified luminosity range.  
Figure~\ref{fig:fhl_zrei} shows the fraction of haloes that host galaxies brighter than $M_V=0$, $f_{\rm h}(M_V<0)$, as a function of peak halo mass in the fiducial model with three choices of reionization modelling: $\zrei=9$, $\zrei=6$, and no reionization. The latter model is not realistic but provides a useful baseline for gauging effects of reionization. The magnitude threshold of  $M_V=0$ is chosen to be close to the absolute magnitude of Cetus II of the faintest galaxy detected around Milky Way so far \citep[][]{drlica_wagner_etal20}. Although results are qualitatively similar for the fiducial resolution and high-resolution ELVIS haloes for $\Mhp>5\times 10^7\,\Msun$, here we use the high-resolution simulations to minimize simulation resolution effects at the smallest halo masses and to estimate the halo occupation fraction more robustly at masses $\Mhp\lesssim 10^7\,\Msun$. 

The figure shows that the fraction of {\it satellite} subhaloes that host galaxies with $M_V<0$ is predicted to be quite sensitive to reionization redshift at $\Mhp\lesssim 10^9\,\Msun$. This is because gas accretion and star formation in such haloes is shut down after reionization and the ability of such haloes to build stellar mass depends on the amount of time between the epoch when they become sufficiently massive to form stars $\Mh\gtrsim 10^6\,\Msun$ and the local epoch of reionization. 
Given that the mass function of subhaloes predicted in CDM simulations increases towards small masses approximately as $N(>\Mhp)\propto \Mhp^{0.9}$ \citep[e.g.,][]{Kravtsov.etal.2004b}, the significantly different occupation fractions for the models shown in Figure~\ref{fig:fhl_zrei} imply significantly different numbers of detectable ultrafaint galaxies \citep[see also][]{bose_etal18}. These results thus indicate that the local reionization redshift in the Local Group volume can be constrained using abundance of ultrafaint satellites. We will present such a constraint in a companion paper (Manwadkar \& Kravtsov, in prep.).

Note that the occupation fraction of zero does not necessarily mean that haloes at these masses are completely dark, but that they host galaxies dimmer than $M_V=0$. Figures~\ref{fig:fhl_zrei} and  \ref{fig:mhmv_zrei} show that the model does predict existence of galaxies with $M_V>0$. Given that 
such galaxies have not been discovered yet, this is a {\it pre}diction of the model. Identifying such galaxies in observations is challenging due to small number of stars that can be used to measure 
velocity dispersion and metallicity distribution. However, such measurements may become feasible with the advent of next generation large optical telescopes. 

The figure shows that in the fiducial model with $\zrei=6$ galaxies of $M_V<0$ can be hosted in haloes with peak masses as small as $\Mhp\approx 10^7\,\Msun$. This is consistent with recent observational constraints based on the abundance of Milky Way satellites and assumptions about mapping of their luminosity and halo mass \citep[][]{Jethwa.etal.2018,Nadler.etal.2020}, which indicate that the characteristic peak halo mass at which $50\%$ of haloes do not host detectable galaxies, is  $\Mhalf\lesssim 10^8\,\Msun$ at $95\%$ confidence level. Specifically, the 95\% confidence lower limit on $f(M_V<0)$ obtained by \citet[][]{Nadler.etal.2020} based on modelling Milky Way satellite population is shown by red arrow in Fig.~\ref{fig:fhl_zrei}. In the context of our model this lower limit implies that the cosmic neighborhood of the Milky Way was reionized at $\zrei\lesssim 9$. This is consistent with the range of reionization redshifts predicted by the CROC simulations of reionization for host halos with $\Mh\approx 10^{12}\,\Msun$ at $z=0$ \citep[][]{Zhu.etal.2019} and results of the CoDa II simulations, which show that outer regions of the Milky Way and M31 Lagrangian regions are reionized at $z\gtrsim 7$ \citep[][see their Fig. 11 ]{Ocvirk.etal.2020}.

This agreement implies that the luminosity function of Milky Way satellites predicted using our model with $\zrei\lesssim 9$ is in general agreement with observations.  We will present a detailed
comparisons of our model predictions for the satellite luminosity function and other statistics taking into account subhalo disruption processes and proper treatment of galaxy detectability as a function of their surface brightness and distance to the host center in a follow-up paper. 

We also show the median of the $M_V$ distributions in bins of $\Mhp$ for the same three reionization model choices  in Figure~\ref{fig:mhmv_zrei}. The shaded areas for $\zrei=9$ and $\zrei=6$ models show regions containing $95\%$ of absolute magnitudes in a given $\Mhp$ bin. The figure shows that reionization in our model has significant effect on the luminosities of ultrafaint galaxies with $M_V\gtrsim -7$ forming in haloes of a given mass, and has a more modest but still sizeable effect on galaxies of $-13\lesssim M_V\lesssim -7$. Notably, the figure shows that due to effects of reionization the median $\Mhp-M_V$ relation predicted by the models with reionization is not well described by a single power law at $\Mhp\lesssim 5\times 10^9\,\Msun$ or galaxies with $M_V\gtrsim -8$ -- the regime where reionization induced scatter also increases significantly. Earlier reionization produces considerably more scatter. For example, a galaxy with $M_V=0$ can be hosted in haloes with masses of $\Mhp\approx 5\times 10^6-10^9\,\Msun$ depending on the redshift of reionization. 

\section{Discussion}
\label{sec:discussion}

Reasonable qualitative agreement of model galaxy properties with corresponding observed properties of galaxies is relevant for forward modelling populations of dwarf galaxies. For example, our model matches the observed $\rhalf-\Ms$ relation quite well and correspondingly the range of surface brightnesses of observed dwarf galaxies (see Figs.~\ref{fig:msrhalf} and \ref{fig:grmur_comp} above). The model can thus be used to make forecasts for observational surveys or interpretation of observations.

It is intriguing that such a simple model can match a wide range of observed properties of dwarf galaxies. This may indicate that
formation of these galaxies is governed by a relatively straightforward mass exchange and conversion processes. However, we cannot
discard a possibility that real formation processes are more complicated than envisioned in this simple framework, but observations are not sufficiently constraining. For example, we showed that the fiducial model, which does not account for any preventative feedback processes beyond the UV heating after reionization, can match a wide range of observable properties of dwarf galaxies. 
However, as Figures~\ref{fig:msmhwind} and \ref{fig:mszwind} show, the data can also be matched equally well by a model in which gas accretion is suppressed by $40\%$ ($\epsinprev=0.6$) -- the amount similar to that measured for the FIRE simulations (see dashed lines vs. points in Fig.~\ref{fig:mgin_ratio}) -- but with a modest wind metallicity enhancement ($\Zw/\Zg=1.35$) and reduced normalization of the wind mass loading factors $\etanorm=2.7$, which is consistent with recent analysis of \citet{Pandya.etal.2021}.
Thus, models with such suppression (e.g., due to preheating of gas around massive galaxies by shocks and winds) are also consistent with observed dwarf galaxy population. 

The fact that our fiducial model works as well as the model with preventative feedback may be surprising in light of results of several previous studies, which concluded that observations of dwarf galaxies cannot be reproduced without preventative feedback \citep[e.g.,][]{Lu.etal.2015,Lu.etal.2017,Xia.Quinjuan.2019}. In particular, \citet{Lu.etal.2017} have concluded that 
while mass-dependent outflows can reproduce the observed-mass metallicity relation, the stellar mass function of satellite galaxies in the Local Group (and thus $\Ms-\Mh$ relation) cannot be reproduced without preventative feedback due to preheating. This 
seemingly contradicts results in the previous section, where we show that the fiducial model with winds can reproduce both
the observational constraints on the $\Ms-\Mh$ relation and mass-metallicity relation. 

This contradiction may be due to differences
in the gas fractions produced by our model and the model of \citet{Lu.etal.2017}. We showed that gas fractions in the model 
are sensitive to details of $\HH$ model that controls star formation, while in the \citealt{Lu.etal.2017} model star formation is handled differently without modelling molecular gas. Scaling of gas fractions with galaxy mass directly affect the $\Ms-Z$ relation along with the wind mass loading factor. Specifically, gas metallicity is given by equation~\ref{eq:ps11eq} above \citep[][]{Peeples.Shankar.2011} and depends on both the wind mass loading factor {\it and} ratio of ISM gas mass to stellar mass, $\Mg/\Ms$.
Thus, a scaling of $\etaw$ that produces correct $\Ms-\Mh$ relation may not reproduce observed 
$\Ms-Z$ relation if gas fractions in the model are off. \citet{Lu.etal.2017} did not consider gas fractions in their study, but we show here that it is possible to reproduce $\Ms-\Mh$, $\Ms-Z$, and $\Ms-\Mg$ relations simultaneously with only gas accretion due to UV heating, but not due to any additional preheating. The most likely source of the difference between our models is thus in the gas fractions that they predict. 

Equation~\ref{eq:ps11eq} also illustrates that mass scalings of $\zetaw$ and $\etaw$ are degenerate, as only the product $\zetaw\etaw$ affects ISM metallicity \citep[see also][]{Lu.etal.2015b}. This is why the model of \citet{Krumholz.Dekel.2012} in which a constant $\etaw$ but mass-dependent $\zetaw$ were assumed can match observed $\Ms-Z$ reasonably well. However, as we showed in Figure~\ref{fig:msmhwind} such model cannot reproduce the observational constraints on the $\Mh-\Ms$ relation of dwarf galaxies. 

In the context of eq.~\ref{eq:ps11eq} it is also interesting that observations down to ultra-faint dwarf regime are best matched by our model with the slope of the scaling of the mass loading factor with stellar mass, $\etaw\propto \Ms^{-\etapow}$, of  $\etapow\approx 0.35-0.4$. As discussed above, this scaling is consistent with the results of galaxy formation simulations
\citep[][]{Muratov.etal.2015}, but it is also consistent with inferences from observations \citep[][]{Chisholm.etal.2017,Sanders.etal.2020}. In particular, based on their modelling of evolution of the observed $\Ms-\Zg$ relation at $\Ms\gtrsim 10^9\,\Msun$ \citet{Sanders.etal.2020} conclude that observations require scaling  $\zetaw\etaw\propto \Ms^{-0.35\pm 0.02}$, which is consistent with our results for constant $\zetaw$. Observations thus indicate that the $\Ms-Z$ relation of galaxies from $\sim L_\star$ to the ultra-faint dwarfs is shaped by the power law close to $\etaw\propto\Ms^{-\etapow}$. 
At the same time, \citet{Sanders.etal.2020} conclude that evolution of the $\Ms-Z$ relation requires that normalization of $\etaw$ evolves somewhat with redshift, while we find that model with constant normalization can match the relation in the dwarf regime 
at $z=0$. Given the diversity of SFHs of dwarf galaxies (see Section~\ref{sec:sfh}) and the fact that many of them formed most of their stars at high redshifts, the model result and tightness of the observed $\Ms-Z$ correlation indicate that in the dwarf regime galaxies may evolve mostly {\it along} the relation, while results of \citet[][]{Sanders.etal.2020} imply that galaxies of larger masses likely evolve differently.  

The scaling of $\zetaw$ with stellar mass is not well known currently. Measurements in the FIRE simulations show $\zetaw\approx 1.2\pm 0.7$ relatively independent of mass \citep{Muratov.etal.2017}, but with outflow metallicity depending on the distance to the galaxy and redshift \citep[see also][]{Pandya.etal.2021}. Observational measurements of \citet[][]{Chisholm.etal.2018} indicate that $\zetaw$ may scale with stellar mass. Thus, although many models of the type used in this study assume $\Zw=\Zg$, it will be warranted to consider models with $\Zw\ne\Zg$ and explore ways to explore scalings of $\zetaw$ using observations, especially because inclusion of such scaling in the model is straightforward (see Section~\ref{sec:zevo}).  

As we commented in Section~\ref{sec:msmg}, the gas fraction in model dwarf galaxies with $\Ms\lesssim 10^8\,\Msun$ is quite sensitive to details of modelling of star-forming gas fraction or, in the context of our model, of molecular hydrogen (see Fig.~\ref{fig:msmgh2}). This sensitivity implies that measurements of stellar mass and gas mass in dwarf galaxies of the smallest masses can be used as a sensitive 
test of such model. Conversely, this sensitivity implies that one must be careful with modelling $\HH$ and star formation 
in galaxies at these masses both in the frameworks of the kind used in this study and in galaxy formation simulations. This also  motivates further careful efforts to calibrate $\fHH$ models using simulations. 

An interesting aspect of the model is that for $\zrei\lesssim 9$ it predicts high occupation fraction of galaxies with ultra-faint luminosities in haloes of $\lesssim 10^8\,\Msun$ (see Fig.~\ref{fig:fhl_zrei}), which is actually required by the observed abundance of ultra-faint  Milky Way satellites \citep[][]{Jethwa.etal.2018,Graus.etal.2019,Nadler.etal.2020}. This is despite the fact that model includes  processes that strongly limit or suppress build-up of stellar mass in model galaxies, such as gas accretion suppression due to UV heating, strong feedback-driven gas outflows, and inefficiency of dense molecular gas formation in gaseous disks of low surface density. 

Results of cosmological galaxy formation simulations, on the other hand, were often used to argue that UV heating efficiently suppresses galaxy formation in haloes with masses $\lesssim 10^9\,\Msun$ \citep[e.g.,][]{Shen.etal.2014,Sawala.etal.2015,Sawala.etal.2016,Ocvirk.etal.2016,Fitts.etal.2017,Jahn.etal.2019}. Given the strong dependence of the occupation fraction on the reionization redshift shown in Figure~\ref{fig:fhl_zrei}, such predictions can perhaps be explained in part by an earlier reionization assumed in these simulations. Numerical resolution effects may also play a role because it is generally extremely challenging to model dynamics and thermodynamics in the ISM and star formation at the scales of the smallest dwarf systems while modelling them in cosmological volumes \citep[see also discussion in Section 4.1 of][]{Munshi.etal.2021}. For example, results in Fig. 2 of \citet{Sawala.etal.2016} show that galaxy formation is suppressed in haloes of $\lesssim 5\times 10^8\,\Msun$ even without reionization. 

Indeed, high-resolution simulations of galaxies at high redshifts \citep[][see also Fig. 13 in \citealt{Cote.etal.2018} for a compilation of recent simulation results]{Ricotti.Gnedin.2005,Bovill.Ricotti.2009} show that  galaxies with stellar masses of $\sim 10^3\,\Msun$ do form in haloes as small as $\Mh\approx 10^7\,\Msun$ before UV heating suppresses gas accretion onto such haloes after reionization. Likewise, recent high-resolution simulations of dwarf galaxies show that haloes of $\Mh\sim 10^7-10^8\,\Msun$ build up a substantial stellar mass at high redshifts \citep[][]{Katz.etal.2020,Agertz.etal.2020}. This implies that the physics of cooling and feedback processes does not prevent formation of detectable galaxies in such small haloes. This conclusion was also reached by \citet{Bland_Hawthorn.etal.2015} who used high-resolution simulations of small-mass haloes to show that as long as these have some gas mass at the beginning even haloes $\Mh\approx 3\times 10^6\,\Msun$ are able to retain gas and continue star formation following ionization and mechanical feedback by young massive stars \citep[see also][]{Webster.etal.2014}. 

The observational constraints and our model results imply that simulations aiming to model faintest dwarf galaxies will need to target haloes of masses $\lesssim 10^8\,\Msun$, considerably smaller than $\Mh\sim 10^9-10^{10}\,\Msun$ that was often assumed in previous studies. 

Also, as noted in Section~\ref{sec:sfmodel}, in this study we neglect to consider contribution of mergers to the build-up of stellar mass of galaxies. Although this is a justified assumption when considering bulk properties of galaxies \citep[see, e.g.][]{fitts_etal18}, mergers may be important for modelling stellar population components of dwarf \citep[e.g.,][]{tarumi_etal21,chiti_etal21a} and Milky Way sized galaxies \citep[e.g.,][]{chiti_etal21}.

Note that the only environmental effect that is included in our model implicitly is shutting down of gas accretion when halo mass of a halo stops growing due to tidal truncation or stripping due to nearby massive haloes. When this happens model galaxy will experience strangulation and its star formation will decrease exponentially.  In this study we did not {\it explicitly} model other environmental effects thought to be important for dwarf satellite galaxies, such as  gas removal by ram pressure stripping \citep[e.g.,][]{Simpson.etal.2018}, or star formation bursts following gas compression during pericentric passages \citep[][]{DiCintio.etal.2021}. This is a deliberate choice as the model without these effects provides useful baseline predictions for the field dwarf galaxy population. In this regard it is interesting that the colors of dwarf galaxies in such baseline model are consistent with the colors of the reddest satellites of nearby galaxies (see Fig.~\ref{fig:grmur_comp}). 

This fact does not leave much room for environmental effects suppressing star formation in satellite galaxies. This is puzzling, as it is generally expected that dense hot haloes of massive galaxies should strip satellite galaxies of their gas, as seems to have happened for the M31 and MW satellites. 
Given that halo tracks contain spatial trajectories of haloes relative to massive neighbors a model for the environmental effects can be included and we plan to explore such models in a future study. 

\section{Summary and conclusions}
\label{sec:summary}

We presented a simple regulator-type framework designed specifically for modelling formation of galaxies of luminosities up to $\sim L_\star$. The model uses halo mass accretion histories (MAHs) as input and a set of prescriptions for size of the gas disk, molecular fraction of the interstellar medium (ISM), and gas inflows and outflows to evolve gas mass of the ISM, as well as stellar mass and metallicity of the ISM and stars. 

when coupled with realistic mass accretion histories of haloes from simulations and reasonable modelling choices,

We showed that despite its simplicity, the model using realistic mass accretion histories of haloes from simulations and reasonable choices for model parameter values can reproduce a remarkably broad range of observed properties of dwarf galaxies over seven orders of magnitude in stellar mass. 
In particular, we show that the model can {\it simultaneously} match observational constraints on the stellar mass--halo mass relation, as well as observed relations between stellar mass and gas phase and stellar metallicities, gas mass, size, and star formation rate, as well as general form and diversity of  star formation histories (SFHs) of observed dwarf galaxies.
 
Our main results and conclusions can be summarized as follows. 

\begin{itemize}
\item[(i)] We explored sensitivity of model predictions for the stellar mass--halo mass and stellar mass--metallicity relations to different modelling choices and parameter values (Sections~\ref{sec:msmh} and \ref{sec:msz}) and showed that the model with values for the wind parameters close to the values motivated by energy-driven wind model and results of the FIRE simulations can reproduce existing observational constraints for the $\Ms-\Mh$, $\Ms-\Zs$, and $\Ms-\Zg$ relations.
Interestingly, in contrast with some previous studies we find that this can be achieved either with or without preventative feedback. \\
\item[(ii)] The shape of the $\Ms-\Mh$ relation, and correspondingly $L_V-\Mh$ relation, on dwarf scales is controlled by the mass-dependence of the mass loading factor of outflows, $\etaw$, and gas accretion suppression in small-mass haloes after reionization. The power law dependence of $\etaw$ on mass controls the slope of the $\Ms-\Mh$ relation at $\Ms\approx 10^6-10^{10}\,\Msun$ (or $M_V\lesssim -8$), while at small masses post-reionization gas accretion suppression leads to significant deviations from the power law form.\\ 
\item[(iii)] Suppression of gas accretion due to UV heating after reionization results in steepening of $\Ms-\Mh$ ($L_V-\Mh$) relation at $\Ms\approx 10^4-10^6\,\Msun$ ($-8\lesssim M_V\lesssim -1$) and flattening at $\Ms\lesssim 10^4\,\Msun$ ($M_V\gtrsim -1$). The deviations from the power law form increase with increasing redshift of reionization.\\
\item[(iv)] Halo occupation by galaxies with $M_V<0$ is consistent with observational constraints in our model with $\zrei\lesssim 9$, which indicates that a relatively late reionization of the LG volume is favoured. \\
\item[(v)] Our fiducial model that reproduces the observed $\Ms-\Mh$, $\Ms-\Zs$, and $\Ms-\Zg$ relations reproduces $\Ms-\rhalf$ relation of dwarf galaxies in the Local Group. Correspondingly, the surface brigthness distribution of observed dwarfs is reproduced (see lower panel in Fig.~\ref{fig:grmur_comp}).
Interestingly, stellar half mass radii of dwarf galaxies are reproduced with a simple model in which gaseous disks are proportional to $\Rtwoh$ and stars are forming in molecular gas with a constant molecular depletion time. 
\\
\item[(vi)] The fiducial model also matches the current data for the $\Ms-\Mg$ relation, although we find that the gas content of galaxies with $\Ms\lesssim 10^7\,\Msun$ is quite sensitive to modelling of molecular hydrogen, which is used to control star formation rates of our model galaxies. \\
\item[(vii)] Model galaxies of a given $\Ms$ or $\Mg$ have star formation rates that match SFRs of low-surface brightness dwarfs, but underestimate SFR in blue compact dwarfs. We interpret this a signature of strong burstiness of star formation in dwarf galaxies. Modelling such burstiness is likely important to match the entire range of specific star formation rates of observed dwarf galaxies. \\
\item[(viii)] Galaxies in the fiducial model have diverse star formation histories in qualitative agreement with SFHs of Local Group dwarfs derived from their stellar color-magnitude diagrams. Furthermore, we show that the model galaxies occupy the same region in the plane of $\tau_{50}-\tau_{90}$, where $\tau_{50}$ and $\tau_{90}$ are the lookback times at which galaxies formed $50\%$ and $90\%$ of their stars, as observed galaxies. \\
\item[(ix)] Results of our model indicate that several properties of dwarf galaxies and their distributions are signatures of reionization. For example, galaxies with  $t_{50}>12$ Gyr and $t_{90}\lesssim 8$ Gyr are a clear signature of reionization and existence of Local Group galaxies with $t_{50}>12$ Gyr and $t_{90}\lesssim 5$ Gyr favors late local reionization redshift ($\zrei\approx 6-9$; see Fig.~\ref{fig:t50t90_rei}). Also, the occupation fraction of detectable UFDs in haloes of $\Mh\lesssim 10^8\,\Msun$ is sensitive to $\zrei$ and current constraints also favor late local reionization redshift of $\zrei\lesssim 9$. 
\end{itemize}

Success of the model presented in this study in matching observational properties of dwarf galaxies makes it a useful tool for estimating photometric properties of dwarf galaxies hosted by dark matter haloes in $N$-body simulations, such as colors, surface brightnesses, and mass-to-light ratios and to forward model observations of dwarf galaxies.  The latter allows matching observational selection effects and applying realistic detection thresholds, thereby avoiding unnecessary assumptions and corrections when comparing model results to observations. At the same time, simplicity of the model and a relatively small set of parameters can be used to interpret both observations and results of cosmological simulations in terms of physical processes that drive galaxy evolution.  

As an illustration, we show that $g-r$ colors and $r$-band surface brightnesses of model dwarf galaxies are similar to the distribution of these quantities for dwarf satellite galaxies in the SAGA2 \citep[][]{Mao.etal.2021} and the \citet{Carlsten.etal.2020} samples. We also find that the model reproduces the trend of quenched fraction observed in the SAGA2 survey with stellar mass. This is interesting as the fiducial model used in this study includes only one environmental effect (strangulation). At the same time, most of the model galaxies with $\Ms\gtrsim 10^5\,\Msun$ are predicted to have neutral gas in stark contast with satellite dwarf galaxies of the MW and M31. This indicates both a likely diversity of dwarf satellite properties around massive galaxies and important of environment effects such as ram pressure stripping. We plan to explore these questions in follow-up studies. 

\section*{Acknowledgements}
We would like to thank Shea Garrison-Kimmel and Michael Boylan-Kolchin for providing halo tracks of the ELVIS simulations, Nickolay Gnedin for useful discussions of $\HH$ mass fraction models and providing numerical results of $f_{\rm H_2}$ in electronic form, and Ethan Nadler, Nickolay Gnedin and Phil Mansfield for useful comments on the manuscript. AK is  grateful to Vasily Belokurov and Kristen McQuinn for discussions about observations and properties of dwarf galaxies and to Gabriel Torrealba for providing his compilation of dwarf galaxy size measurements in electronic form. This work was supported by the National Science Foundation grants AST-1714658 and AST-1911111 and NASA ATP grant 80NSSC20K0512.

This research made use of data from the SAGA Survey (\href{http://sagasurvey.org}{\tt sagasurvey.org}), which was supported by NSF collaborative grants AST-1517148 and AST-1517422 and by Heising–Simons Foundation grant 2019-1402. 
Analyses presented in this paper were greatly aided by the following free software packages: {\tt NumPy} \citep{numpy_ndarray}, {\tt SciPy} \citep{scipy}, {\tt Matplotlib} \citep{matplotlib}, and \href{https://github.com/}{\tt GitHub}. We have also used the Astrophysics Data Service (\href{http://adsabs.harvard.edu/abstract_service.html}{\tt ADS}) and \href{https://arxiv.org}{\tt arXiv} preprint repository extensively during this project and the writing of the paper.

\section*{Data availability}

Halo catalogs from the ELVIS and Phat ELVIS simulations can be obtained upon request from Shea Garrison-Kimmel and Michael Boylan-Kolchin. 
The \texttt{GRUMPY} model pipeline is available at \url{https://github.com/kibokov/GRUMPY}.

\bibliographystyle{mnras}
\bibliography{main}

\begin{thebibliography}{}
\makeatletter
\relax
\def\mn@urlcharsother{\let\do\@makeother \do\$\do\&\do\#\do\^\do\_\do\%\do\~}
\def\mn@doi{\begingroup\mn@urlcharsother \@ifnextchar [ {\mn@doi@}
  {\mn@doi@[]}}
\def\mn@doi@[#1]#2{\def\@tempa{#1}\ifx\@tempa\@empty \href
  {http://dx.doi.org/#2} {doi:#2}\else \href {http://dx.doi.org/#2} {#1}\fi
  \endgroup}
\def\mn@eprint#1#2{\mn@eprint@#1:#2::\@nil}
\def\mn@eprint@arXiv#1{\href {http://arxiv.org/abs/#1} {{\tt arXiv:#1}}}
\def\mn@eprint@dblp#1{\href {http://dblp.uni-trier.de/rec/bibtex/#1.xml}
  {dblp:#1}}
\def\mn@eprint@#1:#2:#3:#4\@nil{\def\@tempa {#1}\def\@tempb {#2}\def\@tempc
  {#3}\ifx \@tempc \@empty \let \@tempc \@tempb \let \@tempb \@tempa \fi \ifx
  \@tempb \@empty \def\@tempb {arXiv}\fi \@ifundefined
  {mn@eprint@\@tempb}{\@tempb:\@tempc}{\expandafter \expandafter \csname
  mn@eprint@\@tempb\endcsname \expandafter{\@tempc}}}

\bibitem[\protect\citeauthoryear{{Agertz} \& {Kravtsov}}{{Agertz} \&
  {Kravtsov}}{2016}]{Agertz.Kravtsov.2016}
{Agertz} O.,  {Kravtsov} A.~V.,  2016, \mn@doi [\apj]
  {10.3847/0004-637X/824/2/79}, \href
  {https://ui.adsabs.harvard.edu/abs/2016ApJ...824...79A} {824, 79}

\bibitem[\protect\citeauthoryear{{Agertz} et~al.,}{{Agertz}
  et~al.}{2020}]{Agertz.etal.2020}
{Agertz} O.,  et~al., 2020, \mn@doi [\mnras] {10.1093/mnras/stz3053}, \href
  {https://ui.adsabs.harvard.edu/abs/2020MNRAS.491.1656A} {491, 1656}

\bibitem[\protect\citeauthoryear{{Angl{\'e}s-Alc{\'a}zar},
  {Faucher-Gigu{\`e}re}, {Kere{\v{s}}}, {Hopkins}, {Quataert}  \&
  {Murray}}{{Angl{\'e}s-Alc{\'a}zar} et~al.}{2017}]{AnglesAlcazar.etal.2017}
{Angl{\'e}s-Alc{\'a}zar} D.,  {Faucher-Gigu{\`e}re} C.-A.,  {Kere{\v{s}}} D.,
  {Hopkins} P.~F.,  {Quataert} E.,   {Murray} N.,  2017, \mn@doi [\mnras]
  {10.1093/mnras/stx1517}, \href
  {https://ui.adsabs.harvard.edu/abs/2017MNRAS.470.4698A} {470, 4698}

\bibitem[\protect\citeauthoryear{{Applebaum}, {Brooks}, {Christensen},
  {Munshi}, {Quinn}, {Shen}  \& {Tremmel}}{{Applebaum}
  et~al.}{2021}]{Applebaum.etal.2021}
{Applebaum} E.,  {Brooks} A.~M.,  {Christensen} C.~R.,  {Munshi} F.,  {Quinn}
  T.~R.,  {Shen} S.,   {Tremmel} M.,  2021, \mn@doi [\apj]
  {10.3847/1538-4357/abcafa}, \href
  {https://ui.adsabs.harvard.edu/abs/2021ApJ...906...96A} {906, 96}

\bibitem[\protect\citeauthoryear{{Asplund}, {Grevesse}, {Sauval}  \&
  {Scott}}{{Asplund} et~al.}{2009}]{Asplund.etal.2009}
{Asplund} M.,  {Grevesse} N.,  {Sauval} A.~J.,   {Scott} P.,  2009, \mn@doi
  [\araa] {10.1146/annurev.astro.46.060407.145222}, \href
  {https://ui.adsabs.harvard.edu/abs/2009ARA&A..47..481A} {47, 481}

\bibitem[\protect\citeauthoryear{{Barat}, {D'Eugenio}, {Colless}, {Sweet},
  {Groves}  \& {Cortese}}{{Barat} et~al.}{2020}]{Barat.etal.2020}
{Barat} D.,  {D'Eugenio} F.,  {Colless} M.,  {Sweet} S.~M.,  {Groves} B.,
  {Cortese} L.,  2020, \mn@doi [\mnras] {10.1093/mnras/staa2716}, \href
  {https://ui.adsabs.harvard.edu/abs/2020MNRAS.498.5885B} {498, 5885}

\bibitem[\protect\citeauthoryear{{Behroozi}, {Wechsler}, {Wu}, {Busha},
  {Klypin}  \& {Primack}}{{Behroozi} et~al.}{2013a}]{Behroozi.etal.2013}
{Behroozi} P.~S.,  {Wechsler} R.~H.,  {Wu} H.-Y.,  {Busha} M.~T.,  {Klypin}
  A.~A.,   {Primack} J.~R.,  2013a, \mn@doi [\apj]
  {10.1088/0004-637X/763/1/18}, \href
  {https://ui.adsabs.harvard.edu/abs/2013ApJ...763...18B} {763, 18}

\bibitem[\protect\citeauthoryear{{Behroozi}, {Wechsler}  \&
  {Conroy}}{{Behroozi} et~al.}{2013b}]{Behroozi.etal.2013c}
{Behroozi} P.~S.,  {Wechsler} R.~H.,   {Conroy} C.,  2013b, \mn@doi [\apj]
  {10.1088/0004-637X/770/1/57}, \href
  {https://ui.adsabs.harvard.edu/abs/2013ApJ...770...57B} {770, 57}

\bibitem[\protect\citeauthoryear{{Benson}}{{Benson}}{2010}]{Benson.2010}
{Benson} A.~J.,  2010, \mn@doi [\physrep] {10.1016/j.physrep.2010.06.001},
  \href {https://ui.adsabs.harvard.edu/abs/2010PhR...495...33B} {495, 33}

\bibitem[\protect\citeauthoryear{{Benson}, {Lacey}, {Baugh}, {Cole}  \&
  {Frenk}}{{Benson} et~al.}{2002a}]{Benson.etal.2002a}
{Benson} A.~J.,  {Lacey} C.~G.,  {Baugh} C.~M.,  {Cole} S.,   {Frenk} C.~S.,
  2002a, \mn@doi [\mnras] {10.1046/j.1365-8711.2002.05387.x}, \href
  {https://ui.adsabs.harvard.edu/abs/2002MNRAS.333..156B} {333, 156}

\bibitem[\protect\citeauthoryear{{Benson}, {Frenk}, {Lacey}, {Baugh}  \&
  {Cole}}{{Benson} et~al.}{2002b}]{Benson.etal.2002b}
{Benson} A.~J.,  {Frenk} C.~S.,  {Lacey} C.~G.,  {Baugh} C.~M.,   {Cole} S.,
  2002b, \mn@doi [\mnras] {10.1046/j.1365-8711.2002.05388.x}, \href
  {https://ui.adsabs.harvard.edu/abs/2002MNRAS.333..177B} {333, 177}

\bibitem[\protect\citeauthoryear{{Berg} et~al.,}{{Berg}
  et~al.}{2012}]{Berg.etal.2012}
{Berg} D.~A.,  et~al., 2012, \mn@doi [\apj] {10.1088/0004-637X/754/2/98}, \href
  {https://ui.adsabs.harvard.edu/abs/2012ApJ...754...98B} {754, 98}

\bibitem[\protect\citeauthoryear{{Berg}, {Skillman}, {Henry}, {Erb}  \&
  {Carigi}}{{Berg} et~al.}{2016}]{Berg.etal.2016}
{Berg} D.~A.,  {Skillman} E.~D.,  {Henry} R. B.~C.,  {Erb} D.~K.,   {Carigi}
  L.,  2016, \mn@doi [\apj] {10.3847/0004-637X/827/2/126}, \href
  {https://ui.adsabs.harvard.edu/abs/2016ApJ...827..126B} {827, 126}

\bibitem[\protect\citeauthoryear{{Bernardi}, {Meert}, {Sheth}, {Vikram},
  {Huertas-Company}, {Mei}  \& {Shankar}}{{Bernardi}
  et~al.}{2013}]{Bernardi.2013}
{Bernardi} M.,  {Meert} A.,  {Sheth} R.~K.,  {Vikram} V.,  {Huertas-Company}
  M.,  {Mei} S.,   {Shankar} F.,  2013, \mn@doi [\mnras]
  {10.1093/mnras/stt1607}, \href
  {https://ui.adsabs.harvard.edu/abs/2013MNRAS.436..697B} {436, 697}

\bibitem[\protect\citeauthoryear{{Bigiel} \& {Blitz}}{{Bigiel} \&
  {Blitz}}{2012}]{Bigiel.Blitz.2012}
{Bigiel} F.,  {Blitz} L.,  2012, \mn@doi [\apj] {10.1088/0004-637X/756/2/183},
  \href {https://ui.adsabs.harvard.edu/abs/2012ApJ...756..183B} {756, 183}

\bibitem[\protect\citeauthoryear{{Bigiel}, {Leroy}, {Walter}, {Brinks}, {de
  Blok}, {Madore}  \& {Thornley}}{{Bigiel} et~al.}{2008}]{Bigiel.etal.2008}
{Bigiel} F.,  {Leroy} A.,  {Walter} F.,  {Brinks} E.,  {de Blok} W.~J.~G.,
  {Madore} B.,   {Thornley} M.~D.,  2008, \mn@doi [\aj]
  {10.1088/0004-6256/136/6/2846}, \href
  {https://ui.adsabs.harvard.edu/abs/2008AJ....136.2846B} {136, 2846}

\bibitem[\protect\citeauthoryear{{Bigiel} et~al.,}{{Bigiel}
  et~al.}{2011}]{Bigiel.etal.2011}
{Bigiel} F.,  et~al., 2011, \mn@doi [\apjl] {10.1088/2041-8205/730/2/L13},
  \href {https://ui.adsabs.harvard.edu/abs/2011ApJ...730L..13B} {730, L13}

\bibitem[\protect\citeauthoryear{{Birnboim} \& {Dekel}}{{Birnboim} \&
  {Dekel}}{2003}]{Birnboim.Dekel.2003}
{Birnboim} Y.,  {Dekel} A.,  2003, \mn@doi [\mnras]
  {10.1046/j.1365-8711.2003.06955.x}, \href
  {https://ui.adsabs.harvard.edu/abs/2003MNRAS.345..349B} {345, 349}

\bibitem[\protect\citeauthoryear{{Birrer}, {Lilly}, {Amara}, {Paranjape}  \&
  {Refregier}}{{Birrer} et~al.}{2014}]{Birrer.etal.2014}
{Birrer} S.,  {Lilly} S.,  {Amara} A.,  {Paranjape} A.,   {Refregier} A.,
  2014, \mn@doi [\apj] {10.1088/0004-637X/793/1/12}, \href
  {https://ui.adsabs.harvard.edu/abs/2014ApJ...793...12B} {793, 12}

\bibitem[\protect\citeauthoryear{{Bland-Hawthorn}, {Sutherland}  \&
  {Webster}}{{Bland-Hawthorn} et~al.}{2015}]{Bland_Hawthorn.etal.2015}
{Bland-Hawthorn} J.,  {Sutherland} R.,   {Webster} D.,  2015, \mn@doi [\apj]
  {10.1088/0004-637X/807/2/154}, \href
  {https://ui.adsabs.harvard.edu/abs/2015ApJ...807..154B} {807, 154}

\bibitem[\protect\citeauthoryear{{Bland-Hawthorn}, {Maloney}, {Stephens},
  {Zovaro}  \& {Popping}}{{Bland-Hawthorn} et~al.}{2017}]{Bland.Hawthorne.2017}
{Bland-Hawthorn} J.,  {Maloney} P.~R.,  {Stephens} A.,  {Zovaro} A.,
  {Popping} A.,  2017, \mn@doi [\apj] {10.3847/1538-4357/aa8f45}, \href
  {https://ui.adsabs.harvard.edu/abs/2017ApJ...849...51B} {849, 51}

\bibitem[\protect\citeauthoryear{{Bochkarev} \& {Sunyaev}}{{Bochkarev} \&
  {Sunyaev}}{1977}]{Bochkarev.Sunyaev.1977}
{Bochkarev} N.~G.,  {Sunyaev} R.~A.,  1977, \sovast, \href
  {https://ui.adsabs.harvard.edu/abs/1977SvA....21..542B} {21, 542}

\bibitem[\protect\citeauthoryear{{Bolatto} et~al.,}{{Bolatto}
  et~al.}{2011}]{Bolatto.etal.2011}
{Bolatto} A.~D.,  et~al., 2011, \mn@doi [\apj] {10.1088/0004-637X/741/1/12},
  \href {https://ui.adsabs.harvard.edu/abs/2011ApJ...741...12B} {741, 12}

\bibitem[\protect\citeauthoryear{{Bolatto} et~al.,}{{Bolatto}
  et~al.}{2017}]{Bolatto.etal.2017}
{Bolatto} A.~D.,  et~al., 2017, \mn@doi [\apj] {10.3847/1538-4357/aa86aa},
  \href {https://ui.adsabs.harvard.edu/abs/2017ApJ...846..159B} {846, 159}

\bibitem[\protect\citeauthoryear{{Bose}, {Deason}  \& {Frenk}}{{Bose}
  et~al.}{2018}]{bose_etal18}
{Bose} S.,  {Deason} A.~J.,   {Frenk} C.~S.,  2018, \mn@doi [\apj]
  {10.3847/1538-4357/aacbc4}, \href
  {https://ui.adsabs.harvard.edu/abs/2018ApJ...863..123B} {863, 123}

\bibitem[\protect\citeauthoryear{{Bouch{\'e}} et~al.,}{{Bouch{\'e}}
  et~al.}{2010}]{Bouche.etal.2010}
{Bouch{\'e}} N.,  et~al., 2010, \mn@doi [\apj] {10.1088/0004-637X/718/2/1001},
  \href {https://ui.adsabs.harvard.edu/abs/2010ApJ...718.1001B} {718, 1001}

\bibitem[\protect\citeauthoryear{{Bovill} \& {Ricotti}}{{Bovill} \&
  {Ricotti}}{2009}]{Bovill.Ricotti.2009}
{Bovill} M.~S.,  {Ricotti} M.,  2009, \mn@doi [\apj]
  {10.1088/0004-637X/693/2/1859}, \href
  {https://ui.adsabs.harvard.edu/abs/2009ApJ...693.1859B} {693, 1859}

\bibitem[\protect\citeauthoryear{{Bower}, {Schaye}, {Frenk}, {Theuns},
  {Schaller}, {Crain}  \& {McAlpine}}{{Bower} et~al.}{2017}]{Bower.etal.2017}
{Bower} R.~G.,  {Schaye} J.,  {Frenk} C.~S.,  {Theuns} T.,  {Schaller} M.,
  {Crain} R.~A.,   {McAlpine} S.,  2017, \mn@doi [\mnras]
  {10.1093/mnras/stw2735}, \href
  {https://ui.adsabs.harvard.edu/abs/2017MNRAS.465...32B} {465, 32}

\bibitem[\protect\citeauthoryear{{Bullock} \& {Boylan-Kolchin}}{{Bullock} \&
  {Boylan-Kolchin}}{2017}]{bullock_boylankolchin17}
{Bullock} J.~S.,  {Boylan-Kolchin} M.,  2017, \mn@doi [\araa]
  {10.1146/annurev-astro-091916-055313}, \href
  {https://ui.adsabs.harvard.edu/abs/2017ARA&A..55..343B} {55, 343}

\bibitem[\protect\citeauthoryear{{Bullock}, {Kravtsov}  \&
  {Weinberg}}{{Bullock} et~al.}{2000}]{Bullock.etal.2000}
{Bullock} J.~S.,  {Kravtsov} A.~V.,   {Weinberg} D.~H.,  2000, \mn@doi [\apj]
  {10.1086/309279}, \href
  {https://ui.adsabs.harvard.edu/abs/2000ApJ...539..517B} {539, 517}

\bibitem[\protect\citeauthoryear{{Bullock}, {Dekel}, {Kolatt}, {Kravtsov},
  {Klypin}, {Porciani}  \& {Primack}}{{Bullock}
  et~al.}{2001}]{Bullock.etal.2001}
{Bullock} J.~S.,  {Dekel} A.,  {Kolatt} T.~S.,  {Kravtsov} A.~V.,  {Klypin}
  A.~A.,  {Porciani} C.,   {Primack} J.~R.,  2001, \mn@doi [\apj]
  {10.1086/321477}, \href
  {https://ui.adsabs.harvard.edu/abs/2001ApJ...555..240B} {555, 240}

\bibitem[\protect\citeauthoryear{{Carlsten}, {Greco}, {Beaton}  \&
  {Greene}}{{Carlsten} et~al.}{2020}]{Carlsten.etal.2020}
{Carlsten} S.~G.,  {Greco} J.~P.,  {Beaton} R.~L.,   {Greene} J.~E.,  2020,
  \mn@doi [\apj] {10.3847/1538-4357/ab7758}, \href
  {https://ui.adsabs.harvard.edu/abs/2020ApJ...891..144C} {891, 144}

\bibitem[\protect\citeauthoryear{{Chabrier}}{{Chabrier}}{2003}]{Chabrier.2003}
{Chabrier} G.,  2003, \mn@doi [\pasp] {10.1086/376392}, \href
  {https://ui.adsabs.harvard.edu/abs/2003PASP..115..763C} {115, 763}

\bibitem[\protect\citeauthoryear{{Chisholm}, {Tremonti}, {Leitherer}  \&
  {Chen}}{{Chisholm} et~al.}{2017}]{Chisholm.etal.2017}
{Chisholm} J.,  {Tremonti} C.~A.,  {Leitherer} C.,   {Chen} Y.,  2017, \mn@doi
  [\mnras] {10.1093/mnras/stx1164}, \href
  {https://ui.adsabs.harvard.edu/abs/2017MNRAS.469.4831C} {469, 4831}

\bibitem[\protect\citeauthoryear{{Chisholm}, {Tremonti}  \&
  {Leitherer}}{{Chisholm} et~al.}{2018}]{Chisholm.etal.2018}
{Chisholm} J.,  {Tremonti} C.,   {Leitherer} C.,  2018, \mn@doi [\mnras]
  {10.1093/mnras/sty2380}, \href
  {https://ui.adsabs.harvard.edu/abs/2018MNRAS.481.1690C} {481, 1690}

\bibitem[\protect\citeauthoryear{{Chiti} et~al.,}{{Chiti}
  et~al.}{2021a}]{chiti_etal21a}
{Chiti} A.,  et~al., 2021a, \mn@doi [Nature Astronomy]
  {10.1038/s41550-020-01285-w}, \href
  {https://ui.adsabs.harvard.edu/abs/2021NatAs...5..392C} {5, 392}

\bibitem[\protect\citeauthoryear{{Chiti}, {Mardini}, {Frebel}  \&
  {Daniel}}{{Chiti} et~al.}{2021b}]{chiti_etal21}
{Chiti} A.,  {Mardini} M.~K.,  {Frebel} A.,   {Daniel} T.,  2021b, \mn@doi
  [\apjl] {10.3847/2041-8213/abd629}, \href
  {https://ui.adsabs.harvard.edu/abs/2021ApJ...911L..23C} {911, L23}

\bibitem[\protect\citeauthoryear{{Conroy} \& {Gunn}}{{Conroy} \&
  {Gunn}}{2010}]{Conroy.Gunn.2010}
{Conroy} C.,  {Gunn} J.~E.,  2010, \mn@doi [\apj]
  {10.1088/0004-637X/712/2/833}, \href
  {https://ui.adsabs.harvard.edu/abs/2010ApJ...712..833C} {712, 833}

\bibitem[\protect\citeauthoryear{{Conroy}, {Gunn}  \& {White}}{{Conroy}
  et~al.}{2009}]{Conroy.etal.2009}
{Conroy} C.,  {Gunn} J.~E.,   {White} M.,  2009, \mn@doi [\apj]
  {10.1088/0004-637X/699/1/486}, \href
  {https://ui.adsabs.harvard.edu/abs/2009ApJ...699..486C} {699, 486}

\bibitem[\protect\citeauthoryear{{C{\^o}t{\'e}}, {Silvia}, {O'Shea}, {Smith}
  \& {Wise}}{{C{\^o}t{\'e}} et~al.}{2018}]{Cote.etal.2018}
{C{\^o}t{\'e}} B.,  {Silvia} D.~W.,  {O'Shea} B.~W.,  {Smith} B.,   {Wise}
  J.~H.,  2018, \mn@doi [\apj] {10.3847/1538-4357/aabe8f}, \href
  {https://ui.adsabs.harvard.edu/abs/2018ApJ...859...67C} {859, 67}

\bibitem[\protect\citeauthoryear{{Dawoodbhoy} et~al.,}{{Dawoodbhoy}
  et~al.}{2018}]{Dawoodbhoy.etal.2018}
{Dawoodbhoy} T.,  et~al., 2018, \mn@doi [\mnras] {10.1093/mnras/sty1945}, \href
  {https://ui.adsabs.harvard.edu/abs/2018MNRAS.480.1740D} {480, 1740}

\bibitem[\protect\citeauthoryear{{De Lucia}, {Xie}, {Fontanot}  \&
  {Hirschmann}}{{De Lucia} et~al.}{2020}]{DeLucia.etal.2020}
{De Lucia} G.,  {Xie} L.,  {Fontanot} F.,   {Hirschmann} M.,  2020, \mn@doi
  [\mnras] {10.1093/mnras/staa2556}, \href
  {https://ui.adsabs.harvard.edu/abs/2020MNRAS.498.3215D} {498, 3215}

\bibitem[\protect\citeauthoryear{{Dekel} \& {Birnboim}}{{Dekel} \&
  {Birnboim}}{2006}]{Dekel.Birnboim.2006}
{Dekel} A.,  {Birnboim} Y.,  2006, \mn@doi [\mnras]
  {10.1111/j.1365-2966.2006.10145.x}, \href
  {https://ui.adsabs.harvard.edu/abs/2006MNRAS.368....2D} {368, 2}

\bibitem[\protect\citeauthoryear{{Dekel} \& {Silk}}{{Dekel} \&
  {Silk}}{1986}]{Dekel.Silk.1986}
{Dekel} A.,  {Silk} J.,  1986, \mn@doi [\apj] {10.1086/164050}, \href
  {https://ui.adsabs.harvard.edu/abs/1986ApJ...303...39D} {303, 39}

\bibitem[\protect\citeauthoryear{{Dekel} et~al.,}{{Dekel}
  et~al.}{2009}]{Dekel.etal.2009}
{Dekel} A.,  et~al., 2009, \mn@doi [\nat] {10.1038/nature07648}, \href
  {https://ui.adsabs.harvard.edu/abs/2009Natur.457..451D} {457, 451}

\bibitem[\protect\citeauthoryear{{Di Cintio}, {Mostoghiu}, {Knebe}  \&
  {Navarro}}{{Di Cintio} et~al.}{2021}]{DiCintio.etal.2021}
{Di Cintio} A.,  {Mostoghiu} R.,  {Knebe} A.,   {Navarro} J.~F.,  2021, \mn@doi
  [\mnras] {10.1093/mnras/stab1682}, \href
  {https://ui.adsabs.harvard.edu/abs/2021MNRAS.506..531D} {506, 531}

\bibitem[\protect\citeauthoryear{{Diemer}, {More}  \& {Kravtsov}}{{Diemer}
  et~al.}{2013}]{Diemer.etal.2013}
{Diemer} B.,  {More} S.,   {Kravtsov} A.~V.,  2013, \mn@doi [\apj]
  {10.1088/0004-637X/766/1/25}, \href
  {https://ui.adsabs.harvard.edu/abs/2013ApJ...766...25D} {766, 25}

\bibitem[\protect\citeauthoryear{{Dooley}, {Peter}, {Carlin}, {Frebel},
  {Bechtol}  \& {Willman}}{{Dooley} et~al.}{2017}]{Dooley.etal.2017}
{Dooley} G.~A.,  {Peter} A. H.~G.,  {Carlin} J.~L.,  {Frebel} A.,  {Bechtol}
  K.,   {Willman} B.,  2017, \mn@doi [\mnras] {10.1093/mnras/stx2001}, \href
  {https://ui.adsabs.harvard.edu/abs/2017MNRAS.472.1060D} {472, 1060}

\bibitem[\protect\citeauthoryear{{Dou} et~al.,}{{Dou}
  et~al.}{2021}]{Dou.etal.2020}
{Dou} J.,  et~al., 2021, \mn@doi [\apj] {10.3847/1538-4357/abd17c}, \href
  {https://ui.adsabs.harvard.edu/abs/2021ApJ...907..114D} {907, 114}

\bibitem[\protect\citeauthoryear{{Drlica-Wagner} et~al.,}{{Drlica-Wagner}
  et~al.}{2015a}]{Drlica_Wagner.etal.2015}
{Drlica-Wagner} A.,  et~al., 2015a, \mn@doi [\apj]
  {10.1088/0004-637X/813/2/109}, \href
  {https://ui.adsabs.harvard.edu/abs/2015ApJ...813..109D} {813, 109}

\bibitem[\protect\citeauthoryear{{Drlica-Wagner} et~al.,}{{Drlica-Wagner}
  et~al.}{2015b}]{Drlica-Wagner.etal.2015}
{Drlica-Wagner} A.,  et~al., 2015b, \mn@doi [\apj]
  {10.1088/0004-637X/813/2/109}, \href
  {https://ui.adsabs.harvard.edu/abs/2015ApJ...813..109D} {813, 109}

\bibitem[\protect\citeauthoryear{{Drlica-Wagner} et~al.,}{{Drlica-Wagner}
  et~al.}{2020}]{drlica_wagner_etal20}
{Drlica-Wagner} A.,  et~al., 2020, \mn@doi [\apj] {10.3847/1538-4357/ab7eb9},
  \href {https://ui.adsabs.harvard.edu/abs/2020ApJ...893...47D} {893, 47}

\bibitem[\protect\citeauthoryear{{Durbala}, {Finn}, {Crone Odekon}, {Haynes},
  {Koopmann}  \& {O'Donoghue}}{{Durbala} et~al.}{2020}]{Durbala.etal.2020}
{Durbala} A.,  {Finn} R.~A.,  {Crone Odekon} M.,  {Haynes} M.~P.,  {Koopmann}
  R.~A.,   {O'Donoghue} A.~A.,  2020, \mn@doi [\aj] {10.3847/1538-3881/abc018},
  \href {https://ui.adsabs.harvard.edu/abs/2020AJ....160..271D} {160, 271}

\bibitem[\protect\citeauthoryear{{Efstathiou}}{{Efstathiou}}{1992}]{Efstathiou.1992}
{Efstathiou} G.,  1992, \mn@doi [\mnras] {10.1093/mnras/256.1.43P}, \href
  {https://ui.adsabs.harvard.edu/abs/1992MNRAS.256P..43E} {256, 43P}

\bibitem[\protect\citeauthoryear{{Efstathiou}}{{Efstathiou}}{2000}]{Efstathiou.2000}
{Efstathiou} G.,  2000, \mn@doi [\mnras] {10.1046/j.1365-8711.2000.03665.x},
  \href {https://ui.adsabs.harvard.edu/abs/2000MNRAS.317..697E} {317, 697}

\bibitem[\protect\citeauthoryear{{Fall} \& {Efstathiou}}{{Fall} \&
  {Efstathiou}}{1980}]{Fall.Efstathiou.1980}
{Fall} S.~M.,  {Efstathiou} G.,  1980, \mn@doi [\mnras]
  {10.1093/mnras/193.2.189}, \href
  {https://ui.adsabs.harvard.edu/abs/1980MNRAS.193..189F} {193, 189}

\bibitem[\protect\citeauthoryear{{Faucher-Gigu{\`e}re}}{{Faucher-Gigu{\`e}re}}{2018}]{Faucher_Giguere.2018}
{Faucher-Gigu{\`e}re} C.-A.,  2018, \mn@doi [\mnras] {10.1093/mnras/stx2595},
  \href {https://ui.adsabs.harvard.edu/abs/2018MNRAS.473.3717F} {473, 3717}

\bibitem[\protect\citeauthoryear{{Feldmann}}{{Feldmann}}{2013}]{Feldmann.2013}
{Feldmann} R.,  2013, \mn@doi [\mnras] {10.1093/mnras/stt851}, \href
  {https://ui.adsabs.harvard.edu/abs/2013MNRAS.433.1910F} {433, 1910}

\bibitem[\protect\citeauthoryear{{Felten} \& {Bergeron}}{{Felten} \&
  {Bergeron}}{1969}]{Felten.Bergeron.1969}
{Felten} J.~E.,  {Bergeron} J.,  1969, \aplett, \href
  {https://ui.adsabs.harvard.edu/abs/1969ApL.....4..155F} {4, 155}

\bibitem[\protect\citeauthoryear{{Finlator} \& {Dav{\'e}}}{{Finlator} \&
  {Dav{\'e}}}{2008}]{Finlator.Dave.2008}
{Finlator} K.,  {Dav{\'e}} R.,  2008, \mn@doi [\mnras]
  {10.1111/j.1365-2966.2008.12991.x}, \href
  {https://ui.adsabs.harvard.edu/abs/2008MNRAS.385.2181F} {385, 2181}

\bibitem[\protect\citeauthoryear{{Fitts} et~al.,}{{Fitts}
  et~al.}{2017}]{Fitts.etal.2017}
{Fitts} A.,  et~al., 2017, \mn@doi [\mnras] {10.1093/mnras/stx1757}, \href
  {https://ui.adsabs.harvard.edu/abs/2017MNRAS.471.3547F} {471, 3547}

\bibitem[\protect\citeauthoryear{{Fitts} et~al.,}{{Fitts}
  et~al.}{2018}]{fitts_etal18}
{Fitts} A.,  et~al., 2018, \mn@doi [\mnras] {10.1093/mnras/sty1488}, \href
  {https://ui.adsabs.harvard.edu/abs/2018MNRAS.479..319F} {479, 319}

\bibitem[\protect\citeauthoryear{{Flores Vel{\'a}zquez} et~al.,}{{Flores
  Vel{\'a}zquez} et~al.}{2021}]{FloresVelazquez.etal.2021}
{Flores Vel{\'a}zquez} J.~A.,  et~al., 2021, \mn@doi [\mnras]
  {10.1093/mnras/staa3893}, \href
  {https://ui.adsabs.harvard.edu/abs/2021MNRAS.501.4812F} {501, 4812}

\bibitem[\protect\citeauthoryear{{Font} et~al.,}{{Font}
  et~al.}{2020}]{Font.etal.2020}
{Font} A.~S.,  et~al., 2020, \mn@doi [\mnras] {10.1093/mnras/staa2463}, \href
  {https://ui.adsabs.harvard.edu/abs/2020MNRAS.498.1765F} {498, 1765}

\bibitem[\protect\citeauthoryear{{Font}, {McCarthy}  \& {Belokurov}}{{Font}
  et~al.}{2021}]{Font.etal.2021}
{Font} A.~S.,  {McCarthy} I.~G.,   {Belokurov} V.,  2021, \mn@doi [\mnras]
  {10.1093/mnras/stab1332}, \href
  {https://ui.adsabs.harvard.edu/abs/2021MNRAS.505..783F} {505, 783}

\bibitem[\protect\citeauthoryear{{Forbes}, {Krumholz}  \& {Speagle}}{{Forbes}
  et~al.}{2019}]{Forbes.etal.2019}
{Forbes} J.~C.,  {Krumholz} M.~R.,   {Speagle} J.~S.,  2019, \mn@doi [\mnras]
  {10.1093/mnras/stz1473}, \href
  {https://ui.adsabs.harvard.edu/abs/2019MNRAS.487.3581F} {487, 3581}

\bibitem[\protect\citeauthoryear{{Furlanetto}, {Mirocha}, {Mebane}  \&
  {Sun}}{{Furlanetto} et~al.}{2017}]{Furlanetto.etal.2017}
{Furlanetto} S.~R.,  {Mirocha} J.,  {Mebane} R.~H.,   {Sun} G.,  2017, \mn@doi
  [\mnras] {10.1093/mnras/stx2132}, \href
  {https://ui.adsabs.harvard.edu/abs/2017MNRAS.472.1576F} {472, 1576}

\bibitem[\protect\citeauthoryear{{Garnett}}{{Garnett}}{2002}]{Garnett.2002}
{Garnett} D.~R.,  2002, \mn@doi [\apj] {10.1086/344301}, \href
  {https://ui.adsabs.harvard.edu/abs/2002ApJ...581.1019G} {581, 1019}

\bibitem[\protect\citeauthoryear{{Garrison-Kimmel}, {Boylan-Kolchin}, {Bullock}
   \& {Lee}}{{Garrison-Kimmel} et~al.}{2014}]{GarrisonKimmel.etal.2014}
{Garrison-Kimmel} S.,  {Boylan-Kolchin} M.,  {Bullock} J.~S.,   {Lee} K.,
  2014, \mn@doi [\mnras] {10.1093/mnras/stt2377}, \href
  {https://ui.adsabs.harvard.edu/abs/2014MNRAS.438.2578G} {438, 2578}

\bibitem[\protect\citeauthoryear{{Garrison-Kimmel} et~al.,}{{Garrison-Kimmel}
  et~al.}{2019}]{Garrison_Kimmel.etal.2019}
{Garrison-Kimmel} S.,  et~al., 2019, \mn@doi [\mnras] {10.1093/mnras/stz2507},
  \href {https://ui.adsabs.harvard.edu/abs/2019MNRAS.489.4574G} {489, 4574}

\bibitem[\protect\citeauthoryear{{Geha} et~al.,}{{Geha}
  et~al.}{2017}]{Geha.etal.2017}
{Geha} M.,  et~al., 2017, \mn@doi [\apj] {10.3847/1538-4357/aa8626}, \href
  {https://ui.adsabs.harvard.edu/abs/2017ApJ...847....4G} {847, 4}

\bibitem[\protect\citeauthoryear{{Gnedin}}{{Gnedin}}{2000}]{Gnedin.2000}
{Gnedin} N.~Y.,  2000, \mn@doi [\apj] {10.1086/317042}, \href
  {https://ui.adsabs.harvard.edu/abs/2000ApJ...542..535G} {542, 535}

\bibitem[\protect\citeauthoryear{{Gnedin} \& {Draine}}{{Gnedin} \&
  {Draine}}{2014}]{Gnedin.Draine.2014}
{Gnedin} N.~Y.,  {Draine} B.~T.,  2014, \mn@doi [\apj]
  {10.1088/0004-637X/795/1/37}, \href
  {https://ui.adsabs.harvard.edu/abs/2014ApJ...795...37G} {795, 37}

\bibitem[\protect\citeauthoryear{{Gnedin} \& {Draine}}{{Gnedin} \&
  {Draine}}{2016}]{Gnedin.Draine.2016}
{Gnedin} N.~Y.,  {Draine} B.~T.,  2016, \mn@doi [\apj]
  {10.3847/0004-637X/830/1/54}, \href
  {https://ui.adsabs.harvard.edu/abs/2016ApJ...830...54G} {830, 54}

\bibitem[\protect\citeauthoryear{{Gnedin} \& {Kaurov}}{{Gnedin} \&
  {Kaurov}}{2014}]{Gnedin.Kaurov.2014}
{Gnedin} N.~Y.,  {Kaurov} A.~A.,  2014, \mn@doi [\apj]
  {10.1088/0004-637X/793/1/30}, \href
  {https://ui.adsabs.harvard.edu/abs/2014ApJ...793...30G} {793, 30}

\bibitem[\protect\citeauthoryear{{Gnedin} \& {Kravtsov}}{{Gnedin} \&
  {Kravtsov}}{2006}]{Gnedin.Kravtsov.2006}
{Gnedin} N.~Y.,  {Kravtsov} A.~V.,  2006, \mn@doi [\apj] {10.1086/504404},
  \href {https://ui.adsabs.harvard.edu/abs/2006ApJ...645.1054G} {645, 1054}

\bibitem[\protect\citeauthoryear{{Graus}, {Bullock}, {Kelley},
  {Boylan-Kolchin}, {Garrison-Kimmel}  \& {Qi}}{{Graus}
  et~al.}{2019}]{Graus.etal.2019}
{Graus} A.~S.,  {Bullock} J.~S.,  {Kelley} T.,  {Boylan-Kolchin} M.,
  {Garrison-Kimmel} S.,   {Qi} Y.,  2019, \mn@doi [\mnras]
  {10.1093/mnras/stz1992}, \href
  {https://ui.adsabs.harvard.edu/abs/2019MNRAS.488.4585G} {488, 4585}

\bibitem[\protect\citeauthoryear{{Greif}, {Glover}, {Bromm}  \&
  {Klessen}}{{Greif} et~al.}{2010}]{Greif.etal.2010}
{Greif} T.~H.,  {Glover} S. C.~O.,  {Bromm} V.,   {Klessen} R.~S.,  2010,
  \mn@doi [\apj] {10.1088/0004-637X/716/1/510}, \href
  {https://ui.adsabs.harvard.edu/abs/2010ApJ...716..510G} {716, 510}

\bibitem[\protect\citeauthoryear{{Hahn} \& {Abel}}{{Hahn} \&
  {Abel}}{2011}]{hahn_abel11}
{Hahn} O.,  {Abel} T.,  2011, \mn@doi [\mnras]
  {10.1111/j.1365-2966.2011.18820.x}, \href
  {https://ui.adsabs.harvard.edu/abs/2011MNRAS.415.2101H} {415, 2101}

\bibitem[\protect\citeauthoryear{{Hargis} et~al.,}{{Hargis}
  et~al.}{2020}]{Hargis.etal.2020}
{Hargis} J.~R.,  et~al., 2020, \mn@doi [\apj] {10.3847/1538-4357/ab58d2}, \href
  {https://ui.adsabs.harvard.edu/abs/2020ApJ...888...31H} {888, 31}

\bibitem[\protect\citeauthoryear{{Haydon}, {Kruijssen}, {Chevance}, {Hygate},
  {Krumholz}, {Schruba}  \& {Longmore}}{{Haydon}
  et~al.}{2020}]{Haydon.etal.2020}
{Haydon} D.~T.,  {Kruijssen} J.~M.~D.,  {Chevance} M.,  {Hygate} A. P.~S.,
  {Krumholz} M.~R.,  {Schruba} A.,   {Longmore} S.~N.,  2020, \mn@doi [\mnras]
  {10.1093/mnras/staa2430}, \href
  {https://ui.adsabs.harvard.edu/abs/2020MNRAS.498..235H} {498, 235}

\bibitem[\protect\citeauthoryear{{Hidalgo} et~al.,}{{Hidalgo}
  et~al.}{2011}]{Hidalgo.etal.2011}
{Hidalgo} S.~L.,  et~al., 2011, \mn@doi [\apj] {10.1088/0004-637X/730/1/14},
  \href {https://ui.adsabs.harvard.edu/abs/2011ApJ...730...14H} {730, 14}

\bibitem[\protect\citeauthoryear{{Homma} et~al.,}{{Homma}
  et~al.}{2019}]{Homma.etal.2019}
{Homma} D.,  et~al., 2019, \mn@doi [\pasj] {10.1093/pasj/psz076}, \href
  {https://ui.adsabs.harvard.edu/abs/2019PASJ...71...94H} {71, 94}

\bibitem[\protect\citeauthoryear{{Huang} et~al.,}{{Huang}
  et~al.}{2017}]{Huang.etal.2017}
{Huang} K.-H.,  et~al., 2017, \mn@doi [\apj] {10.3847/1538-4357/aa62a6}, \href
  {https://ui.adsabs.harvard.edu/abs/2017ApJ...838....6H} {838, 6}

\bibitem[\protect\citeauthoryear{Hunter}{Hunter}{2007}]{matplotlib}
Hunter J.~D.,  2007, \mn@doi [Computing In Science \& Engineering]
  {10.1109/MCSE.2007.55}, 9, 90

\bibitem[\protect\citeauthoryear{{Hutter}, {Dayal}, {Yepes}, {Gottl{\"o}ber},
  {Legrand}  \& {Ucci}}{{Hutter} et~al.}{2021}]{Hutter.etal.2020}
{Hutter} A.,  {Dayal} P.,  {Yepes} G.,  {Gottl{\"o}ber} S.,  {Legrand} L.,
  {Ucci} G.,  2021, \mn@doi [\mnras] {10.1093/mnras/stab602}, \href
  {https://ui.adsabs.harvard.edu/abs/2021MNRAS.503.3698H} {503, 3698}

\bibitem[\protect\citeauthoryear{{Ivezi{\'c}} et~al.,}{{Ivezi{\'c}}
  et~al.}{2019}]{Ivezic.etal.2019}
{Ivezi{\'c}} {\v Z}.,  et~al., 2019, \mn@doi [\apj] {10.3847/1538-4357/ab042c},
  \href {http://adsabs.harvard.edu/abs/2019ApJ...873..111I} {873, 111}

\bibitem[\protect\citeauthoryear{{Jahn}, {Sales}, {Wetzel}, {Boylan-Kolchin},
  {Chan}, {El-Badry}, {Lazar}  \& {Bullock}}{{Jahn}
  et~al.}{2019}]{Jahn.etal.2019}
{Jahn} E.~D.,  {Sales} L.~V.,  {Wetzel} A.,  {Boylan-Kolchin} M.,  {Chan}
  T.~K.,  {El-Badry} K.,  {Lazar} A.,   {Bullock} J.~S.,  2019, \mn@doi
  [\mnras] {10.1093/mnras/stz2457}, \href
  {https://ui.adsabs.harvard.edu/abs/2019MNRAS.489.5348J} {489, 5348}

\bibitem[\protect\citeauthoryear{{James}, {Koposov}, {Stark}, {Belokurov},
  {Pettini}, {Olszewski}  \& {McQuinn}}{{James} et~al.}{2017}]{James.etal.2017}
{James} B.~L.,  {Koposov} S.~E.,  {Stark} D.~P.,  {Belokurov} V.,  {Pettini}
  M.,  {Olszewski} E.~W.,   {McQuinn} K. B.~W.,  2017, \mn@doi [\mnras]
  {10.1093/mnras/stw2962}, \href
  {https://ui.adsabs.harvard.edu/abs/2017MNRAS.465.3977J} {465, 3977}

\bibitem[\protect\citeauthoryear{{Jethwa}, {Erkal}  \& {Belokurov}}{{Jethwa}
  et~al.}{2018}]{Jethwa.etal.2018}
{Jethwa} P.,  {Erkal} D.,   {Belokurov} V.,  2018, \mn@doi [\mnras]
  {10.1093/mnras/stx2330}, \href
  {https://ui.adsabs.harvard.edu/abs/2018MNRAS.473.2060J} {473, 2060}

\bibitem[\protect\citeauthoryear{{Jimmy}, {Tran}, {Saintonge}, {Accurso},
  {Brough}  \& {Oliva-Altamirano}}{{Jimmy} et~al.}{2015}]{Jimmy.etal.2015}
{Jimmy} {Tran} K.-V.,  {Saintonge} A.,  {Accurso} G.,  {Brough} S.,
  {Oliva-Altamirano} P.,  2015, \mn@doi [\apj] {10.1088/0004-637X/812/2/98},
  \href {https://ui.adsabs.harvard.edu/abs/2015ApJ...812...98J} {812, 98}

\bibitem[\protect\citeauthoryear{Jones, Oliphant, Peterson  et~al.}{Jones
  et~al.}{01  }]{scipy}
Jones E.,  Oliphant T.,  Peterson P.,   et~al., 2001--, {SciPy}: Open source
  scientific tools for {Python}, \url {http://www.scipy.org/}

\bibitem[\protect\citeauthoryear{{Katz} et~al.,}{{Katz}
  et~al.}{2020}]{Katz.etal.2020}
{Katz} H.,  et~al., 2020, \mn@doi [\mnras] {10.1093/mnras/staa639}, \href
  {https://ui.adsabs.harvard.edu/abs/2020MNRAS.494.2200K} {494, 2200}

\bibitem[\protect\citeauthoryear{{Kauffmann}, {White}  \&
  {Guiderdoni}}{{Kauffmann} et~al.}{1993}]{Kauffmann.etal.1993}
{Kauffmann} G.,  {White} S.~D.~M.,   {Guiderdoni} B.,  1993, \mn@doi [\mnras]
  {10.1093/mnras/264.1.201}, \href
  {https://ui.adsabs.harvard.edu/abs/1993MNRAS.264..201K} {264, 201}

\bibitem[\protect\citeauthoryear{{Kennicutt} \& {Evans}}{{Kennicutt} \&
  {Evans}}{2012}]{Kennicutt.Evans.2012}
{Kennicutt} R.~C.,  {Evans} N.~J.,  2012, \mn@doi [\araa]
  {10.1146/annurev-astro-081811-125610}, \href
  {https://ui.adsabs.harvard.edu/abs/2012ARA&A..50..531K} {50, 531}

\bibitem[\protect\citeauthoryear{{Kere{\v{s}}}, {Katz}, {Weinberg}  \&
  {Dav{\'e}}}{{Kere{\v{s}}} et~al.}{2005}]{Keres.etal.2005}
{Kere{\v{s}}} D.,  {Katz} N.,  {Weinberg} D.~H.,   {Dav{\'e}} R.,  2005,
  \mn@doi [\mnras] {10.1111/j.1365-2966.2005.09451.x}, \href
  {https://ui.adsabs.harvard.edu/abs/2005MNRAS.363....2K} {363, 2}

\bibitem[\protect\citeauthoryear{{Kere{\v{s}}}, {Katz}, {Fardal}, {Dav{\'e}}
  \& {Weinberg}}{{Kere{\v{s}}} et~al.}{2009}]{Keres.etal.2009}
{Kere{\v{s}}} D.,  {Katz} N.,  {Fardal} M.,  {Dav{\'e}} R.,   {Weinberg} D.~H.,
   2009, \mn@doi [\mnras] {10.1111/j.1365-2966.2009.14541.x}, \href
  {https://ui.adsabs.harvard.edu/abs/2009MNRAS.395..160K} {395, 160}

\bibitem[\protect\citeauthoryear{{Kirby}, {Cohen}  \& {Bellazzini}}{{Kirby}
  et~al.}{2012}]{Kirby.etal.2012}
{Kirby} E.~N.,  {Cohen} J.~G.,   {Bellazzini} M.,  2012, \mn@doi [\apj]
  {10.1088/0004-637X/751/1/46}, \href
  {https://ui.adsabs.harvard.edu/abs/2012ApJ...751...46K} {751, 46}

\bibitem[\protect\citeauthoryear{{Kirby}, {Cohen}, {Guhathakurta}, {Cheng},
  {Bullock}  \& {Gallazzi}}{{Kirby} et~al.}{2013}]{Kirby.etal.2013}
{Kirby} E.~N.,  {Cohen} J.~G.,  {Guhathakurta} P.,  {Cheng} L.,  {Bullock}
  J.~S.,   {Gallazzi} A.,  2013, \mn@doi [\apj] {10.1088/0004-637X/779/2/102},
  \href {https://ui.adsabs.harvard.edu/abs/2013ApJ...779..102K} {779, 102}

\bibitem[\protect\citeauthoryear{{Koposov}, {Belokurov}, {Torrealba}  \&
  {Evans}}{{Koposov} et~al.}{2015}]{Koposov.etal.2015}
{Koposov} S.~E.,  {Belokurov} V.,  {Torrealba} G.,   {Evans} N.~W.,  2015,
  \mn@doi [\apj] {10.1088/0004-637X/805/2/130}, \href
  {https://ui.adsabs.harvard.edu/abs/2015ApJ...805..130K} {805, 130}

\bibitem[\protect\citeauthoryear{{Kravtsov}}{{Kravtsov}}{2010}]{Kravtsov.2010}
{Kravtsov} A.,  2010, \mn@doi [Advances in Astronomy] {10.1155/2010/281913},
  \href {https://ui.adsabs.harvard.edu/abs/2010AdAst2010E...8K} {2010, 281913}

\bibitem[\protect\citeauthoryear{{Kravtsov}}{{Kravtsov}}{2013}]{Kravtsov.2013}
{Kravtsov} A.~V.,  2013, \mn@doi [\apjl] {10.1088/2041-8205/764/2/L31}, \href
  {https://ui.adsabs.harvard.edu/abs/2013ApJ...764L..31K} {764, L31}

\bibitem[\protect\citeauthoryear{{Kravtsov}, {Berlind}, {Wechsler}, {Klypin},
  {Gottl{\"o}ber}, {Allgood}  \& {Primack}}{{Kravtsov}
  et~al.}{2004a}]{Kravtsov.etal.2004b}
{Kravtsov} A.~V.,  {Berlind} A.~A.,  {Wechsler} R.~H.,  {Klypin} A.~A.,
  {Gottl{\"o}ber} S.,  {Allgood} B.,   {Primack} J.~R.,  2004a, \mn@doi [\apj]
  {10.1086/420959}, \href
  {https://ui.adsabs.harvard.edu/abs/2004ApJ...609...35K} {609, 35}

\bibitem[\protect\citeauthoryear{{Kravtsov}, {Gnedin}  \& {Klypin}}{{Kravtsov}
  et~al.}{2004b}]{Kravtsov.etal.2004}
{Kravtsov} A.~V.,  {Gnedin} O.~Y.,   {Klypin} A.~A.,  2004b, \mn@doi [\apj]
  {10.1086/421322}, \href
  {https://ui.adsabs.harvard.edu/abs/2004ApJ...609..482K} {609, 482}

\bibitem[\protect\citeauthoryear{{Kravtsov}, {Vikhlinin}  \&
  {Meshcheryakov}}{{Kravtsov} et~al.}{2018}]{Kravtsov.2018}
{Kravtsov} A.~V.,  {Vikhlinin} A.~A.,   {Meshcheryakov} A.~V.,  2018, \mn@doi
  [Astronomy Letters] {10.1134/S1063773717120015}, \href
  {https://ui.adsabs.harvard.edu/abs/2018AstL...44....8K} {44, 8}

\bibitem[\protect\citeauthoryear{{Krumholz}}{{Krumholz}}{2013}]{Krumholz.2013}
{Krumholz} M.~R.,  2013, \mn@doi [\mnras] {10.1093/mnras/stt1780}, \href
  {https://ui.adsabs.harvard.edu/abs/2013MNRAS.436.2747K} {436, 2747}

\bibitem[\protect\citeauthoryear{{Krumholz} \& {Dekel}}{{Krumholz} \&
  {Dekel}}{2012}]{Krumholz.Dekel.2012}
{Krumholz} M.~R.,  {Dekel} A.,  2012, \mn@doi [\apj]
  {10.1088/0004-637X/753/1/16}, \href
  {https://ui.adsabs.harvard.edu/abs/2012ApJ...753...16K} {753, 16}

\bibitem[\protect\citeauthoryear{{Laevens} et~al.,}{{Laevens}
  et~al.}{2015}]{Laevens.etal.2015}
{Laevens} B. P.~M.,  et~al., 2015, \mn@doi [\apj] {10.1088/0004-637X/813/1/44},
  \href {https://ui.adsabs.harvard.edu/abs/2015ApJ...813...44L} {813, 44}

\bibitem[\protect\citeauthoryear{{Lahav}, {Lilje}, {Primack}  \&
  {Rees}}{{Lahav} et~al.}{1991}]{Lahav.etal.1991}
{Lahav} O.,  {Lilje} P.~B.,  {Primack} J.~R.,   {Rees} M.~J.,  1991, \mn@doi
  [\mnras] {10.1093/mnras/251.1.128}, \href
  {https://ui.adsabs.harvard.edu/abs/1991MNRAS.251..128L} {251, 128}

\bibitem[\protect\citeauthoryear{{Ledinauskas} \& {Zubovas}}{{Ledinauskas} \&
  {Zubovas}}{2018}]{Ledinauskas.Zubovas.2018}
{Ledinauskas} E.,  {Zubovas} K.,  2018, \mn@doi [\aap]
  {10.1051/0004-6361/201832824}, \href
  {https://ui.adsabs.harvard.edu/abs/2018A&A...615A..64L} {615, A64}

\bibitem[\protect\citeauthoryear{{Ledinauskas} \& {Zubovas}}{{Ledinauskas} \&
  {Zubovas}}{2020}]{Ledinauskas.Zubovas.2020}
{Ledinauskas} E.,  {Zubovas} K.,  2020, \mn@doi [\mnras]
  {10.1093/mnras/staa298}, \href
  {https://ui.adsabs.harvard.edu/abs/2020MNRAS.493..638L} {493, 638}

\bibitem[\protect\citeauthoryear{{Lee}, {Skillman}, {Cannon}, {Jackson},
  {Gehrz}, {Polomski}  \& {Woodward}}{{Lee} et~al.}{2006}]{Lee.etal.2006}
{Lee} H.,  {Skillman} E.~D.,  {Cannon} J.~M.,  {Jackson} D.~C.,  {Gehrz} R.~D.,
   {Polomski} E.~F.,   {Woodward} C.~E.,  2006, \mn@doi [\apj]
  {10.1086/505573}, \href
  {https://ui.adsabs.harvard.edu/abs/2006ApJ...647..970L} {647, 970}

\bibitem[\protect\citeauthoryear{{Lee} et~al.,}{{Lee}
  et~al.}{2009}]{Lee.etal.2009}
{Lee} J.~C.,  et~al., 2009, \mn@doi [\apj] {10.1088/0004-637X/706/1/599}, \href
  {https://ui.adsabs.harvard.edu/abs/2009ApJ...706..599L} {706, 599}

\bibitem[\protect\citeauthoryear{{Leitner} \& {Kravtsov}}{{Leitner} \&
  {Kravtsov}}{2011}]{Leitner.Kravtsov.2011}
{Leitner} S.~N.,  {Kravtsov} A.~V.,  2011, \mn@doi [\apj]
  {10.1088/0004-637X/734/1/48}, \href
  {https://ui.adsabs.harvard.edu/abs/2011ApJ...734...48L} {734, 48}

\bibitem[\protect\citeauthoryear{{Leroy} et~al.,}{{Leroy}
  et~al.}{2013}]{Leroy.etal.2013}
{Leroy} A.~K.,  et~al., 2013, \mn@doi [\aj] {10.1088/0004-6256/146/2/19}, \href
  {https://ui.adsabs.harvard.edu/abs/2013AJ....146...19L} {146, 19}

\bibitem[\protect\citeauthoryear{{Lilly}, {Carollo}, {Pipino}, {Renzini}  \&
  {Peng}}{{Lilly} et~al.}{2013}]{Lilly.etal.2013}
{Lilly} S.~J.,  {Carollo} C.~M.,  {Pipino} A.,  {Renzini} A.,   {Peng} Y.,
  2013, \mn@doi [\apj] {10.1088/0004-637X/772/2/119}, \href
  {https://ui.adsabs.harvard.edu/abs/2013ApJ...772..119L} {772, 119}

\bibitem[\protect\citeauthoryear{{Lodders}}{{Lodders}}{2019}]{Lodders.2019}
{Lodders} K.,  2019, arXiv e-prints, \href
  {https://ui.adsabs.harvard.edu/abs/2019arXiv191200844L} {p. arXiv:1912.00844}

\bibitem[\protect\citeauthoryear{{Lu}, {Mo}  \& {Wechsler}}{{Lu}
  et~al.}{2015a}]{Lu.etal.2015}
{Lu} Y.,  {Mo} H.~J.,   {Wechsler} R.~H.,  2015a, \mn@doi [\mnras]
  {10.1093/mnras/stu2215}, \href
  {https://ui.adsabs.harvard.edu/abs/2015MNRAS.446.1907L} {446, 1907}

\bibitem[\protect\citeauthoryear{{Lu}, {Blanc}  \& {Benson}}{{Lu}
  et~al.}{2015b}]{Lu.etal.2015b}
{Lu} Y.,  {Blanc} G.~A.,   {Benson} A.,  2015b, \mn@doi [\apj]
  {10.1088/0004-637X/808/2/129}, \href
  {https://ui.adsabs.harvard.edu/abs/2015ApJ...808..129L} {808, 129}

\bibitem[\protect\citeauthoryear{{Lu}, {Benson}, {Wetzel}, {Mao}, {Tonnesen},
  {Peter}, {Boylan-Kolchin}  \& {Wechsler}}{{Lu} et~al.}{2017}]{Lu.etal.2017}
{Lu} Y.,  {Benson} A.,  {Wetzel} A.,  {Mao} Y.-Y.,  {Tonnesen} S.,  {Peter} A.
  H.~G.,  {Boylan-Kolchin} M.,   {Wechsler} R.~H.,  2017, \mn@doi [\apj]
  {10.3847/1538-4357/aa845e}, \href
  {https://ui.adsabs.harvard.edu/abs/2017ApJ...846...66L} {846, 66}

\bibitem[\protect\citeauthoryear{{Mac Low} \& {Ferrara}}{{Mac Low} \&
  {Ferrara}}{1999}]{MacLow.Ferrara.1999}
{Mac Low} M.-M.,  {Ferrara} A.,  1999, \mn@doi [\apj] {10.1086/306832}, \href
  {https://ui.adsabs.harvard.edu/abs/1999ApJ...513..142M} {513, 142}

\bibitem[\protect\citeauthoryear{{Maloney}}{{Maloney}}{1993}]{Maloney.1993}
{Maloney} P.,  1993, \mn@doi [\apj] {10.1086/173055}, \href
  {https://ui.adsabs.harvard.edu/abs/1993ApJ...414...41M} {414, 41}

\bibitem[\protect\citeauthoryear{{Mao}, {Geha}, {Wechsler}, {Weiner},
  {Tollerud}, {Nadler}  \& {Kallivayalil}}{{Mao} et~al.}{2021}]{Mao.etal.2021}
{Mao} Y.-Y.,  {Geha} M.,  {Wechsler} R.~H.,  {Weiner} B.,  {Tollerud} E.~J.,
  {Nadler} E.~O.,   {Kallivayalil} N.,  2021, \mn@doi [\apj]
  {10.3847/1538-4357/abce58}, \href
  {https://ui.adsabs.harvard.edu/abs/2021ApJ...907...85M} {907, 85}

\bibitem[\protect\citeauthoryear{{McGaugh}, {Schombert}  \& {Lelli}}{{McGaugh}
  et~al.}{2017}]{McGaugh.etal.2017}
{McGaugh} S.~S.,  {Schombert} J.~M.,   {Lelli} F.,  2017, \mn@doi [\apj]
  {10.3847/1538-4357/aa9790}, \href
  {https://ui.adsabs.harvard.edu/abs/2017ApJ...851...22M} {851, 22}

\bibitem[\protect\citeauthoryear{{McQuinn}, {Skillman}, {Dolphin}  \&
  {Mitchell}}{{McQuinn} et~al.}{2015a}]{McQuinn.etal.2015b}
{McQuinn} K. B.~W.,  {Skillman} E.~D.,  {Dolphin} A.~E.,   {Mitchell} N.~P.,
  2015a, \mn@doi [\apj] {10.1088/0004-637X/808/2/109}, \href
  {https://ui.adsabs.harvard.edu/abs/2015ApJ...808..109M} {808, 109}

\bibitem[\protect\citeauthoryear{{McQuinn} et~al.,}{{McQuinn}
  et~al.}{2015b}]{McQuinn.etal.2015}
{McQuinn} K. B.~W.,  et~al., 2015b, \mn@doi [\apjl]
  {10.1088/2041-8205/815/2/L17}, \href
  {https://ui.adsabs.harvard.edu/abs/2015ApJ...815L..17M} {815, L17}

\bibitem[\protect\citeauthoryear{{McQuinn}, {van Zee}  \& {Skillman}}{{McQuinn}
  et~al.}{2019}]{McQuinn.etal.2019}
{McQuinn} K. B.~W.,  {van Zee} L.,   {Skillman} E.~D.,  2019, \mn@doi [\apj]
  {10.3847/1538-4357/ab4c37}, \href
  {https://ui.adsabs.harvard.edu/abs/2019ApJ...886...74M} {886, 74}

\bibitem[\protect\citeauthoryear{{McQuinn} et~al.,}{{McQuinn}
  et~al.}{2020}]{McQuinn.etal.2020}
{McQuinn} K. B.~W.,  et~al., 2020, \mn@doi [\apj] {10.3847/1538-4357/ab7447},
  \href {https://ui.adsabs.harvard.edu/abs/2020ApJ...891..181M} {891, 181}

\bibitem[\protect\citeauthoryear{{Mina}, {Shen}, {Keller}, {Mayer}, {Madau}  \&
  {Wadsley}}{{Mina} et~al.}{2021}]{Mina.etal.2020}
{Mina} M.,  {Shen} S.,  {Keller} B.~W.,  {Mayer} L.,  {Madau} P.,   {Wadsley}
  J.,  2021, \mn@doi [\aap] {10.1051/0004-6361/202039420}, \href
  {https://ui.adsabs.harvard.edu/abs/2021A&A...655A..22M} {655, A22}

\bibitem[\protect\citeauthoryear{{Mirocha}}{{Mirocha}}{2020}]{Mirocha.2020}
{Mirocha} J.,  2020, \mn@doi [\mnras] {10.1093/mnras/staa3150}, \href
  {https://ui.adsabs.harvard.edu/abs/2020MNRAS.499.4534M} {499, 4534}

\bibitem[\protect\citeauthoryear{{Mitchell}, {Schaye}, {Bower}  \&
  {Crain}}{{Mitchell} et~al.}{2020}]{Mitchell.etal.2020}
{Mitchell} P.~D.,  {Schaye} J.,  {Bower} R.~G.,   {Crain} R.~A.,  2020, \mn@doi
  [\mnras] {10.1093/mnras/staa938}, \href
  {https://ui.adsabs.harvard.edu/abs/2020MNRAS.494.3971M} {494, 3971}

\bibitem[\protect\citeauthoryear{{Mo} \& {Mao}}{{Mo} \&
  {Mao}}{2002}]{Mo.Mao.2002}
{Mo} H.~J.,  {Mao} S.,  2002, \mn@doi [\mnras]
  {10.1046/j.1365-8711.2002.05416.x}, \href
  {https://ui.adsabs.harvard.edu/abs/2002MNRAS.333..768M} {333, 768}

\bibitem[\protect\citeauthoryear{{Mo} \& {Mao}}{{Mo} \&
  {Mao}}{2004}]{Mo.Mao.2004}
{Mo} H.~J.,  {Mao} S.,  2004, \mn@doi [\mnras]
  {10.1111/j.1365-2966.2004.08114.x}, \href
  {https://ui.adsabs.harvard.edu/abs/2004MNRAS.353..829M} {353, 829}

\bibitem[\protect\citeauthoryear{{Mo}, {Mao}  \& {White}}{{Mo}
  et~al.}{1998}]{Mo.etal.1998}
{Mo} H.~J.,  {Mao} S.,   {White} S. D.~M.,  1998, \mn@doi [\mnras]
  {10.1046/j.1365-8711.1998.01227.x}, \href
  {https://ui.adsabs.harvard.edu/abs/1998MNRAS.295..319M} {295, 319}

\bibitem[\protect\citeauthoryear{{Munshi}, {Brooks}, {Christensen},
  {Applebaum}, {Holley-Bockelmann}, {Quinn}  \& {Wadsley}}{{Munshi}
  et~al.}{2019}]{Munshi.etal.2019}
{Munshi} F.,  {Brooks} A.~M.,  {Christensen} C.,  {Applebaum} E.,
  {Holley-Bockelmann} K.,  {Quinn} T.~R.,   {Wadsley} J.,  2019, \mn@doi [\apj]
  {10.3847/1538-4357/ab0085}, \href
  {https://ui.adsabs.harvard.edu/abs/2019ApJ...874...40M} {874, 40}

\bibitem[\protect\citeauthoryear{{Munshi}, {Brooks}, {Applebaum},
  {Christensen}, {Sligh}  \& {Quinn}}{{Munshi} et~al.}{2021}]{Munshi.etal.2021}
{Munshi} F.,  {Brooks} A.,  {Applebaum} E.,  {Christensen} C.,  {Sligh} J.~P.,
   {Quinn} T.,  2021, arXiv e-prints, \href
  {https://ui.adsabs.harvard.edu/abs/2021arXiv210105822M} {p. arXiv:2101.05822}

\bibitem[\protect\citeauthoryear{{Muratov}, {Kere{\v{s}}},
  {Faucher-Gigu{\`e}re}, {Hopkins}, {Quataert}  \& {Murray}}{{Muratov}
  et~al.}{2015}]{Muratov.etal.2015}
{Muratov} A.~L.,  {Kere{\v{s}}} D.,  {Faucher-Gigu{\`e}re} C.-A.,  {Hopkins}
  P.~F.,  {Quataert} E.,   {Murray} N.,  2015, \mn@doi [\mnras]
  {10.1093/mnras/stv2126}, \href
  {https://ui.adsabs.harvard.edu/abs/2015MNRAS.454.2691M} {454, 2691}

\bibitem[\protect\citeauthoryear{{Muratov} et~al.,}{{Muratov}
  et~al.}{2017}]{Muratov.etal.2017}
{Muratov} A.~L.,  et~al., 2017, \mn@doi [\mnras] {10.1093/mnras/stx667}, \href
  {https://ui.adsabs.harvard.edu/abs/2017MNRAS.468.4170M} {468, 4170}

\bibitem[\protect\citeauthoryear{{Naab} \& {Ostriker}}{{Naab} \&
  {Ostriker}}{2017}]{Naab.Ostriker.2017}
{Naab} T.,  {Ostriker} J.~P.,  2017, \mn@doi [\araa]
  {10.1146/annurev-astro-081913-040019}, \href
  {https://ui.adsabs.harvard.edu/abs/2017ARA&A..55...59N} {55, 59}

\bibitem[\protect\citeauthoryear{{Nadler} et~al.,}{{Nadler}
  et~al.}{2020}]{Nadler.etal.2020}
{Nadler} E.~O.,  et~al., 2020, \mn@doi [\apj] {10.3847/1538-4357/ab846a}, \href
  {https://ui.adsabs.harvard.edu/abs/2020ApJ...893...48N} {893, 48}

\bibitem[\protect\citeauthoryear{{Naoz}, {Yoshida}  \& {Gnedin}}{{Naoz}
  et~al.}{2013}]{Naoz.etal.2013}
{Naoz} S.,  {Yoshida} N.,   {Gnedin} N.~Y.,  2013, \mn@doi [\apj]
  {10.1088/0004-637X/763/1/27}, \href
  {https://ui.adsabs.harvard.edu/abs/2013ApJ...763...27N} {763, 27}

\bibitem[\protect\citeauthoryear{{Noh} \& {McQuinn}}{{Noh} \&
  {McQuinn}}{2014}]{Noh.McQuinn.2014}
{Noh} Y.,  {McQuinn} M.,  2014, \mn@doi [\mnras] {10.1093/mnras/stu1412}, \href
  {https://ui.adsabs.harvard.edu/abs/2014MNRAS.444..503N} {444, 503}

\bibitem[\protect\citeauthoryear{{Ocvirk} et~al.,}{{Ocvirk}
  et~al.}{2016}]{Ocvirk.etal.2016}
{Ocvirk} P.,  et~al., 2016, \mn@doi [\mnras] {10.1093/mnras/stw2036}, \href
  {https://ui.adsabs.harvard.edu/abs/2016MNRAS.463.1462O} {463, 1462}

\bibitem[\protect\citeauthoryear{{Ocvirk} et~al.,}{{Ocvirk}
  et~al.}{2020}]{Ocvirk.etal.2020}
{Ocvirk} P.,  et~al., 2020, \mn@doi [\mnras] {10.1093/mnras/staa1266}, \href
  {https://ui.adsabs.harvard.edu/abs/2020MNRAS.496.4087O} {496, 4087}

\bibitem[\protect\citeauthoryear{{Oh} \& {Benson}}{{Oh} \&
  {Benson}}{2003}]{Oh.Benson.2003}
{Oh} S.~P.,  {Benson} A.~J.,  2003, \mn@doi [\mnras]
  {10.1046/j.1365-8711.2003.06594.x}, \href
  {https://ui.adsabs.harvard.edu/abs/2003MNRAS.342..664O} {342, 664}

\bibitem[\protect\citeauthoryear{{Okamoto}, {Gao}  \& {Theuns}}{{Okamoto}
  et~al.}{2008}]{Okamoto.etal.2008}
{Okamoto} T.,  {Gao} L.,   {Theuns} T.,  2008, \mn@doi [\mnras]
  {10.1111/j.1365-2966.2008.13830.x}, \href
  {https://ui.adsabs.harvard.edu/abs/2008MNRAS.390..920O} {390, 920}

\bibitem[\protect\citeauthoryear{{Orban}, {Gnedin}, {Weisz}, {Skillman},
  {Dolphin}  \& {Holtzman}}{{Orban} et~al.}{2008}]{Orban.etal.2008}
{Orban} C.,  {Gnedin} O.~Y.,  {Weisz} D.~R.,  {Skillman} E.~D.,  {Dolphin}
  A.~E.,   {Holtzman} J.~A.,  2008, \mn@doi [\apj] {10.1086/591496}, \href
  {https://ui.adsabs.harvard.edu/abs/2008ApJ...686.1030O} {686, 1030}

\bibitem[\protect\citeauthoryear{{Ott} et~al.,}{{Ott}
  et~al.}{2012}]{Ott.etal.2012}
{Ott} J.,  et~al., 2012, \mn@doi [\aj] {10.1088/0004-6256/144/4/123}, \href
  {https://ui.adsabs.harvard.edu/abs/2012AJ....144..123O} {144, 123}

\bibitem[\protect\citeauthoryear{{Pandya} et~al.,}{{Pandya}
  et~al.}{2020}]{Pandya.etal.2020}
{Pandya} V.,  et~al., 2020, \mn@doi [\apj] {10.3847/1538-4357/abc3c1}, \href
  {https://ui.adsabs.harvard.edu/abs/2020ApJ...905....4P} {905, 4}

\bibitem[\protect\citeauthoryear{{Pandya} et~al.,}{{Pandya}
  et~al.}{2021}]{Pandya.etal.2021}
{Pandya} V.,  et~al., 2021, arXiv e-prints, \href
  {https://ui.adsabs.harvard.edu/abs/2021arXiv210306891P} {p. arXiv:2103.06891}

\bibitem[\protect\citeauthoryear{{Peeples} \& {Shankar}}{{Peeples} \&
  {Shankar}}{2011}]{Peeples.Shankar.2011}
{Peeples} M.~S.,  {Shankar} F.,  2011, \mn@doi [\mnras]
  {10.1111/j.1365-2966.2011.19456.x}, \href
  {https://ui.adsabs.harvard.edu/abs/2011MNRAS.417.2962P} {417, 2962}

\bibitem[\protect\citeauthoryear{{Peng} \& {Maiolino}}{{Peng} \&
  {Maiolino}}{2014}]{Peng.Maiolino.2014}
{Peng} Y.-j.,  {Maiolino} R.,  2014, \mn@doi [\mnras] {10.1093/mnras/stu1288},
  \href {https://ui.adsabs.harvard.edu/abs/2014MNRAS.443.3643P} {443, 3643}

\bibitem[\protect\citeauthoryear{{Pontzen} \& {Governato}}{{Pontzen} \&
  {Governato}}{2012}]{Pontzen.Governato.2012}
{Pontzen} A.,  {Governato} F.,  2012, \mn@doi [\mnras]
  {10.1111/j.1365-2966.2012.20571.x}, \href
  {https://ui.adsabs.harvard.edu/abs/2012MNRAS.421.3464P} {421, 3464}

\bibitem[\protect\citeauthoryear{{Purcell}, {Bullock}  \& {Zentner}}{{Purcell}
  et~al.}{2007}]{Purcell.etal.2007}
{Purcell} C.~W.,  {Bullock} J.~S.,   {Zentner} A.~R.,  2007, \mn@doi [\apj]
  {10.1086/519787}, \href
  {https://ui.adsabs.harvard.edu/abs/2007ApJ...666...20P} {666, 20}

\bibitem[\protect\citeauthoryear{{Rahman} et~al.,}{{Rahman}
  et~al.}{2012}]{Rahman.etal.2012}
{Rahman} N.,  et~al., 2012, \mn@doi [\apj] {10.1088/0004-637X/745/2/183}, \href
  {https://ui.adsabs.harvard.edu/abs/2012ApJ...745..183R} {745, 183}

\bibitem[\protect\citeauthoryear{{Read} \& {Erkal}}{{Read} \&
  {Erkal}}{2019}]{Read.Erkal.2019}
{Read} J.~I.,  {Erkal} D.,  2019, \mn@doi [\mnras] {10.1093/mnras/stz1320},
  \href {https://ui.adsabs.harvard.edu/abs/2019MNRAS.487.5799R} {487, 5799}

\bibitem[\protect\citeauthoryear{{Read}, {Iorio}, {Agertz}  \&
  {Fraternali}}{{Read} et~al.}{2017}]{Read.etal.2017}
{Read} J.~I.,  {Iorio} G.,  {Agertz} O.,   {Fraternali} F.,  2017, \mn@doi
  [\mnras] {10.1093/mnras/stx147}, \href
  {https://ui.adsabs.harvard.edu/abs/2017MNRAS.467.2019R} {467, 2019}

\bibitem[\protect\citeauthoryear{{Rey}, {Pontzen}, {Agertz}, {Orkney}, {Read},
  {Saintonge}  \& {Pedersen}}{{Rey} et~al.}{2019}]{Rey.etal.2019}
{Rey} M.~P.,  {Pontzen} A.,  {Agertz} O.,  {Orkney} M. D.~A.,  {Read} J.~I.,
  {Saintonge} A.,   {Pedersen} C.,  2019, \mn@doi [\apjl]
  {10.3847/2041-8213/ab53dd}, \href
  {https://ui.adsabs.harvard.edu/abs/2019ApJ...886L...3R} {886, L3}

\bibitem[\protect\citeauthoryear{{Ricotti} \& {Gnedin}}{{Ricotti} \&
  {Gnedin}}{2005}]{Ricotti.Gnedin.2005}
{Ricotti} M.,  {Gnedin} N.~Y.,  2005, \mn@doi [\apj] {10.1086/431415}, \href
  {https://ui.adsabs.harvard.edu/abs/2005ApJ...629..259R} {629, 259}

\bibitem[\protect\citeauthoryear{{Rodriguez Wimberly}, {Cooper}, {Fillingham},
  {Boylan-Kolchin}, {Bullock}  \& {Garrison-Kimmel}}{{Rodriguez Wimberly}
  et~al.}{2019}]{RodriguezWimberly.etal.2019}
{Rodriguez Wimberly} M.~K.,  {Cooper} M.~C.,  {Fillingham} S.~P.,
  {Boylan-Kolchin} M.,  {Bullock} J.~S.,   {Garrison-Kimmel} S.,  2019, \mn@doi
  [\mnras] {10.1093/mnras/sty3357}, \href
  {https://ui.adsabs.harvard.edu/abs/2019MNRAS.483.4031R} {483, 4031}

\bibitem[\protect\citeauthoryear{{Rosdahl} \& {Blaizot}}{{Rosdahl} \&
  {Blaizot}}{2012}]{Rosdahl.Blaizot.2012}
{Rosdahl} J.,  {Blaizot} J.,  2012, \mn@doi [\mnras]
  {10.1111/j.1365-2966.2012.20883.x}, \href
  {https://ui.adsabs.harvard.edu/abs/2012MNRAS.423..344R} {423, 344}

\bibitem[\protect\citeauthoryear{{Ryden} \& {Gunn}}{{Ryden} \&
  {Gunn}}{1987}]{Ryden.Gunn.1987}
{Ryden} B.~S.,  {Gunn} J.~E.,  1987, \mn@doi [\apj] {10.1086/165349}, \href
  {https://ui.adsabs.harvard.edu/abs/1987ApJ...318...15R} {318, 15}

\bibitem[\protect\citeauthoryear{{Saintonge} et~al.,}{{Saintonge}
  et~al.}{2011}]{Saintonge.etal.2011}
{Saintonge} A.,  et~al., 2011, \mn@doi [\mnras]
  {10.1111/j.1365-2966.2011.18823.x}, \href
  {https://ui.adsabs.harvard.edu/abs/2011MNRAS.415...61S} {415, 61}

\bibitem[\protect\citeauthoryear{{Saintonge} et~al.,}{{Saintonge}
  et~al.}{2016}]{Saintonge.etal.2016}
{Saintonge} A.,  et~al., 2016, \mn@doi [\mnras] {10.1093/mnras/stw1715}, \href
  {https://ui.adsabs.harvard.edu/abs/2016MNRAS.462.1749S} {462, 1749}

\bibitem[\protect\citeauthoryear{{Saintonge} et~al.,}{{Saintonge}
  et~al.}{2017}]{Saintonge.etal.2017}
{Saintonge} A.,  et~al., 2017, \mn@doi [\apjs] {10.3847/1538-4365/aa97e0},
  \href {https://ui.adsabs.harvard.edu/abs/2017ApJS..233...22S} {233, 22}

\bibitem[\protect\citeauthoryear{{Sales}, {Navarro}, {Schaye}, {Dalla Vecchia},
  {Springel}  \& {Booth}}{{Sales} et~al.}{2010}]{Sales.etal.2010}
{Sales} L.~V.,  {Navarro} J.~F.,  {Schaye} J.,  {Dalla Vecchia} C.,  {Springel}
  V.,   {Booth} C.~M.,  2010, \mn@doi [\mnras]
  {10.1111/j.1365-2966.2010.17391.x}, \href
  {https://ui.adsabs.harvard.edu/abs/2010MNRAS.409.1541S} {409, 1541}

\bibitem[\protect\citeauthoryear{{Sanders} et~al.,}{{Sanders}
  et~al.}{2021}]{Sanders.etal.2020}
{Sanders} R.~L.,  et~al., 2021, \mn@doi [\apj] {10.3847/1538-4357/abf4c1},
  \href {https://ui.adsabs.harvard.edu/abs/2021ApJ...914...19S} {914, 19}

\bibitem[\protect\citeauthoryear{{Sawala} et~al.,}{{Sawala}
  et~al.}{2015}]{Sawala.etal.2015}
{Sawala} T.,  et~al., 2015, \mn@doi [\mnras] {10.1093/mnras/stu2753}, \href
  {https://ui.adsabs.harvard.edu/abs/2015MNRAS.448.2941S} {448, 2941}

\bibitem[\protect\citeauthoryear{{Sawala} et~al.,}{{Sawala}
  et~al.}{2016}]{Sawala.etal.2016}
{Sawala} T.,  et~al., 2016, \mn@doi [\mnras] {10.1093/mnras/stv2597}, \href
  {https://ui.adsabs.harvard.edu/abs/2016MNRAS.456...85S} {456, 85}

\bibitem[\protect\citeauthoryear{{Scannapieco}, {Tissera}, {White}  \&
  {Springel}}{{Scannapieco} et~al.}{2008}]{Scannapieco.etal.2008}
{Scannapieco} C.,  {Tissera} P.~B.,  {White} S. D.~M.,   {Springel} V.,  2008,
  \mn@doi [\mnras] {10.1111/j.1365-2966.2008.13678.x}, \href
  {https://ui.adsabs.harvard.edu/abs/2008MNRAS.389.1137S} {389, 1137}

\bibitem[\protect\citeauthoryear{{Shen}, {Madau}, {Conroy}, {Governato}  \&
  {Mayer}}{{Shen} et~al.}{2014}]{Shen.etal.2014}
{Shen} S.,  {Madau} P.,  {Conroy} C.,  {Governato} F.,   {Mayer} L.,  2014,
  \mn@doi [\apj] {10.1088/0004-637X/792/2/99}, \href
  {https://ui.adsabs.harvard.edu/abs/2014ApJ...792...99S} {792, 99}

\bibitem[\protect\citeauthoryear{{Shibuya}, {Ouchi}  \& {Harikane}}{{Shibuya}
  et~al.}{2015}]{Shibuya.etal.2015}
{Shibuya} T.,  {Ouchi} M.,   {Harikane} Y.,  2015, \mn@doi [\apjs]
  {10.1088/0067-0049/219/2/15}, \href
  {https://ui.adsabs.harvard.edu/abs/2015ApJS..219...15S} {219, 15}

\bibitem[\protect\citeauthoryear{{Simon}}{{Simon}}{2019}]{simon19}
{Simon} J.~D.,  2019, \mn@doi [\araa] {10.1146/annurev-astro-091918-104453},
  \href {https://ui.adsabs.harvard.edu/abs/2019ARA&A..57..375S} {57, 375}

\bibitem[\protect\citeauthoryear{{Simpson}, {Grand}, {G{\'o}mez}, {Marinacci},
  {Pakmor}, {Springel}, {Campbell}  \& {Frenk}}{{Simpson}
  et~al.}{2018}]{Simpson.etal.2018}
{Simpson} C.~M.,  {Grand} R. J.~J.,  {G{\'o}mez} F.~A.,  {Marinacci} F.,
  {Pakmor} R.,  {Springel} V.,  {Campbell} D. J.~R.,   {Frenk} C.~S.,  2018,
  \mn@doi [\mnras] {10.1093/mnras/sty774}, \href
  {https://ui.adsabs.harvard.edu/abs/2018MNRAS.478..548S} {478, 548}

\bibitem[\protect\citeauthoryear{{Soko{\l}owska}, {Capelo}, {Fall}, {Mayer},
  {Shen}  \& {Bonoli}}{{Soko{\l}owska} et~al.}{2017}]{Sokolowska.etal.2017}
{Soko{\l}owska} A.,  {Capelo} P.~R.,  {Fall} S.~M.,  {Mayer} L.,  {Shen} S.,
  {Bonoli} S.,  2017, \mn@doi [\apj] {10.3847/1538-4357/835/2/289}, \href
  {https://ui.adsabs.harvard.edu/abs/2017ApJ...835..289S} {835, 289}

\bibitem[\protect\citeauthoryear{{Somerville}}{{Somerville}}{2002}]{Somerville.2002}
{Somerville} R.~S.,  2002, \mn@doi [\apjl] {10.1086/341444}, \href
  {https://ui.adsabs.harvard.edu/abs/2002ApJ...572L..23S} {572, L23}

\bibitem[\protect\citeauthoryear{{Somerville} \& {Dav{\'e}}}{{Somerville} \&
  {Dav{\'e}}}{2015}]{Somerville.Dave.2015}
{Somerville} R.~S.,  {Dav{\'e}} R.,  2015, \mn@doi [\araa]
  {10.1146/annurev-astro-082812-140951}, \href
  {https://ui.adsabs.harvard.edu/abs/2015ARA&A..53...51S} {53, 51}

\bibitem[\protect\citeauthoryear{{Somerville} et~al.,}{{Somerville}
  et~al.}{2018}]{Somerville.etal.2018}
{Somerville} R.~S.,  et~al., 2018, \mn@doi [\mnras] {10.1093/mnras/stx2040},
  \href {https://ui.adsabs.harvard.edu/abs/2018MNRAS.473.2714S} {473, 2714}

\bibitem[\protect\citeauthoryear{{Sparre}, {Hayward}, {Feldmann},
  {Faucher-Gigu{\`e}re}, {Muratov}, {Kere{\v{s}}}  \& {Hopkins}}{{Sparre}
  et~al.}{2017}]{Sparre.etal.2017}
{Sparre} M.,  {Hayward} C.~C.,  {Feldmann} R.,  {Faucher-Gigu{\`e}re} C.-A.,
  {Muratov} A.~L.,  {Kere{\v{s}}} D.,   {Hopkins} P.~F.,  2017, \mn@doi
  [\mnras] {10.1093/mnras/stw3011}, \href
  {https://ui.adsabs.harvard.edu/abs/2017MNRAS.466...88S} {466, 88}

\bibitem[\protect\citeauthoryear{{Springel}}{{Springel}}{2005}]{Springel.2005}
{Springel} V.,  2005, \mn@doi [\mnras] {10.1111/j.1365-2966.2005.09655.x},
  \href {https://ui.adsabs.harvard.edu/abs/2005MNRAS.364.1105S} {364, 1105}

\bibitem[\protect\citeauthoryear{{Starkenburg} et~al.,}{{Starkenburg}
  et~al.}{2013}]{starkenburg_etal13}
{Starkenburg} E.,  et~al., 2013, \mn@doi [\mnras] {10.1093/mnras/sts367}, \href
  {https://ui.adsabs.harvard.edu/abs/2013MNRAS.429..725S} {429, 725}

\bibitem[\protect\citeauthoryear{{Sunyaev}}{{Sunyaev}}{1969}]{Sunyaev.1969}
{Sunyaev} R.~A.,  1969, \aplett, \href
  {https://ui.adsabs.harvard.edu/abs/1969ApL.....3...33S} {3, 33}

\bibitem[\protect\citeauthoryear{{Tacchella}, {Forbes}  \&
  {Caplar}}{{Tacchella} et~al.}{2020}]{Tacchella.etal.2020}
{Tacchella} S.,  {Forbes} J.~C.,   {Caplar} N.,  2020, \mn@doi [\mnras]
  {10.1093/mnras/staa1838}, \href
  {https://ui.adsabs.harvard.edu/abs/2020MNRAS.497..698T} {497, 698}

\bibitem[\protect\citeauthoryear{{Tacconi}, {Genzel}  \& {Sternberg}}{{Tacconi}
  et~al.}{2020}]{Tacconi.etal.2020}
{Tacconi} L.~J.,  {Genzel} R.,   {Sternberg} A.,  2020, \mn@doi [\araa]
  {10.1146/annurev-astro-082812-141034}, \href
  {https://ui.adsabs.harvard.edu/abs/2020ARA&A..58..157T} {58, 157}

\bibitem[\protect\citeauthoryear{{Tarumi}, {Yoshida}  \& {Frebel}}{{Tarumi}
  et~al.}{2021}]{tarumi_etal21}
{Tarumi} Y.,  {Yoshida} N.,   {Frebel} A.,  2021, \mn@doi [\apjl]
  {10.3847/2041-8213/ac024e}, \href
  {https://ui.adsabs.harvard.edu/abs/2021ApJ...914L..10T} {914, L10}

\bibitem[\protect\citeauthoryear{{Tassis}, {Kravtsov}  \& {Gnedin}}{{Tassis}
  et~al.}{2008}]{Tassis.etal.2008}
{Tassis} K.,  {Kravtsov} A.~V.,   {Gnedin} N.~Y.,  2008, \mn@doi [\apj]
  {10.1086/523880}, \href
  {https://ui.adsabs.harvard.edu/abs/2008ApJ...672..888T} {672, 888}

\bibitem[\protect\citeauthoryear{{Tassis}, {Gnedin}  \& {Kravtsov}}{{Tassis}
  et~al.}{2012}]{Tassis.etal.2012}
{Tassis} K.,  {Gnedin} N.~Y.,   {Kravtsov} A.~V.,  2012, \mn@doi [\apj]
  {10.1088/0004-637X/745/1/68}, \href
  {https://ui.adsabs.harvard.edu/abs/2012ApJ...745...68T} {745, 68}

\bibitem[\protect\citeauthoryear{{Teich} et~al.,}{{Teich}
  et~al.}{2016}]{Teich.etal.2016}
{Teich} Y.~G.,  et~al., 2016, \mn@doi [\apj] {10.3847/0004-637X/832/1/85},
  \href {https://ui.adsabs.harvard.edu/abs/2016ApJ...832...85T} {832, 85}

\bibitem[\protect\citeauthoryear{{Torrealba} et~al.,}{{Torrealba}
  et~al.}{2019}]{Torrealba.etal.2019}
{Torrealba} G.,  et~al., 2019, \mn@doi [\mnras] {10.1093/mnras/stz1624}, \href
  {https://ui.adsabs.harvard.edu/abs/2019MNRAS.488.2743T} {488, 2743}

\bibitem[\protect\citeauthoryear{{Torrey} et~al.,}{{Torrey}
  et~al.}{2019}]{Torrey.etal.2019}
{Torrey} P.,  et~al., 2019, \mn@doi [\mnras] {10.1093/mnras/stz243}, \href
  {https://ui.adsabs.harvard.edu/abs/2019MNRAS.484.5587T} {484, 5587}

\bibitem[\protect\citeauthoryear{{Van Der Walt}, {Colbert}  \&
  {Varoquaux}}{{Van Der Walt} et~al.}{2011}]{numpy_ndarray}
{Van Der Walt} S.,  {Colbert} S.~C.,   {Varoquaux} G.,  2011, ArXiv:1102.1523,
  \href {http://adsabs.harvard.edu/abs/2011arXiv1102.1523V} {p.~1}

\bibitem[\protect\citeauthoryear{{Vincenzo}, {Matteucci}, {Belfiore}  \&
  {Maiolino}}{{Vincenzo} et~al.}{2016}]{Vincenzo.etal.2016}
{Vincenzo} F.,  {Matteucci} F.,  {Belfiore} F.,   {Maiolino} R.,  2016, \mn@doi
  [\mnras] {10.1093/mnras/stv2598}, \href
  {https://ui.adsabs.harvard.edu/abs/2016MNRAS.455.4183V} {455, 4183}

\bibitem[\protect\citeauthoryear{{Wang}, {Dutton}, {Stinson}, {Macci{\`o}},
  {Penzo}, {Kang}, {Keller}  \& {Wadsley}}{{Wang}
  et~al.}{2015}]{Wang.etal.2015}
{Wang} L.,  {Dutton} A.~A.,  {Stinson} G.~S.,  {Macci{\`o}} A.~V.,  {Penzo} C.,
   {Kang} X.,  {Keller} B.~W.,   {Wadsley} J.,  2015, \mn@doi [\mnras]
  {10.1093/mnras/stv1937}, \href
  {https://ui.adsabs.harvard.edu/abs/2015MNRAS.454...83W} {454, 83}

\bibitem[\protect\citeauthoryear{{Wang}, {Nadler}, {Mao}, {Adhikari},
  {Wechsler}  \& {Behroozi}}{{Wang} et~al.}{2021}]{Wang.etal.2021}
{Wang} Y.,  {Nadler} E.~O.,  {Mao} Y.-Y.,  {Adhikari} S.,  {Wechsler} R.~H.,
  {Behroozi} P.,  2021, \mn@doi [\apj] {10.3847/1538-4357/ac024a}, \href
  {https://ui.adsabs.harvard.edu/abs/2021ApJ...915..116W} {915, 116}

\bibitem[\protect\citeauthoryear{{Webster}, {Sutherland}  \&
  {Bland-Hawthorn}}{{Webster} et~al.}{2014}]{Webster.etal.2014}
{Webster} D.,  {Sutherland} R.,   {Bland-Hawthorn} J.,  2014, \mn@doi [\apj]
  {10.1088/0004-637X/796/1/11}, \href
  {https://ui.adsabs.harvard.edu/abs/2014ApJ...796...11W} {796, 11}

\bibitem[\protect\citeauthoryear{{Wechsler} \& {Tinker}}{{Wechsler} \&
  {Tinker}}{2018}]{Wechsler.Tinker.2018}
{Wechsler} R.~H.,  {Tinker} J.~L.,  2018, \mn@doi [\araa]
  {10.1146/annurev-astro-081817-051756}, \href
  {https://ui.adsabs.harvard.edu/abs/2018ARA&A..56..435W} {56, 435}

\bibitem[\protect\citeauthoryear{{Weisz} et~al.,}{{Weisz}
  et~al.}{2011}]{Weisz.etal.2011}
{Weisz} D.~R.,  et~al., 2011, \mn@doi [\apj] {10.1088/0004-637X/739/1/5}, \href
  {https://ui.adsabs.harvard.edu/abs/2011ApJ...739....5W} {739, 5}

\bibitem[\protect\citeauthoryear{{Weisz}, {Dolphin}, {Skillman}, {Holtzman},
  {Gilbert}, {Dalcanton}  \& {Williams}}{{Weisz} et~al.}{2014a}]{weisz_etal14}
{Weisz} D.~R.,  {Dolphin} A.~E.,  {Skillman} E.~D.,  {Holtzman} J.,  {Gilbert}
  K.~M.,  {Dalcanton} J.~J.,   {Williams} B.~F.,  2014a, \mn@doi [\apj]
  {10.1088/0004-637X/789/2/147}, \href
  {https://ui.adsabs.harvard.edu/abs/2014ApJ...789..147W} {789, 147}

\bibitem[\protect\citeauthoryear{{Weisz}, {Dolphin}, {Skillman}, {Holtzman},
  {Gilbert}, {Dalcanton}  \& {Williams}}{{Weisz}
  et~al.}{2014b}]{Weisz.etal.2014}
{Weisz} D.~R.,  {Dolphin} A.~E.,  {Skillman} E.~D.,  {Holtzman} J.,  {Gilbert}
  K.~M.,  {Dalcanton} J.~J.,   {Williams} B.~F.,  2014b, \mn@doi [\apj]
  {10.1088/0004-637X/789/2/147}, \href
  {https://ui.adsabs.harvard.edu/abs/2014ApJ...789..147W} {789, 147}

\bibitem[\protect\citeauthoryear{{Weisz} et~al.,}{{Weisz}
  et~al.}{2019}]{Weisz.etal.2019}
{Weisz} D.~R.,  et~al., 2019, \mn@doi [\apjl] {10.3847/2041-8213/ab4b52}, \href
  {https://ui.adsabs.harvard.edu/abs/2019ApJ...885L...8W} {885, L8}

\bibitem[\protect\citeauthoryear{{Wetzel}, {Hopkins}, {Kim},
  {Faucher-Gigu{\`e}re}, {Kere{\v{s}}}  \& {Quataert}}{{Wetzel}
  et~al.}{2016}]{Wetzel.etal.2016}
{Wetzel} A.~R.,  {Hopkins} P.~F.,  {Kim} J.-h.,  {Faucher-Gigu{\`e}re} C.-A.,
  {Kere{\v{s}}} D.,   {Quataert} E.,  2016, \mn@doi [\apjl]
  {10.3847/2041-8205/827/2/L23}, \href
  {https://ui.adsabs.harvard.edu/abs/2016ApJ...827L..23W} {827, L23}

\bibitem[\protect\citeauthoryear{{Wheeler} et~al.,}{{Wheeler}
  et~al.}{2019}]{Wheeler.etal.2019}
{Wheeler} C.,  et~al., 2019, \mn@doi [\mnras] {10.1093/mnras/stz2887}, \href
  {https://ui.adsabs.harvard.edu/abs/2019MNRAS.490.4447W} {490, 4447}

\bibitem[\protect\citeauthoryear{{Willman}, {Geha}, {Strader}, {Strigari},
  {Simon}, {Kirby}, {Ho}  \& {Warres}}{{Willman}
  et~al.}{2011}]{Willman.etal.2011}
{Willman} B.,  {Geha} M.,  {Strader} J.,  {Strigari} L.~E.,  {Simon} J.~D.,
  {Kirby} E.,  {Ho} N.,   {Warres} A.,  2011, \mn@doi [\aj]
  {10.1088/0004-6256/142/4/128}, \href
  {https://ui.adsabs.harvard.edu/abs/2011AJ....142..128W} {142, 128}

\bibitem[\protect\citeauthoryear{{Wise}, {Turk}, {Norman}  \& {Abel}}{{Wise}
  et~al.}{2012}]{Wise.etal.2012}
{Wise} J.~H.,  {Turk} M.~J.,  {Norman} M.~L.,   {Abel} T.,  2012, \mn@doi
  [\apj] {10.1088/0004-637X/745/1/50}, \href
  {https://ui.adsabs.harvard.edu/abs/2012ApJ...745...50W} {745, 50}

\bibitem[\protect\citeauthoryear{{Woo}, {Courteau}  \& {Dekel}}{{Woo}
  et~al.}{2008}]{woo_etal08}
{Woo} J.,  {Courteau} S.,   {Dekel} A.,  2008, \mn@doi [\mnras]
  {10.1111/j.1365-2966.2008.13770.x}, \href
  {https://ui.adsabs.harvard.edu/abs/2008MNRAS.390.1453W} {390, 1453}

\bibitem[\protect\citeauthoryear{{Xia} \& {Yu}}{{Xia} \&
  {Yu}}{2019}]{Xia.Quinjuan.2019}
{Xia} M.,  {Yu} Q.,  2019, \mn@doi [\apj] {10.3847/1538-4357/ab2628}, \href
  {https://ui.adsabs.harvard.edu/abs/2019ApJ...880....5X} {880, 5}

\bibitem[\protect\citeauthoryear{{Zavala}, {Okamoto}  \& {Frenk}}{{Zavala}
  et~al.}{2008}]{Zavala.etal.2008}
{Zavala} J.,  {Okamoto} T.,   {Frenk} C.~S.,  2008, \mn@doi [\mnras]
  {10.1111/j.1365-2966.2008.13243.x}, \href
  {https://ui.adsabs.harvard.edu/abs/2008MNRAS.387..364Z} {387, 364}

\bibitem[\protect\citeauthoryear{{Zhu}, {Avestruz}  \& {Gnedin}}{{Zhu}
  et~al.}{2019}]{Zhu.etal.2019}
{Zhu} H.,  {Avestruz} C.,   {Gnedin} N.~Y.,  2019, \mn@doi [\apj]
  {10.3847/1538-4357/ab3794}, \href
  {https://ui.adsabs.harvard.edu/abs/2019ApJ...882..152Z} {882, 152}

\bibitem[\protect\citeauthoryear{{van Loon}, {Mitchell}  \& {Schaye}}{{van
  Loon} et~al.}{2021}]{VanLoon.etal.2021}
{van Loon} M.~L.,  {Mitchell} P.~D.,   {Schaye} J.,  2021, \mn@doi [\mnras]
  {10.1093/mnras/stab1254}, \href
  {https://ui.adsabs.harvard.edu/abs/2021MNRAS.504.4817V} {504, 4817}

\bibitem[\protect\citeauthoryear{{von Steiger} \& {Zurbuchen}}{{von Steiger} \&
  {Zurbuchen}}{2016}]{VonSteiger.Zurbuchen.2016}
{von Steiger} R.,  {Zurbuchen} T.~H.,  2016, \mn@doi [\apj]
  {10.3847/0004-637X/816/1/13}, \href
  {https://ui.adsabs.harvard.edu/abs/2016ApJ...816...13V} {816, 13}

\makeatother
\end{thebibliography}



\appendix


\section{Approximation for the dependence of characteristic mass for UV suppression on $\zrei$}
\label{app:mc_oka}

\begin{figure}
    \centering
    \includegraphics[width=\columnwidth]{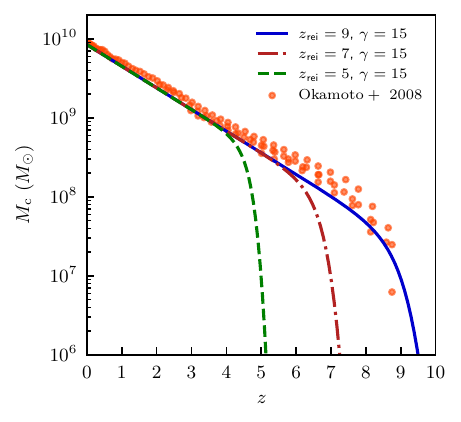}
    \caption{Evolution of the characteristic mass scale $M_c$ as a function of redshift. The data points shown are simulation results from Figure 3 in \citet{Okamoto.etal.2008}. The colored lines show the evolution of characteristic function for reionizaton redshifts $\zrei = 5,7,9$ according to our analytical parametrization in Equation~\ref{eq:Mc_zre}. }
    \label{fig:mc_oka}
\end{figure}

We use the simulations of \citet{Okamoto.etal.2008} to obtain the redshift evolution for characteristic mass scale  $M_{\rm c}$ at reionization redshift $\zrei = 9$. Figure~\ref{fig:mc_oka} shows the data from \citet{Okamoto.etal.2008} simulations over plotted with our approximating function (blue line) described in eq.~\ref{eqn:Mcz}. The plot is in log-linear scale and thus approximately linear behaviour of $\log M_{\rm c}$ with $z$ at $z\lesssim 8$ indicates that this dependence can be approximated by the exponential function. At redshifts $z\gtrsim \zrei=9$ $M_{\rm c}$ decreases sharply with increasing $z$. Note that the points include simulations at different resolutions, but the lowest points at high $z$ correspond to the highest resolution simulation, which our analytic equation aims to approximate. 

The characteristic mass scale at time of reionization ($M_{\rm c,rei}$) according to our approximation function (eq.~\ref{eqn:Mcz}) is 
$M_{\rm c,rei} = M_{\rm c}(z = 9) \approx 9.3 \times 10^{6}\, \Msol$.
The general expression for evolution for characteristic mass scale is given by eq.~\ref{eq:Mc_zre} where $\beta$ controls \textit{when} the sharp increase in characteristic mass scale occurs and $\gamma$ controls the sharpness of that increase. To be able to flexibly model the characteristic mass scale evolution for different reionization redshifts, we we apply the constraint that the corresponding characteristic mass scale must be equal to $M_{\rm c, rei}$ at the assumed $\zrei$. Furthermore, the characteristic mass scale must be equal to that seen in \citet{Okamoto.etal.2008} simulations at $z=0$. Keeping this sharpness parameter $\gamma$ fixed, the above constraints give $M_{\rm c, rei}=M_{\rm c}(\zrei)$ with $M_{\rm c}(z)$ given by eq.~\ref{eq:Mc_zre}. We can also use this equation to get the expression for $\beta$ (Equation~\ref{eq:betarei}) for a given $\zrei$ and $\gamma$. Figure~\ref{fig:mc_oka} shows the evolution of characteristic mass scale for different $\zrei$.

\section{Sensitivity of gas masses of dwarf galaxies to $\HH$ model}
\label{app:mhh2}

\begin{figure*}
    \centering
    \includegraphics[width=0.49\textwidth]{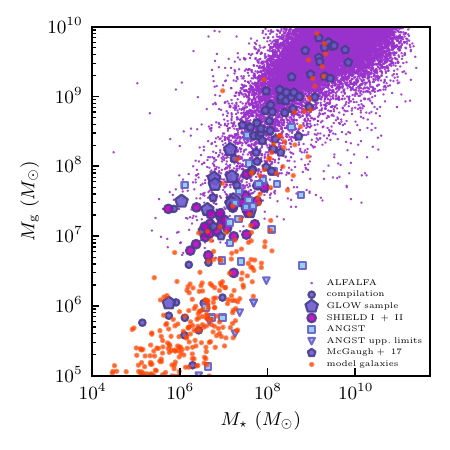}
        \includegraphics[width=0.49\textwidth]{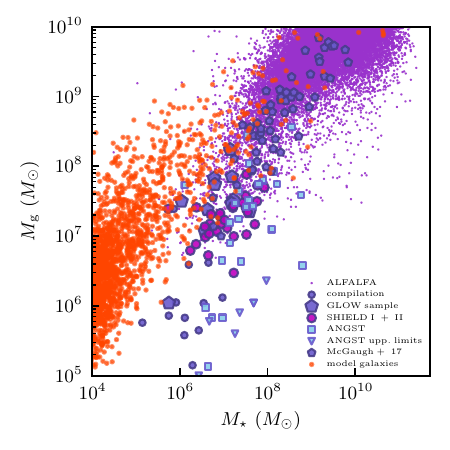}
    \caption{Comparison of the $\Ms-\Mg$ relation for model galaxies (red dots) to observations similar to that shown in Fig.~\ref{fig:msmgfid} for the fiducial model. The two panels shows two models with all parameters identical to the fiducial model, but with different models for molecular fraction $f_{\rm H_2}$: in the model shown in the left panel  the modified Gnedin-Draine model given by eqs.~8a-10a in \citet{Gnedin.Draine.2016} is used, while in the right panel the model described in Section 2 of \citet[][namely eqs 18-21 with $c=5$]{Krumholz.Dekel.2012}.  } 
    \label{fig:msmgh2}
\end{figure*}

Figure~\ref{fig:msmgh2} shows comparison of the $\Ms-\Mg$ relation for model galaxies (red dots) to observations similar to that shown in Fig.~\ref{fig:msmgfid} for the fiducial model. The two panels two models with all parameters identical to the fiducial model, but with different models for molecular fraction $f_{\rm H_2}$: the modified Gnedin-Draine model given by eqs.~8a-10a in \citet{Gnedin.Draine.2016} in the left panel, and the model described in Section 2 of \citet[][namely eqs 18-21 with $c=5$]{Krumholz.Dekel.2012}.

\begin{figure*}
    \centering
    \includegraphics[width=0.49\textwidth]{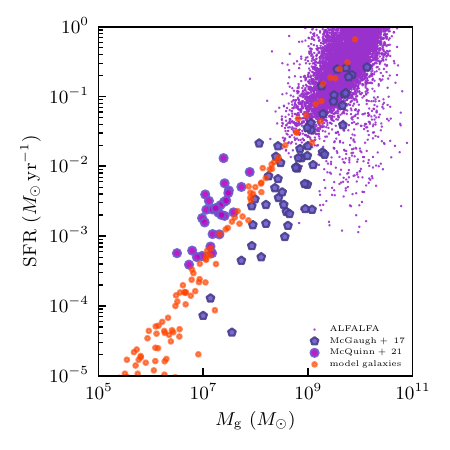}
        \includegraphics[width=0.49\textwidth]{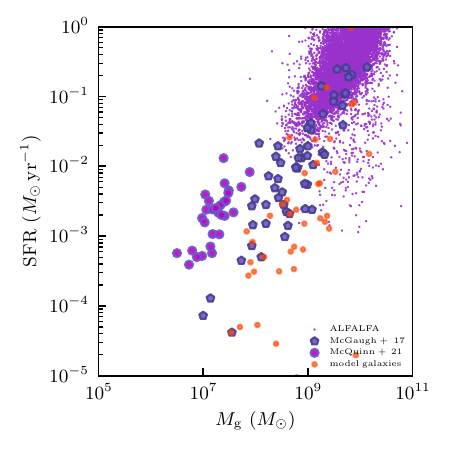}
    \caption{Comparison of the $\Mg-\sfr$ relation for model galaxies (red dots) to observations similar to that shown in Fig.~\ref{fig:msfrfid} for the fiducial model. The two panels shows two models with all parameters identical to the fiducial model, but with different models for molecular fraction $f_{\rm H_2}$: in the model shown in the left panel  the modified Gnedin-Draine model given by eqs.~8a-10a in \citet{Gnedin.Draine.2016} is used, while in the right panel the model described in Section 2 of \citet[][namely eqs 18-21 with $c=5$]{Krumholz.Dekel.2012}.  } 
    \label{fig:mgsfrh2}
\end{figure*}

Figure~\ref{fig:mgsfrh2} shows results of these models for the $\Mg-\sfr$ correlation. Compared 
to the fiducial model results shown in Figure~\ref{fig:msfrfid}, the star formation rates in the modified Gnedin-Draine model are larger compared to fiducial model, while SFRs in the KD12 model are systematically and considerably lower. The number of model galaxies is also noticeably smaller in the right panel because in this model a large fraction of dwarf galaxies does not have $\HH$ and thus no ongoing star formation. 

These figures show that gas masses and star formation rates in model galaxies of $\Ms\lesssim 10^9\,\Msun$ are very sensitive to the details of $f_{\HH}$ model. To be fair the model used KD12 is based on equilibrium assumption, which is not expected to be valid at low metallicities \citep[see][]{Krumholz.2013}. The latter paper discusses extensions of the model to dwarf galaxy regime, but at the expense of additional assumptions. Given the sensitivity of gas masses to the details of such model, it is not clear that predictions would be robust. 

The differences that can be seen in Figures~\ref{fig:msmgfid} and left panel of Figure~\ref{fig:msmgh2} for the models that use different fitting formulae \citep[][]{Gnedin.Draine.2016} to {\it the same} numerical simulation results of \citet{Gnedin.Draine.2014} show that even the current calibrations based on simulation results are likely not robust.

\bsp	
\label{lastpage}
\end{document}